\documentclass[times]{qjrms4}

\newcommand\BibTeX{{\rmfamily B\kern-.05em \textsc{i\kern-.025em b}\kern-.08em
T\kern-.1667em\lower.7ex\hbox{E}\kern-.125emX}}

\usepackage{moreverb}
\usepackage{natbib}
\usepackage{balance}
\usepackage[normalem]{ulem}
\usepackage{color}
\usepackage{subfig}
\usepackage{bigints}
\usepackage{amsmath}
\def\volumeyear{2018}
\def\volumenumber{144}
\begin{document}
%\linenumbers
\runningheads{P.~L.~Read et al.}{Atmospheric circulation regimes and energy budgets}

\title{Comparative terrestrial atmospheric circulation regimes in simplified global circulation models: II. energy budgets and spectral transfers}

\author{\corrauth P.~L.~Read\affil{a},   F. Tabataba-Vakili\affil{a,b}, Y. Wang\affil{a}, P. Augier\affil{c}, E. Lindborg\affil{d}, A. Valeanu\affil{a}, R. M. B. Young\affil{a,e} 
}

\address{\affilnum{a}Atmospheric, Oceanic \& Planetary Physics, Department of Physics, University of Oxford, Clarendon Laboratory, Parks Road, Oxford, OX1~3PU, UK \\
\affilnum{b} Now at Jet Propulsion Laboratory, Pasadena, USA \\
\affilnum{c} LEGI, BP 53, 38041 Grenoble Cedex 9, France \\
\affilnum{d} Linn\'e Flow Centre, KTH Mechanics, Stockholm, Sweden \\
\affilnum{e} Now at Laboratoire de M\'et\'eorologie Dynamique/Institut Pierre-Simon Laplace (LMD/IPSL), Sorbonne Universit\'es, UPMC Univ Paris 06, PSL Research University, \'Ecole Normale Sup\'erieure, Universit\'e Paris-Saclay, \'Ecole Polytechnique, Centre National de la Recherche Scientifique, 75005 Paris, France
}

\corraddr{Atmospheric, Oceanic \& Planetary Physics, Clarendon Laboratory, Parks Road, Oxford, OX1~3PU, UK.}

\begin{abstract}
The energetics of possible global atmospheric circulation patterns in an Earth-like atmosphere are explored using a simplified GCM based on the University of Hamburg's Portable University Model for the Atmosphere (designated here as PUMA-S), {forced by linear relaxation towards a prescribed temperature field and subject to Rayleigh surface drag and hyperdiffusive dissipation}. Results from a series of simulations, obtained by varying planetary rotation rate $\Omega$ with an imposed equator-to-pole temperature difference, were analysed to determine the structure and magnitude of the heat transport and other contributions to the energy budget for the time-averaged, equilibrated flow. These show clear trends with rotation rate, with the most intense Lorenz energy cycle for an Earth-sized planet occurring with a rotation rate around half that of the present day Earth (i.e. $\Omega^\ast = \Omega/\Omega_E = 1/2$, where $\Omega_E$ is the rotation rate of the Earth). KE and APE spectra, $E_K(n)$ and $E_A(n)$ (where $n$ is total spherical wavenumber), also show clear trends with rotation rate, with $n^{-3}$ enstrophy-dominated spectra around $\Omega^\ast =1$ and steeper ($\sim n^{-5}$) slopes in the zonal mean flow with little evidence for the $n^{-5/3}$ spectrum anticipated for an inverse KE cascade. Instead, both KE and APE spectra become almost flat at scales larger than the internal Rossby radius, $L_d$, and exhibit near-equipartition at high wavenumbers. At $\Omega^\ast << 1$, the spectrum becomes dominated by KE with $E_K(n) \sim {(2-3)} E_A(n)$ at most wavenumbers and a slope that tends towards $n^{-5/3}$ across most of the spectrum. Spectral flux calculations show that enstrophy and APE are almost always cascaded downscale, regardless of rotation rate. KE cascades are more complicated, however, with downscale transfers across almost all wavenumbers, dominated by horizontally divergent modes, for $\Omega^\ast \lesssim 1/4$. At higher rotation rates, transfers of KE become increasingly dominated by rotational (horizontally non-divergent) components with strong upscale transfers (dominated by eddy-zonal flow interactions) for scales larger than $L_d$ and weaker downscale transfers for scales smaller than $L_d$.
\end{abstract}

\keywords{Energy budget, General circulation model experiments, global, dynamics, atmosphere}

\maketitle

\section{Introduction} \label{sec:introduction}
%The efficiency by which heat is transferred from the warm tropical regions of an Earth-like planetary atmosphere to the cooler mid- and high-latitudes is an important characteristic of the global atmospheric circulation. It largely determines how the climate varies from place to place, and in particular sets the mean temperature contrast between the tropics and polar regions. For the Earth, how this efficiency depends upon key parameters relating to the thermodynamic driving of the atmospheric circulation is important for understanding and quantifying the response of the climate system e.g. to changing amounts of greenhouse gases in the atmosphere. For other planets, heat transfer efficiency may affect their potential habitability and long term evolution, while also strongly influencing the strength of the meridional and zonal winds. 
Atmospheric circulation can be said to occur because of the propensity of fluid motion to transfer heat energy from regions of net heating to regions of net cooling. 
Horizontal heat transfer is also part of the overall processing and conversion of energy within the atmospheric ``heat engine'', in which local imbalances between incoming radiant energy from the parent star and thermal emission tend to modulate the internal energy and increase the potential energy of the atmosphere. Dynamical processes then act to convert such potential energy into various bulk forms of motion in the form of kinetic energy before dissipative processes ultimately reconvert this back to heat again. Precisely how potential energy is transformed into kinetic energy depends strongly on the dynamical constraints governing atmospheric motion, and this in turn depends on a number of external factors, such as the planetary size, mass and rotation, the overall mass of the atmosphere and its composition. Heat, momentum and other tracers may be transported in latitude either by direct meridional overturning in an axisymmetric Hadley-type circulation, or via non-axisymmetric eddies through systematic covariances between meridional velocity and temperature fluctuations.

Heat transport, of course, represents only part of the overall cycle of energy conversion within a planetary atmosphere. %A more complete picture is provided by a diagnostic analysis of both the thermodynamic and bulk dynamical components of the atmospheric energetics, which can be achieved robustly given a full numerical model simulation that represents the full dynamically consistent set of atmospheric variables. 
A common approach to the analysis of energy conversions is the one based on the work of \citet{lorenz1955}, in which energy reservoirs and exchanges are partitioned between kinetic and (available) potential energy, and between zonally averaged and eddy components. Energy exchanges within the Earth's global circulation have been analysed in this way for many years \citep{Peixoto1974,li2007,boer2008}, although very few studies have examined the Lorenz energy cycle for other planets \citep[e.g.][]{lee2010,pascale2013,schubert2014,tabataba2015}. In the context of an exploration of how the global circulation regime changes within a simplified model atmosphere, it is of significant interest to examine how the cycle of energy conversions changes throughout parameter space. This is investigated in the present study for an Earth-like planet at various rotation rates, based on the set of simulations presented by \citet{wang2016} using the Hamburg PUMA model, and the results presented in Section \ref{sec:lorenz_heat}. 

The Lorenz approach provides insight into how the atmospheric heat engine transfers energy from the planetary scale, zonally-symmetric flow into non-axisymmetric "eddies". But this is only a crude measure of how energy passes from scales that are directly energised by solar heating and radiative cooling into other scales of motion, that takes little account of the macroturbulent processes that distribute energy from the forcing scales towards those affected by dissipation. In this context, the concept of geostrophic turbulence, first introduced by \cite{Charney1971}, has been an important paradigm for theories of the large-scale planetary atmospheric and oceanic circulations. 

The flow in geostrophic turbulence tends to be a highly chaotic, quasi-2D (horizontal), quasi-geostrophic flow, typically featuring an inverse energy cascade if small-scale forcing is present. Planetary rotation, large aspect ratio (between horizontal 
vertical scales) and statically stable stratification all act to bring planetary atmospheric flows into 
quasi-horizontal (quasi-2D) motion. Small-scale forcing is usually envisaged as being provided either by baroclinic 
instability occuring at scales comparable to the Rossby deformation radius ($L_D$) or by small-scale 
convection, as is possibly in the case of Jovian planets \citep[see e.g.][]{Ingersoll2000,Read2007,Read2015b}).
The energy generated through such processes then becomes a small-scale \lq\lq agitator\rq\rq\ of the inverse
energy cascade in the barotropic mode, though the precise mechanism for energising this mode is still not well
understood. It is a typical feature of 2-D isotropic (in a 2-D planar sense, or 
horizontally isotropic) turbulence that energy goes from small scale to large scale through a spectrally-local
inverse cascade. The direct consequence of such an inverse energy cascade is the emergence of large circular 
eddies with no preferred directionality. In the presence of a non-negligible background vorticity gradient 
(e.g. $\beta$-effect), however, it was shown by \cite{Rhines1975} that such large-scale eddies becomes 
anisotropic, causing an elongation of structures in the zonal direction and ultimately leading to the formation
of zonal jets. %The characteristic latitudinal length scale of these jets can be estimated by the Rhines scale, 
%as introduced in the previous section. 

% zonostrophic regime
Galperin and Sukoriansky (\cite{Sukoriansky2002}, \cite{Galperin2006}) recently proposed the paradigm of 
\emph{zonostrophic turbulence} as an attempt to characterise universally the regime of eddy-driven multiple zonal
jets on a $\beta$-plane. Under a strong $\beta$-effect, it is proposed that flows can develop into the regime of 
zonostrophic turbulence which is characterised by a strongly anisotropic KE spectrum with a steep ($-5$) slope 
for the zonally symmetric flow component and a classic Kolmogorov-Batchelor-Kraichnan (KBK) $-5/3$ slope in the 
non-axisymmetric eddy/residual modes. % (as shown in Fig. \ref{fig:zono-diagram}). 
The segments of the spectra 
in this regime take the universal form (when appropriately non-dimensionalised):
\begin{subequations}
 \begin{align}
  E_Z(n) =C_Z\beta^2n^{-5}, C_Z\sim 0.5, \label{eq:zono-z}\\
  E_R(n) = C_K\epsilon^{2/3}n^{-5/3}, C_K\sim 5, \label{eq:zono-r}
 \end{align}
\end{subequations}
where $\epsilon$ is the energy pumping rate of the small-scale excitation (which, in previous 2-D numerical
studies of zonostrophic turbulence, is represented as an artificial energy input at a specific wavenumber
$n_{\xi}$, see e.g. \cite{huang2001} and \cite{Galperin2004}. In a real planetary atmosphere, this can be due to
barotropic or baroclinic eddies, or in the case of gas giants, possibly from small-scale moist convection as 
well.). $C_K$ is the universal Kolmogorov-Kraichnan constant, while barotropic simulations \citep[e.g.][]{Chekhlov1996,huang2001} suggest that $C_Z$ can vary between $0.1$ and $1.0$.

Precisely how these and other regimes emerge and under what conditions has been largely unexplored in general until recently, leaving open many questions as to the nature of the circulation of various planets within and beyond our Solar System. In the present work, therefore, we analyse a set of numerical model simulations, obtained using a simplified global circulation model in which atmospheric flows in an Earth-like planetary atmosphere are driven by simple linear relaxation towards a prescribed (steady) zonally-symmetric temperature field (on a timescale $\tau_R$) and dissipated by a linear Rayleigh drag (with prescribed timescale $\tau_{fr}$. We vary various planetary parameters (especially the planetary rotation rate, but also the surface friction timescale) and allow the simulation to equilibrate over a timescale of order 20 Earth years. The basic model and the phenomenology of the circulation regimes were described in a companion paper \citep{wang2016}, which clearly demonstrated a systematic sequence as $\Omega$ was varied from $\Omega^\ast = 1/16$ to $\Omega^\ast = 8$. The regimes obtained ranged from a super-rotating, barotropically unstable cyclostrophic atmosphere at the lowest values of $\Omega^\ast$ to a highly geostrophically turbulent circulation with multiple zonal jets at $\Omega^\ast >> 1$ via more Earth-like, geostrophic states with simpler patterns of jets and baroclinic eddies that were either regular and periodic or chaotic in nature. 

In this paper, however, we focus on analysing the budgets of kinetic and potential energy and associated heat transport. We begin with an analysis of the global exchanges of energy within the well known framework of the Lorenz energy cycle, but then extend the analysis to consider the more detailed exchange of energy and enstrophy between different scales via the spectra of kinetic and available potential energy and the principal spectral fluxes as a function of spherical harmonic total wavenumber. The computation of spectral fluxes provides arguably the most detailed and precise means of evaluating the direction and intensity of turbulent cascades by directly computing the exchanges of various forms of energy and enstrophy between different scales, as represented in a decomposition of flow structure projected onto a spectrum of spherical harmonics. This approach has been applied for several years to studies of kinetic energy exchanges within the Earth's atmospheric circulation in numerical simulations and assimilated analyses \citep[e.g.][]{Boer1983,Shepherd1987,koshyk2001,burgess2013}, but relatively few such analyses have been extended to include potential energy exchanges and conversions \citep{lambert1984,augier2013,malardel2016}. They have proved able to provide important insights, however, into how the atmosphere transfers key properties between different scales through nonlinear interactions, and in particular the potential impacts of various parameterization schemes on the modelled cascades of energy and enstrophy \citep{malardel2016}. A similar approach has recently been applied to the Earth's oceans, at least on a local scale in the context of mesoscale eddies \citep{scott2005,scott2007}, and even to the kinetic energy budget of Jupiter's atmosphere \citep{young2017}, in which both systems reveal the existence of a double cascade (involving both up- and down-scale segments), energised on scales close to the Rossby deformation radius. 

Section \ref{sec:analysis} presents the framework for analysis of the budgets of kinetic and potential energy and the spectral transfers of energy and enstrophy. Results for the various terms in the Lorenz energy budget as a function of planetary rotation rate are presented and discussed in Section \ref{sec:lorenz_heat} while Section \ref{sec:turb-ke-spectra} provides an overview of trends in the spectra of kinetic and potential energy. The spectral fluxes of energy and enstrophy as a function of $\Omega^\ast$ are presented in Section \ref{sec:turb-fluxes} and the overall results are discussed in Section \ref{sec:conclusions}.

{
\section{Model setup and experiment design} \label{sec:setup}
The model used is the Portable University Model of the Atmosphere
(PUMA; e.g. see \citet{fraedrich1998,frisius1998,vonhardenberg2000}), consisting of a spectral dynamical core solving the dry primitive equations on a sphere, based on the code developed by \citet{hoskins1975}. Temperature, divergence, vorticity, and $\ln p_s$ (where $p_s$ is the surface pressure) are the prognostic variables. The model domain uses finite difference discretization in the vertical using 10 equally spaced $\sigma$ levels (where $\sigma=p/p_s$). The integration in time was carried out with a filtered leap-frog semi-implicit scheme \citep{robert1966}.

Thermal forcing was applied via a linear Newtonian relaxation towards a prescribed (axisymmetric) temperature field which was constant in time, with a relaxation timescale $\tau_R$. The complete restoration temperature field (with equator-to-pole temperature difference of $60\ \text{K}$) was intended to represent a distribution similar to the Earth and is shown in Fig~\ref{fig:4}.
\begin{figure}%[!ht]
 \centering
 \includegraphics[bb=0 0 456 238,width=0.9\columnwidth,clip=true]{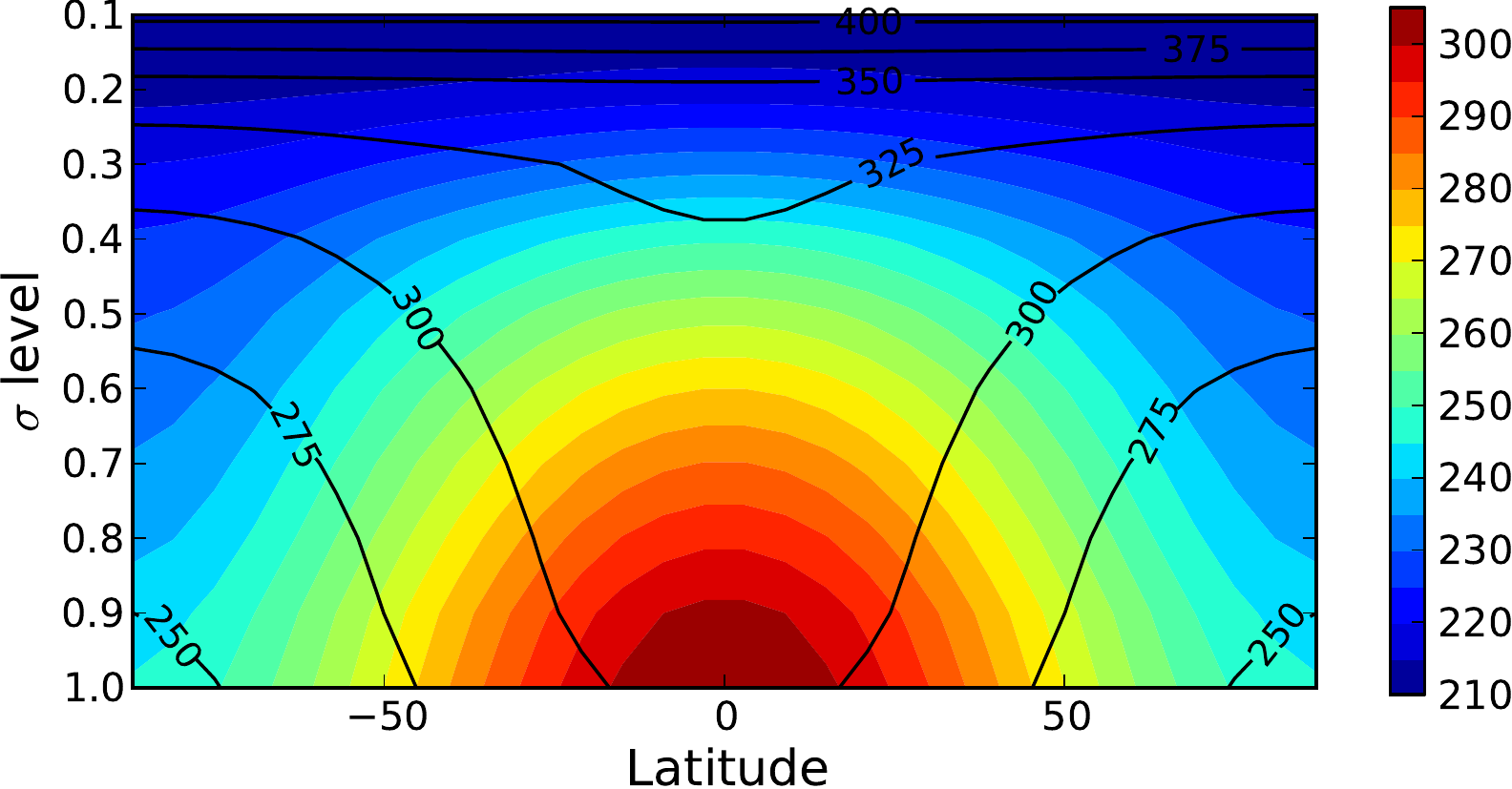}
 \caption{Restoration temperature (colour) and potential temperature (contour) field 
 with equator-to-pole temperature difference of $60\ \text{K}$.}\label{fig:4}
\end{figure}
\noindent The radiative timescale, $\tau_R$, was set to 30 Earth days in the free atmosphere, decreasing to $2.5$ Earth days at $\sigma=1.0$. 

A smooth, spherical planet was assumed in each case with no surface topography. Dissipation consisted of a combination of linear Rayleigh drag towards rest in the lowest two model levels (with a timescale decreasing from zero to $\tau_F = 0.6$ Earth day at the surface) and a $\nabla^8$ hyperdiffusion acting separately on temperature, vorticity and divergence. 

Simulations were run from an isothermal state at rest at a series of planetary rotation rates, from $\Omega^\ast=\Omega/\Omega_E = 1/16$ to $\Omega^\ast=8$, for a period equivalent to 10 Earth years.  Horizontal resolution was set to T42 for slowly rotating simulations ($\Omega^{\ast}\leq 1$), T127 for faster rotating simulations with $\Omega^{\ast}=1, 2, 4$, and with T170 reserved for simulations with $\Omega^{\ast}=8$. The computed diagnostics were averaged over the final model year from each run. Further details on the model setup and experiment design are presented by \citet{wang2016}.}

\section{Analysis of energy budgets} \label{sec:analysis}

The kinetic energy, $E_K$, (KE) and available potential energy, $E_A$, (APE) of an atmosphere can be defined \citep[e.g.][]{augier2013} in pressure coordinates by
 \begin{eqnarray}
E_K(p) & = & \widehat{ |{\bf u}|^2}/2 \label{eqn:ke1}\\
E_A(p) & = & \gamma(p)\widehat{ \theta'^2}/2, \label{eqn:ape1}
\end{eqnarray}
\noindent where ${\bf u}$ is the horizontal component of the total velocity ${\bf v} = ({\bf u},\omega)$, $\omega = Dp/Dt$ is the vertical velocity in pressure coordinates, $\theta$ is potential temperature, $\widehat{(.)}$ denotes an average over an entire pressure level and $(.)'$ departures therefrom. $\gamma(p)$ is defined as
\begin{equation}
\gamma(p) = R / [\Lambda(p) p \partial_p \widehat{\theta}],
 \end{equation}
\noindent where $R$ is the gas constant, 
%\begin{equation}
$\Lambda(p)=(p_R/p)^\kappa$, 
%\end{equation}
%\noindent 
$p_R$ is a reference pressure and $\kappa = R/c_p$. 

%$E_K(p) = \langle |\bf u|^2 \rangle/2$ and $E_A(p)=\gamma(p)\langle \theta'^2\rangle/2$, with $\gamma(p) = R / [\Lambda(p) p \partial_p \langle\theta\rangle]$.
The energy budget of atmospheric KE and APE can be derived from combining Eqns.  (\ref{eqn:ke1}) and (\ref{eqn:ape1}) with equations of motion and conservation of potential temperature \citep[e.g.][]{augier2013} to obtain
%Eqns. (\ref{eqn:aab1}) and (\ref{eqn:aab2})
 \begin{align}
\partial_t E_K(p) & = C(p) + \partial_p F_{K\uparrow} (p) - D_K(p) +S(p) \label{eqn:aa1}\\
\partial_t E_A(p) & = G(p) - C(p) + \partial_p F_{A\uparrow} (p) - D_A (p) +J(p) \label{eqn:aa2}
 \end{align}
Here $G(p)$ is an APE generation term e.g. due to differential heating, $C(p)$ is the conversion from APE to KE,  $F_{K\uparrow} (p)$ and $F_{A\uparrow} (p)$ are vertical fluxes of KE and APE, respectively, and $D_K(p)$, $D_A (p)$ are diffusion terms with:
 \begin{align}
G(p)&=\gamma \widehat{ \theta' Q'_\theta}\\
C(p)&=-\widehat{\omega\alpha_{\rho}}\\
F_{A\uparrow} (p)&=-\gamma (p) \widehat{ \omega\theta'^2}/2\\
F_{K\uparrow} (p)&=-\widehat{ \omega  |{\bf u}|^2}/2 -\widehat{ \omega \Phi}\\
S(p) & = - \delta_{ps} \partial_t \widehat{ (p_s \Phi_s)}\\
J(p) & = - (\partial_p \gamma) \widehat{ \omega \theta'^2} /2 -  \widehat{ \omega} \widehat{ \alpha_\rho}.
%J(p) & - (\partial_p \log \gamma F_{A\uparrow} (p)) 
\end{align}
$p_s$ and $\Phi_s$ are surface pressure and surface geopotential, respectively, and $\delta_{ps}$ is one when $p=p_s$ and zero otherwise.
The $S(p)$ and $J(p)$ terms are adiabatic processes but which do not conserve total available energy $E_K + E_A$. However, these terms have been shown to be negligible in the global mean \citep{siegmund1994,augier2013}, and so will not be considered further in this analysis.

\subsection{Formulation of the Lorenz energy cycle}
\begin{figure}
 \centering
  \includegraphics[width=0.9\columnwidth]{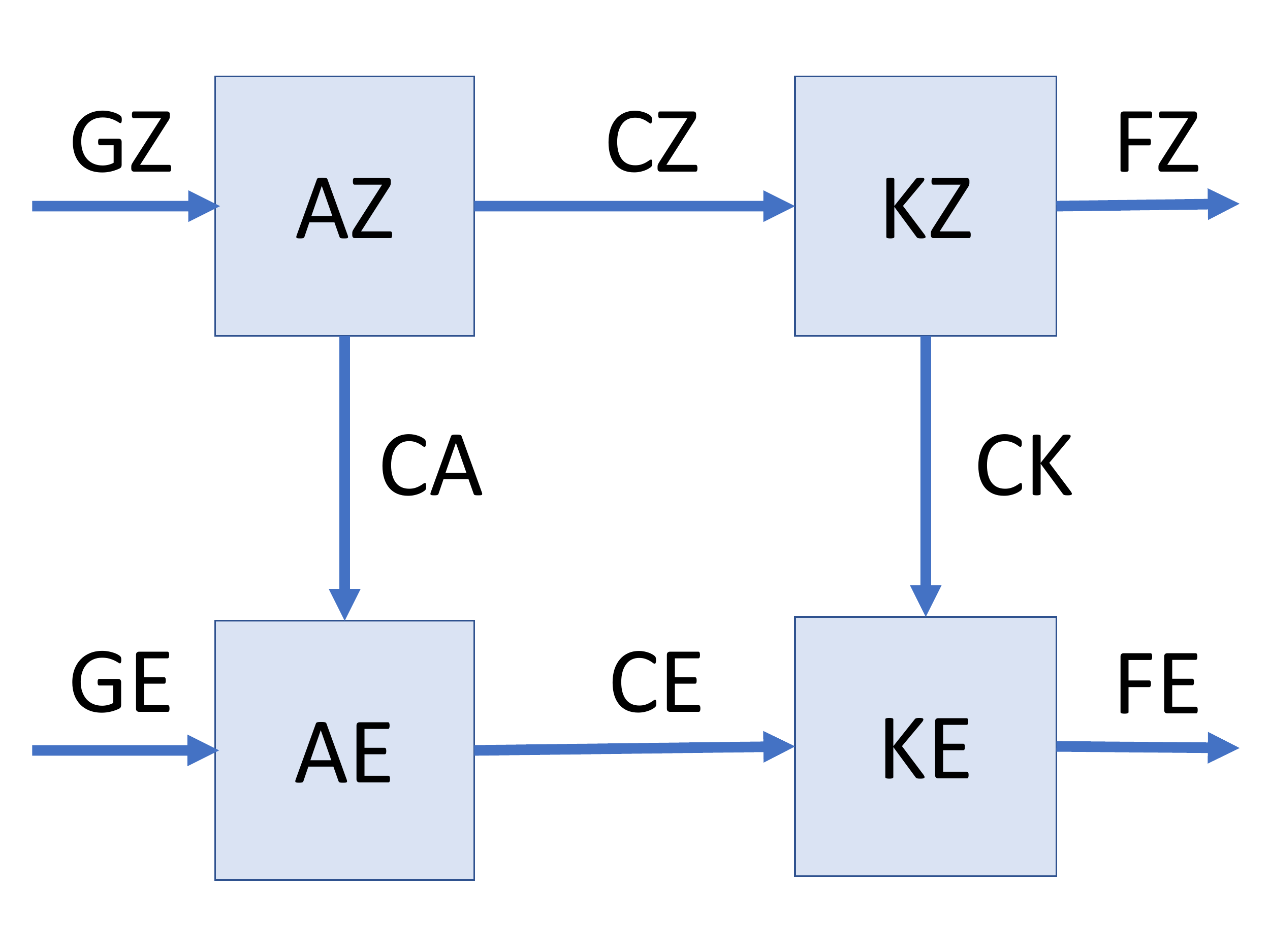}
 \caption{Schematic of Lorenz energy cycle. AZ: zonal mean available potential energy; AE: eddy available
 potential energy; KZ: zonal mean kinetic energy; KE: eddy kinetic energy. The generation, conversion and dissipation terms are shown as G*, C* and F* respectively (where * is either Z or E). }
 \label{lorenz_example}
\end{figure}

The generation and growth of non-axisymmetric waves and other disturbances (designated as ``eddies") within terrestrial planetary atmospheres requires the conversion into eddy kinetic and potential energy from other forms of energy in the background environment, the ultimate source of which is solar or stellar irradiation. This process of energy conversion can be illustrated and quantified most simply with the classical Lorenz energy cycle (\cite{lorenz1955}) which provides a framework for formulating a global mean atmospheric kinetic energy and available potential energy budget, as well as the conversion rates between the zonal mean (indicated by $[.]$ and ``eddy" ($.^*$) components of these energy forms. In addition, we designate time-averaged quantities by $\overline{(.)}$ and mass-weighted, vertically-integrated areal averages (e.g. of a quantity $Q$) by
\begin{equation}
\frac{1}{4\pi a^2 g} \int \int \int_0^{p_s} Q {\rm d}p\, {\rm d}x\, {\rm d}y = \dfrac{1}{4\pi a^2}\iiint Q \text{d}m = \Big<Q\Big>.
\end{equation}
Following e.g. \cite{Peixoto1974} and \cite{James1995}, the conservation equations for kinetic and potential energy can be written as 
\begin{subequations}
\begin{align}
&\frac{\text{dAZ}}{\text{d}t} = \text{GZ} -\text{CZ} - \text{CA}, \\
&\frac{\text{dAE}}{\text{d}t} = \text{GE} + \text{CA} - {\text{CE}}, \\
&\frac{\text{dKZ}}{\text{d}t} = \text{CZ} - \text{CK} - \text{FZ}, \\
&\frac{\text{dKE}}{\text{d}t} = \text{CK} + \text{CE} - {\text{FE}},
\end{align}
\label{eq:energy}
\end{subequations}
where KZ, KE, AZ, and AE refer to zonal mean kinetic energy, eddy kinetic energy, zonal mean available 
potential energy, and eddy available potential energy respectively. The conversion rates among these components are as shown in Fig. \ref{lorenz_example} and are as defined in pressure coordinates e.g. by \citet{James1995}. % QG version
$GZ$ and $GE$ are diabatic generation terms (if positive) for AZ and AE, while $FZ$ and $FE$ are dissipation terms for KZ and KE.

There are at least two principal mechanisms of eddy generation within planetary atmospheres: barotropic 
instability and baroclinic instability. The physical pictures of these two different eddy-generation 
mechanisms can be clearly distinguished from the viewpoint of energy conversion. Eddies generated through
barotropic instability are fed directly by the zonal mean kinetic energy, implying the conversion route 
of KZ$\rightarrow$KE to be important. Baroclinic instability, on the other hand, converts available potential
energy into eddy kinetic energy through the so-called \lq sloping convection\rq , which corresponds to the 
conversion route of AZ$\rightarrow$AE$\rightarrow$KE. Therefore, the dominating mechanism of eddy generation
can be appreciated by comparing the relative intensity and direction of CK and CE. 

\subsection{Spherical harmonic transformation}
The horizontal structure of the flow and other quantities (such as energy conversion rates) can be further decomposed into spectra by projection onto suitable sets of eigenfunctions.  A scalar function (e.g. $\theta'$) on a sphere can be transformed into spherical harmonic spectral space via
 \begin{align}
\theta'({\bf x}_h,p) = \sum_{n\ge0} \sum_{-n\le m\le n} \theta'_{nm} (p) Y_{nm} ({\bf x}_h)
 \end{align}
where $n$ is the total and $m$ the zonal wavenumber {indices}. $Y_{nm}$ are spherical eigenfunctions %(see Eqn.~\ref{eqn:leg}) 
with $\nabla_h^2 Y_{nm} = - n(n+1)Y_{nm}/a^2 $ {(where $\nabla_h$ is the horizontal gradient operator and $\nabla_h^2$ the corresponding Laplacian)}. The horizontal mean of the product of two scalar variables is then 
 \begin{align}
\langle \omega \Phi \rangle = \sum_{n\ge0} \sum_{-n\le m\le n} (\omega,\Phi)_{nm} \label{eqn:sca}
 \end{align}
with 
 \begin{align}
(\omega,\Phi)_{nm}  =  \text{Re}\{ \omega^\dagger_{nm} \Phi_{nm} \}
 \end{align}
where $\text{Re}\{X\}$ is the real part and $X^\dagger$ is the complex conjugate of a complex number $X$ \citep{augier2013}.

%\nomenclature{$Y_{km}$}{spherical eigenfunctions}

For the horizontal velocity field, $\bf u$, a decomposition into divergent and rotational (non-divergent) components can be performed via the Helmholtz decomposition
 \begin{align}
{\bf u} & = {\bf \nabla}_h \wedge (\psi {\bf e}_z) + {\bf \nabla}_h \chi = {\bf u}_r + {\bf u}_d
 \end{align}
with reference to the horizontal streamfunction $\psi({\bf x}_h, p)$ and 
%the spherical potential vorticity 
the horizontal velocity potential 
$\chi ({\bf x}_h, p)$. Using this decomposition, we can obtain the vorticity $\zeta$ and the horizontal divergence $\delta$:
 \begin{align}
\zeta& = {\rm rot}_h\,({\bf u}) = {\bf e}_z \cdot ({\bf \nabla} \wedge {\bf u}) = {\bf \nabla}_h^2\psi  \\
\delta &= {\rm div}_h\,({\bf u}) = {\bf \nabla}_h \cdot {\bf u} = {\bf \nabla}_h^2\chi .
 \end{align}
This decomposition is then used to calculate the horizontal mean of a scalar product between two horizontal vector fields ${\bf a}$ and ${\bf b}$ 
 \begin{align}
\widehat{ {\bf a} \cdot {\bf b}} = \sum_{n\ge0} \sum_{-n\le m\le n} ( {\bf a}, {\bf b})_{nm} \label{eqn:vec}
 \end{align}
with 
 \begin{align}
({\bf a,  b})_{nm} & = & \frac{a^2}{n(n+1)}  \text{Re}\{  {\rm rot}_h\,({\bf a})^\dagger_{nm} {\rm rot}_h\,({\bf b})_{nm} + \nonumber \\
& &   {\rm div}_h\,({\bf a})^\dagger_{nm} {\rm div}_h\,({\bf b})_{nm}  \},
 \end{align}
\noindent {\citep[see e.g.][]{augier2013}}.

Using Eqn. \ref{eqn:sca} for scalars and Eqn. \ref{eqn:vec} for vector fields, the spectral versions of APE and KE can be obtained respectively as:
 \begin{align}
E_A^{nm} & = \gamma(p) \frac{(\theta', \theta')_{nm}}{2} = \gamma(p) \frac{|\theta'_{nm}(p)|^2}{2} \label{eqn:ea}\\
E_K^{nm} & = \frac{({\bf u}, {\bf u})_{nm}}{2} = \frac{a^2 ( |\zeta_{nm}|^2 + |\delta_{nm}|^2 )}{2n(n+1)}  \label{eqn:ek}
 \end{align}

%%%

The APE or KE spectrum can be further decomposed into a zonal mean spectrum and an eddy (or residual) spectrum 
as the following (e.g. for KE):{ 
%\begin{equation}
 \begin{align}
&E_{KZ}^n(t) = \dfrac{1}{4}\dfrac{a^2}{n(n+1)} \big(|\zeta_n(m=0,t)|^2  +|\delta_n(m=0,t)|^2\big),\\
 &E_{KE}^n(t) = \dfrac{1}{4}\dfrac{a^2}{n(n+1)} \sum_{m=-n}^{n}\big(|\zeta_n(m,t)|^2  +|\delta_n(m,t)|^2\big); m\neq0, 
 \end{align}
 }
%\end{equation}
where $E_{KZ}^n$ and $E_{KE}^n$ represent the zonal and eddy (residual) part of the spectrum respectively, such that $E_K^n=E_{KZ}^n+E_{KE}^n$.

\subsection{Calculation of spectral enstrophy fluxes}
%Although power-law slopes observed in the APE and KE spectra provide evidence which might suggest that the 
%model atmospheric circulations are demonstrating some features of the inertial ranges in idealised 2-D turbulence, it should be 
%noted that the existence of a power-law spectrum alone cannot provide conclusive insights into the direction of 
%energy/enstrophy cascade. 
The nonlinear spectral enstrophy transfer flux (\cite{Boer1983}, 
\cite{Shepherd1987}, \cite{burgess2013})%, for example, 
can provide more detailed insights into the enstrophy 
redistribution among different wavenumbers through nonlinear eddy-eddy interactions. Starting from the 
vorticity equation
\begin{equation}
 \frac{\partial \zeta}{\partial t} = -(\mathbf{u}_r\cdot\mathbf{\nabla})\zeta -\mathcal{D}
\end{equation}
where $\mathbf{u}_r=(u_r,v_r)$ is the rotational velocity and $\mathcal{D}$ represents the effects on vorticity evolution due 
to divergence and other vorticity sources and sinks.
Multiply by $\zeta$ to obtain the equation for enstrophy ($G=\frac{1}{2}\zeta^2$)
\begin{equation}
 \frac{\partial G}{\partial t} = -\zeta(\mathbf{u}_r\cdot\mathbf{\nabla})\zeta-\zeta \mathcal{D}.
\end{equation}
In spectral space this can be rewritten as
\begin{equation}
 \frac{\partial G^n}{\partial t} = J^n +\mathcal{D}^n
\end{equation}
where the interaction term $J^n$ is
\begin{equation}
 J^n = -\frac{1}{4}\sum_{m=-n}^{n}\big[\zeta^{nm^\dagger}(\mathbf{u}_r\cdot\mathbf{\nabla}\zeta)^{nm}+\zeta^{nm}(\mathbf{u}_r\cdot\mathbf{\nabla}\zeta)^{nm^\dagger}\big]
 \label{eq:enst_int}
\end{equation}
%The corresponding kinetic energy interaction term $I_n$ can be defined similarly and is related to the 
%enstrophy interaction term $J_n$ by
%\begin{equation}
 %I_n = \frac{a^2}{n(n+1)}J_n
%\end{equation}
%where $a$ is the planetary radius.
Note that interaction terms only redistribute enstrophy among wavenumbers so
$$\sum_{n=0}^{N}J^n =0.$$
We can then define the enstrophy spectral flux as 
\begin{equation}
% \displaystyle \mathcal{F}_{n+1} = -\sum_{l=0}^{n} I_l, \ \text{and}\ 
\displaystyle \mathcal{H}^{n}  = \sum_{l=N}^{n} J^l,
\end{equation}
%respectively, 
where the sign is adopted conventionally such that a positive value corresponds to a 
forward cascade while a negative value corresponds to an inverse cascade.

%The key to obtaining %$\mathcal{F}$ and 
%$\mathcal{H}$ is the calculation of $J^n$ as in Eq. \eqref{eq:enst_int}.
%Model output of streamfunction was used to calculate the rotational component of velocity. The SPHEREPACK 3.2 
%code package from CISL (Computational and Information Systems Laboratory)/NCAR was employed for the spherical 
%harmonic transformation and spherical gradient calculations.

The interaction terms and spectral fluxes of %KE and 
enstrophy can be further decomposed into contributions from
eddy-eddy interactions and eddy-zonal flow interactions (\cite{burgess2013}). Nonlinear interaction terms of 
enstrophy due to purely eddy-eddy interactions, $J^{n(e)}$, can be 
{obtained from Eq (\ref{eq:enst_int}) but carrying out the sum in $m$ for $m\neq 0$ only. }
%written as
%\begin{equation}
 %\centering
 %J^{n(e)} = -\dfrac{1}{4}\sum_{m=-n}^{n} \big[(\zeta^{\ast})^{nm^\dagger} J(\psi^{\ast},
 %\zeta^{\ast})^{nm}+\textbf{c}.\textbf{c}.  \big]
%\end{equation}
%where %all primed quantities are eddy components with respect to zonal mean flow, 
%$J(\psi^{\ast},\zeta^{\ast})$
%is the Jacobian of geostrophic (eddy) streamfunction $\psi^{\ast}$ and vorticity $\zeta^{\ast}$, and 
%$\textbf{c}.\textbf{c}.$ represents the complex conjugate of the first term in square bracket. 
The contribution to 
$J^n$ through eddy-zonal mean interactions is then simply $J^{n(z)}=J^n-J^{n(e)}$. %Similarly, interaction terms of
%KE can be decomposed in the form of $I_n = I_n^{(z)}+I_n^{(e)}$. 
In this way, the spectral flux of %KE and 
enstrophy can be decomposed as %$\mathcal{F}_n = \mathcal{F}_n^{(z)}+\mathcal{F}_n^{(e)}$ and
$\mathcal{H}^n=\mathcal{H}^{n(z)}+\mathcal{H}^{n(e)}$.

\subsection{Spectral energy budget} \label{sec:sum}
The spectrally-resolved energy budget can be obtained by 
inserting Eqn.~\ref{eqn:ea} and Eqn.~\ref{eqn:ek} into Eqns.~\ref{eqn:aa1} and \ref{eqn:aa2}, resulting in \citep{augier2013}:
 \begin{eqnarray}
\partial_t E_K^{nm} (p) & = & C^{nm} (p) + T_K^{nm} (p) + L^{nm} (p) \nonumber \\ 
 & & + \partial_p F_{K\uparrow}^{nm} (p) - D_K^{nm} (p) \label{eqn:aa3}\\
\partial_t E_A^{nm} (p) & = & G^{nm} (p) - C^{nm} (p) + T_A^{nm} (p) \nonumber \\
 & & + \partial_p F_{A\uparrow}^{nm} (p) - D_A^{nm} (p) \label{eqn:aa4}.
 \end{eqnarray}
where $G^{nm}$ is the spectral APE generation term,  $C^{nm}$ is the spectral conversion term, $T_K^{nm}$ and $T_A^{nm}$ are the spectral transfer terms (of KE and APE, respectively) due to non-linear interactions. $L^{nm}$ is a spectral transfer term due to Coriolis forces and $F_{K\uparrow}^{nm}$ and $F_{A\uparrow}^{nm}$ are vertical fluxes. $D_K^{nm}$ and $D_A^{nm}$ are diffusion terms. These terms are computed via
 \begin{align}
C^{nm} (p) &	= - (\omega,\alpha_\rho)_{nm} \\
T_K^{nm} (p)&	= - ({\bf u, \bf v \cdot \bf \nabla \bf u})_{nm} + \partial_p ({\bf u, \omega \bf u})_{nm}/2\\
T_A^{nm} (p)& 	= - \gamma(\theta',{\bf v \cdot \nabla} \theta')_{nm} + \gamma \partial_p (\theta', \omega \theta')_{nm}/2\\
L^{nm} (p)&   	= - ({\bf u}, f[\phi] {\bf e}_z \wedge {\bf u})_{nm}\\
F_{A\uparrow}^{nm} (p) &	= - \gamma (\theta',\omega\theta')_{nm}/2\\
F_{K\uparrow}^{nm} (p) &	= - (\omega,\Phi)_{nm} - \partial_p({\bf u, \omega \bf u})_{nm} /2 \\
G^{nm} (p) &	= \gamma(\theta',Q'_\theta)_{nm}\\
D_A^{nm} (p) & = - \gamma (\theta', D_\theta[\theta])_{nm} .
 \end{align}

%\subsubsection{Wavenumber summation and vertical integration} \label{sec:sum}

The spectral energy and tendency terms obtained in the previous sections are functions of time, zonal wavenumber $m$, total wavenumber $n$, and pressure $p$. To obtain a one dimensional spectrum or spectral flux from these terms, a dependence upon $n$ alone would be preferable for reasons of simplicity of interpretation. The time dependency is removed by averaging the resulting spectral quantities over multiple time steps. Following a summation over zonal wavenumbers and a vertical integration over a pressure range from { $p_b$ at the lowest level (usually the surface) to $p_t$ at the top level}, the vertically integrated KE spectrum is obtained via
 \begin{align}
E_K[n]^{p_b}_{p_t} = \int^{p_b}_{p_t}  \frac{dp}{g} \sum_{-n\le m\le n} E_K^{nm} (p)
 \end{align}
and the vertically integrated KE spectral flux via
 \begin{align}
\Pi_K[n]^{p_b}_{p_t} =\sum_{k\ge n}  \int^{p_b}_{p_t}  \frac{dp}{g} \sum_{-k\le m\le k} T_K^{km} (p) \label{eqn:cumu}
 \end{align}
where $\sum_{k\ge n} \sum_{-k\le m\le k}$ denotes a cumulative sum (from large to small wavenumbers). Other spectral quantities can be similarly vertically integrated and summed. %  so that in total   Eqns.~\ref{eqn:aa1} and \ref{eqn:aa2}  become
% \begin{align}
%\partial_t \mathcal{E}_A[n]^{p_b}_{p_t} & = \mathcal{C}[n]^{p_b}_{p_t} + \Pi_K[n]^{p_b}_{p_t}  + \mathcal{L}[n]^{p_b}_{p_t} +  \mathcal{F}_{K\uparrow}[n](p_b) \nonumber \\
%&  - \mathcal{F}_{K\uparrow}[n](p_l) - \mathcal{D}_K[n]^{p_b}_{p_t} \label{eqn:aa5}\\
%\partial_t \mathcal{E}_K[n]^{p_b}_{p_t}  & = \mathcal{G}[n]^{p_b}_{p_t} - \mathcal{C}[n]^{p_b}_{p_t} + \Pi_A[n]^{p_b}_{p_t} +  \mathcal{F}_{A\uparrow}[n] (p_b) \nonumber \\
% & -  \mathcal{F}_{A\uparrow}[n] (p_t)  - \mathcal{D}_A[n]^{p_b}_{p_t} \label{eqn:aa6}
% \end{align}
%where $\mathcal{F_\uparrow}[n](p)=\sum_{n\ge n} F_{\uparrow}[n]$ are termed cumulative vertical fluxes and the other terms are integrated cumulative fluxes of the terms detailed in the previous section (e.g. cumulative kinetic energy $\mathcal{E}_K[n]^{p_b}_{p_t}= \sum_{k\ge n} E_K[k]^{p_b}_{p_t}$ and cumulative conversion $\mathcal{C}[n]^{p_b}_{p_t}= \sum_{k\ge n} C[k]^{p_b}_{p_t}$). 
Note that the cumulative summation is performed from large wavenumbers to small wavenumbers and that all spectral fluxes (barring conversion and vertical fluxes) are conserved, meaning the cumulative sum over all wavenumbers $n$ should add up to zero.

%\section{Lorenz energy cycles}\label{sec:lorenz_heat}

\section{Lorenz energy cycles as a function of $\Omega^\ast$} \label{sec:lorenz_heat}

\begin{figure*}
 \centering
  \includegraphics[width=\columnwidth]{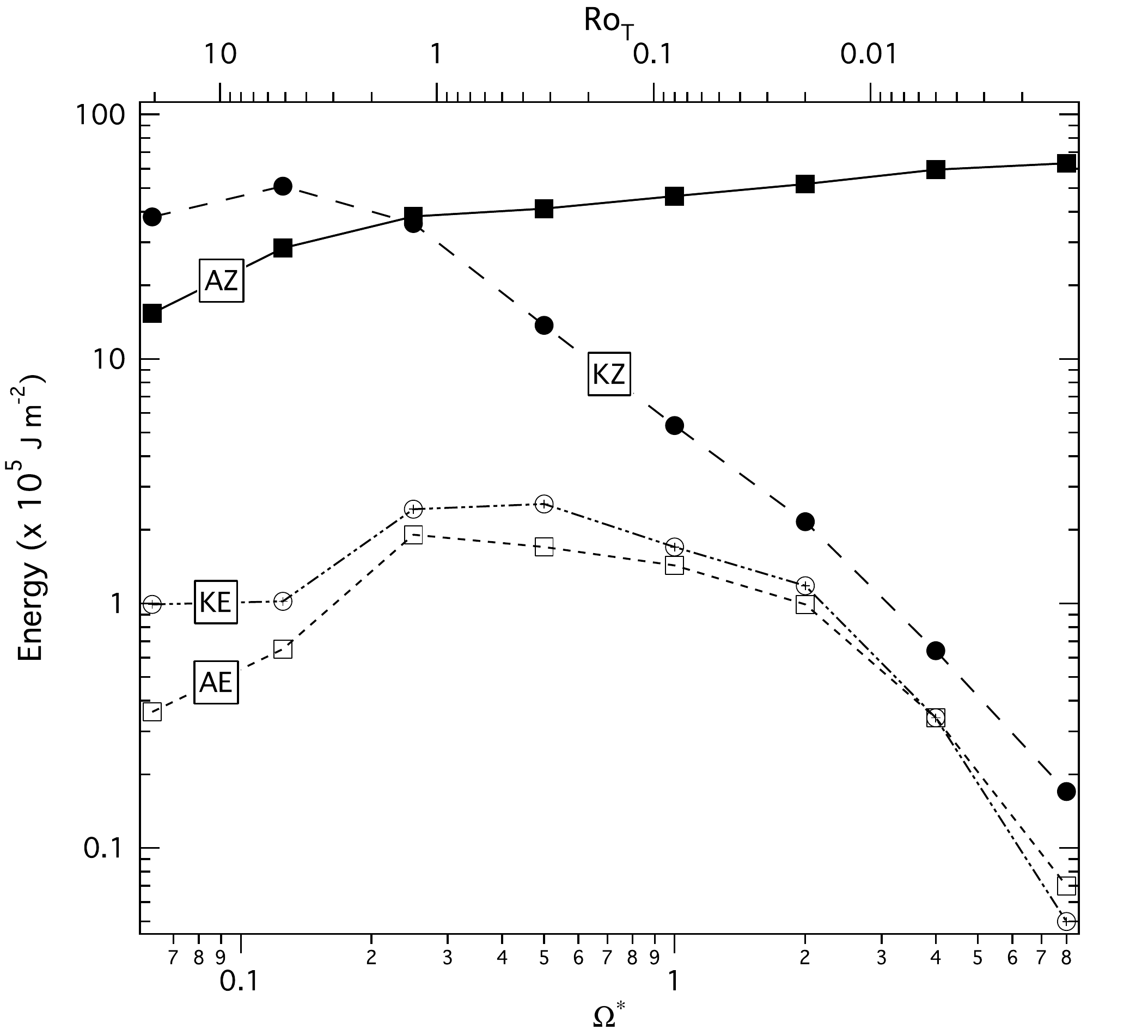}
    \includegraphics[width=\columnwidth]{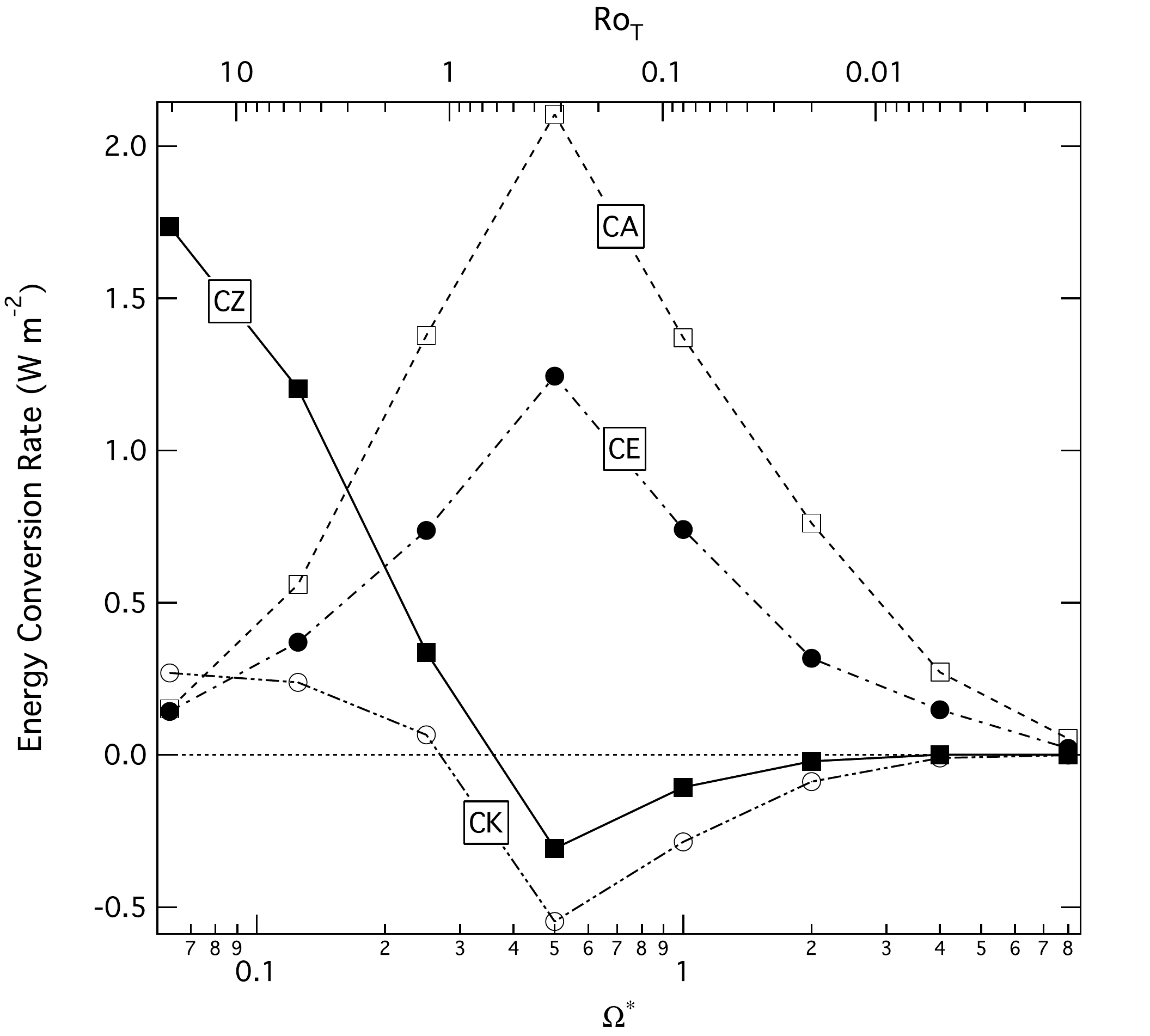}\\
    (a) \hskip 8 cm (b) %\\
%      \includegraphics[width=0.45\textwidth]{for_peter/Lorenz3.pdf} \\
%      (c)
  \caption{Terms in the Lorenz energy budgets for the series of PUMA-S simulations as a function of $\Omega^*$ and thermal Rossby number: (a) globally averaged energies (in $10^5 \text{J}\text{m}^{-2}$); (b) the main energy conversion rates CZ, CA, CE and CK.%; and (c) the combinations CE$+$CZ (representing the total APE to KE conversion rate) and CA$+$CK (representing the total conversion from zonal mean to eddy energy.  
Conversion rates are in unit of $\text{W}\text{m}^{-2}$.}
 \label{lorenz_profiles}
\end{figure*}

\begin{figure}
 \centering
% \includegraphics[scale=1.2]{fig_chap3/lorenz-multi.eps}
%  \includegraphics[width=0.45\textwidth]{for_peter/Lorenz1.pdf}
 %   \includegraphics[width=0.45\textwidth]{for_peter/Lorenz2.pdf}\\
 %   (a) \hskip 8 cm (b) %\\
      \includegraphics[width=\columnwidth]{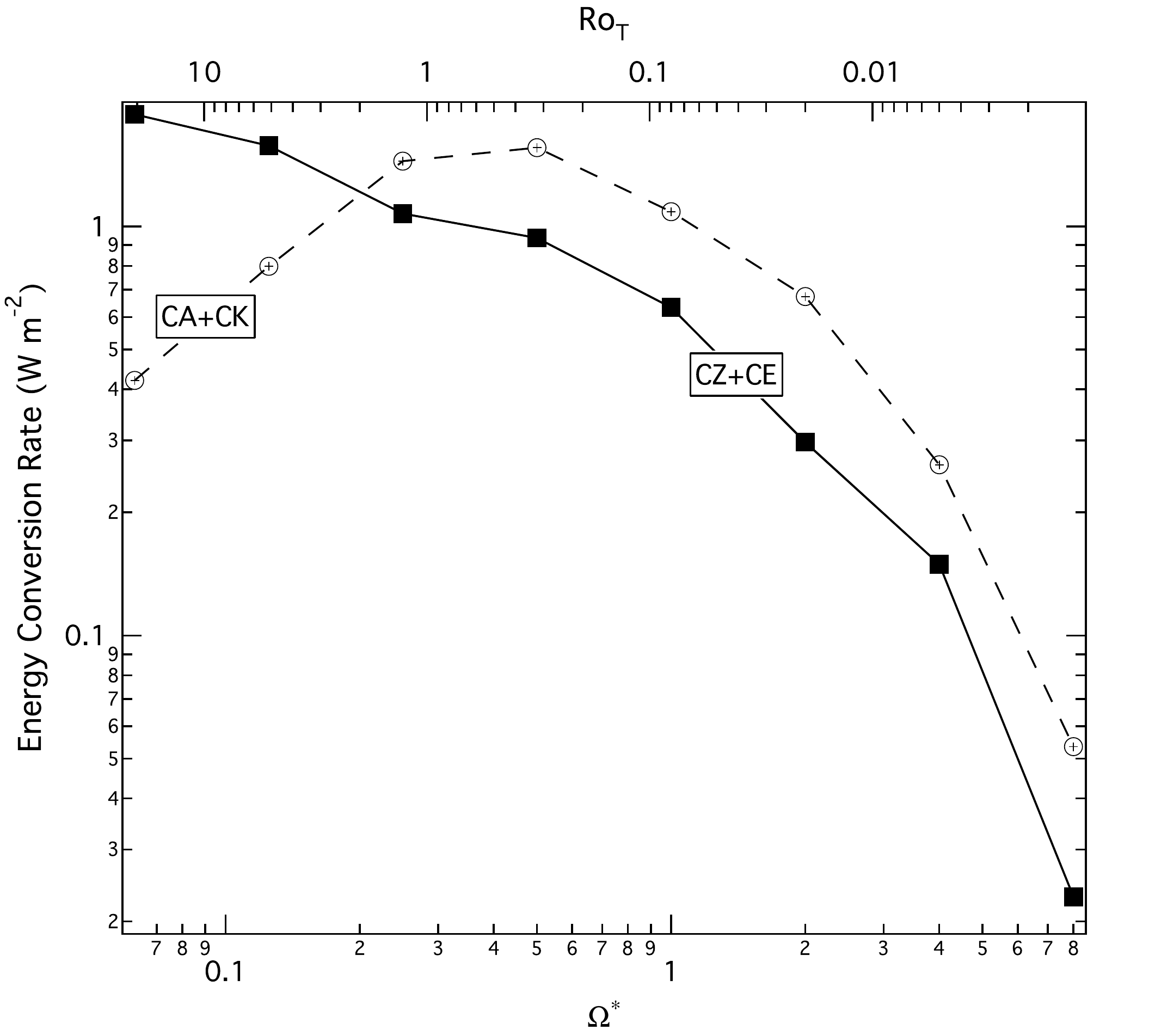} \\
%      (c)
  \caption{Terms in the Lorenz energy budgets for the series of PUMA-S simulations as a function of $\Omega^*$ and thermal Rossby number, showing the combinations CE$+$CZ (representing the total APE to KE conversion rate) and CA$+$CK (representing the total conversion from zonal mean to eddy energy.  Conversion rates are in unit of $\text{W}\text{m}^{-2}$.}
 \label{lorenz_profiles2}
\end{figure}

In this section we compute the various terms in the Lorenz energy budget for each of the 8 rotation rate simulations spanning $1/16 \leq \Omega^\ast \leq 8$ and explore the main trends in energies and conversion rates. Fig. \ref{lorenz_profiles} and Fig. \ref{lorenz_profiles2} show how the terms in the globally- and time-averaged Lorenz energy cycles vary with rotation rate, $\Omega^\ast$, and thermal Rossby number, $\mathcal{R}o_T$, {defined as in \citet{wang2016} by
\begin{equation}
 \mathcal{R}o_T = \frac{R\Delta \theta_h}{\Omega^2 a^2}, \label{eq:Ro}
\end{equation}
\noindent where $\Delta \theta_h$ is the equator-to-pole potential temperature difference, $a$ the planetary radius and $R$ the specific gas constant.} Fig. \ref{lorenz_profiles}(a) plots the magnitudes of the various energy reservoirs, expressed in units of 100 kJ m$^{-2}$. This shows the zonal mean potential energy, AZ, increasing monotonically with $\Omega^\ast$ and decreasing with $\mathcal{R}o_T$, though with only a shallow variation over the range computed. KZ, on the other hand, exhibits a maximum around $\Omega^\ast = 1/8$ but then decreasing rapidly with $\Omega^\ast$ for higher rotation rates. The maximum in KZ corresponds to $\mathcal{R}o_T \simeq 5$. The eddy kinetic and available potential energies follow similar trends to each other, rising with $\Omega^\ast$ to a shallow maximum around $\Omega^\ast = 1/2$ ($\mathcal{R}o_T\simeq 0.3$ and then decreasing with $\Omega^\ast$ quite sharply at higher rotation rates. KE is seen to dominate over AE by a factor $\sim 2-3$ until $\Omega^\ast = 4$, beyond which AE $>$ KE. 

Variations in the mean energy conversion rates (in units of W m$^{-2}$ are shown in Fig. \ref{lorenz_profiles}(b). CZ represents the direct conversion of zonally-averaged APE to KE by zonal mean overturning circulations, and essentially reflect the relative strengths of the (thermally-direct) Hadley cells and the (thermally-indirect) Ferrel cells within the global circulation. This behaves more or less as one might expect, with strongly positive values of CZ at low rotation rates ($\Omega^\ast \lesssim 0.4$; $\mathcal{R}o_T \gtrsim 1$) where the direct Hadley circulation is dominant, but becoming negative at higher rotation rates, where the circulation becomes geostrophic and the Ferrel cells become stronger. This largely reflects the increasing dominance of baroclinic eddies at high rotation rates. The barotropic conversion term, CK, also undergoes a similar reversal of sign around ($\Omega^\ast \simeq 0.3$; $\mathcal{R}o_T \simeq 1$), indicative of a transition from barotropically energised eddies (with CK $> 0$) at low rotation rates to baroclinically energised eddies (with CE $>0$) at higher rotation rates. This interpretation is consistent with $\mathcal{R}o_T \simeq 1$ as the criterion since $\mathcal{R}o_T \lesssim 1$ is also a criterion for strong baroclinic instability. Both conversion rates (CZ and CK) {evidently} become vanishingly small as $\Omega^\ast$ % \rightarrow \infty$. 
{becomes large. This likely reflects the tendency for geostrophic velocities to decrease with increasing $\Omega$, together with the corresponding vertical velocities and the most energetic length scales for eddies.  }

%The direct conversion between AZ and KZ also changes sign at $\Omega^{\ast}=1/4$, which reveals the strengthening
%of Hadley cells as rotation rate is decreased. The Hadley cell, also known as the \lq thermally direct\rq\ cell
%(driven directly by differential heating), converts the zonal mean available potential energy into kinetic
%energy of the zonal mean flow. Weak Coriolis force caused by slow rotation rate enables Hadley cells to expand 
%to higher latitudes and become dominant over the \lq indirect\rq\ Ferrel cells, leading to the global mean
%conversion from AZ to KZ. 
As anticipated in the previous subsection, the strength of the baroclinic conversion rates, CE and CA, reach their maximum at
$\Omega^{\ast}=1/2$. This also agrees with results from \cite{DelGenio1987} in which baroclinic eddies peak 
in energy conversion efficiency at $\Omega^{\ast}=1/2$, and is also consistent with the peak of meridional
eddy heat flux at $\Omega^{\ast}=1/2$ as suggested by the peak in CA $+$ CK in Fig. \ref{lorenz_profiles2} (see also Fig. 7 of \citet{wang2016}). \cite{pascale2013}, however, found that the rotation rate corresponding to this peak in CE was sensitive also to other parameters, notably the strength of the bottom friction, which may account for slight differences in this peak being seen in other studies (e.g. \cite{kaspi2015}). 

%Note that 
%the Rhines scale starts to exceed the planetary radius $a$ at $\Omega^{\ast}=1/4$ as rotation rate decreases,
%as shown in Table \ref{tab.rhines}. As the Rhines scale represents the characteristic meridional width of baroclinic zone, 
%atmospheres with rotation rate of $\Omega^{\ast}=1/4$ or slower tend to have less well-developed baroclinic
%eddies, thus leading to the weaker eddy heat transport.

\section{Kinetic and available potential energy spectra}\label{sec:turb-ke-spectra}

In the following section, the KE and APE spectra are employed as a diagnostic to reveal insights into processes such as the 
jet formation mechanism and the transfer of energy and enstrophy across different scales. %Since current theories relevant to atmospheric macroturbulence are predominantly 
%2-D (as discussed in the previous section), we will focus for some purposes on the barotropic component of the global KE 
%spectrum. Following \cite{Koshyk1999}, the vertically and temporally averaged horizontal KE on a certain
%pressure level can be calculated by
%\begin{equation}
% \overline{E_n(p,t)} = \dfrac{1}{4}\dfrac{a^2}{n(n+1)} \sum_{m=-n}^{n}\big(\overline{|\zeta_n^m|^2} +
% \overline{|\delta_n^m|^2}\big),
%\end{equation}
%where $a$ is the planetary radius, $\zeta_n^m$ and $\delta_n^m$ the spherical harmonic coeffients of vorticity
%$\zeta$ and divergence $\delta$ respectively, $n$ and $m$ the total and zonal wavenumber respectively, and the
%overbar denotes the time average. 

\begin{figure*}[]
\centering
%\hspace{-4.5cm}\textbf{c)} $\Omega^*=\frac{1}{4}$, T42 Resolution \hspace{5cm} \textbf{d)} $\Omega^*=\frac{1}{8}$, T42 Resolution \\
\hspace{-4.5cm}\textbf{a)} $\Omega^*=\frac{1}{16}$, T42 Resolution \hspace{5cm} \textbf{b)} $\Omega^*=\frac{1}{8}$, T42 Resolution \\
\makebox[1.0\textwidth][c]{
\includegraphics[width=0.9\columnwidth]{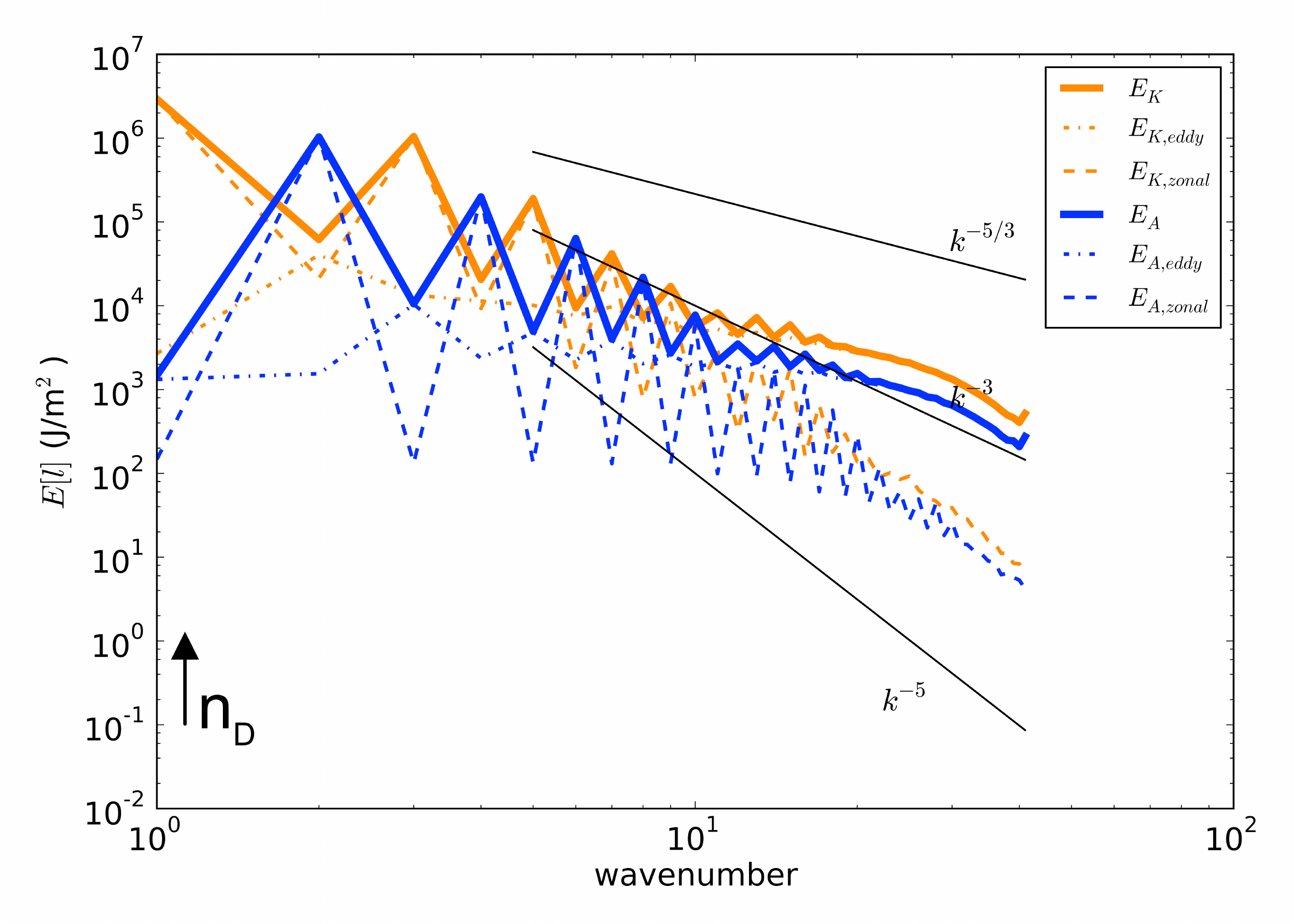}
\includegraphics[width=0.9\columnwidth]{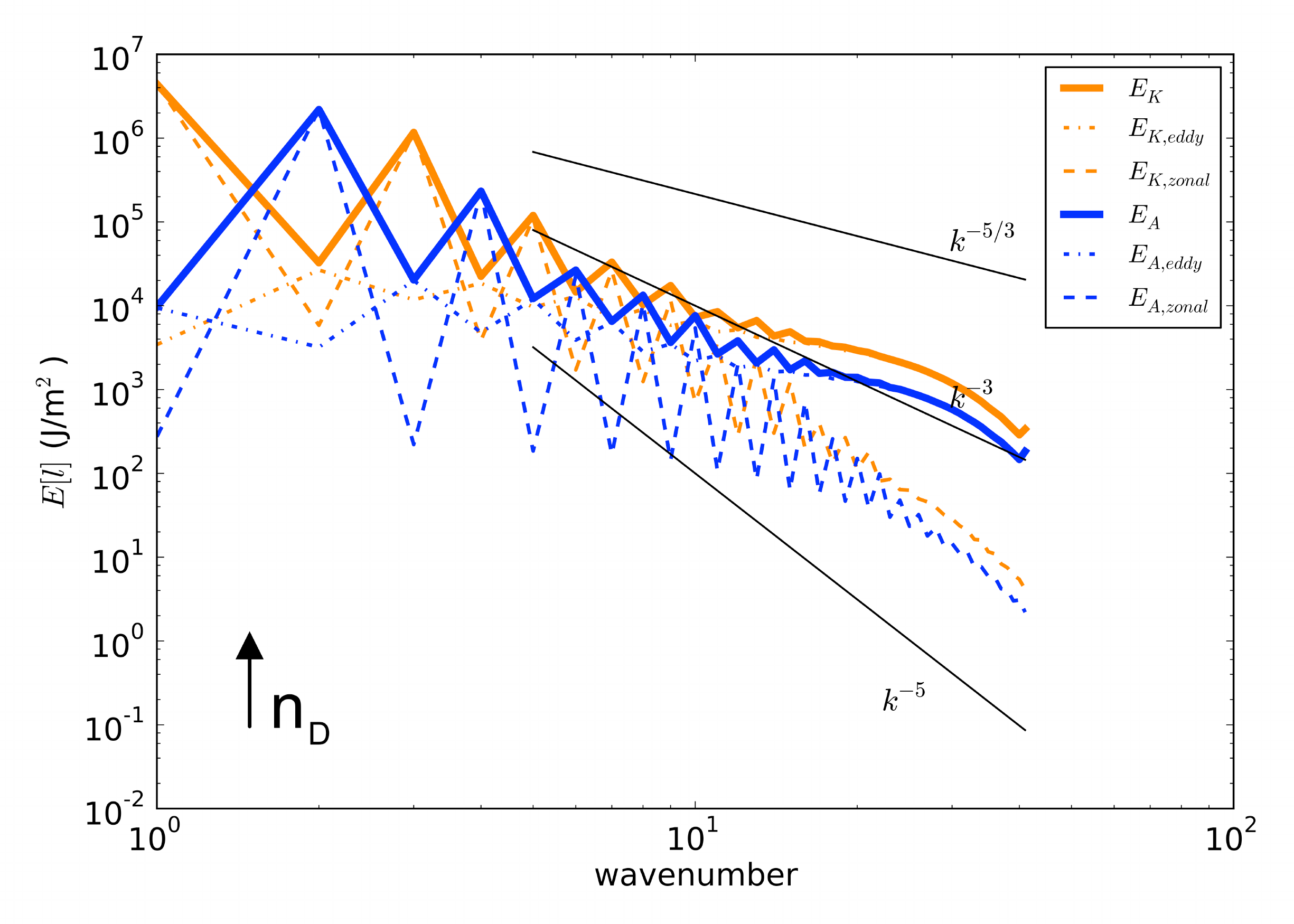}
}\\
\hspace{-4.5cm}\textbf{c)} $\Omega^*=\frac{1}{4}$, T42 Resolution \hspace{5cm} \textbf{d)} $\Omega^*=\frac{1}{2}$, T42 Resolution \\
\makebox[1.0\textwidth][c]{
\includegraphics[width=0.9\columnwidth]{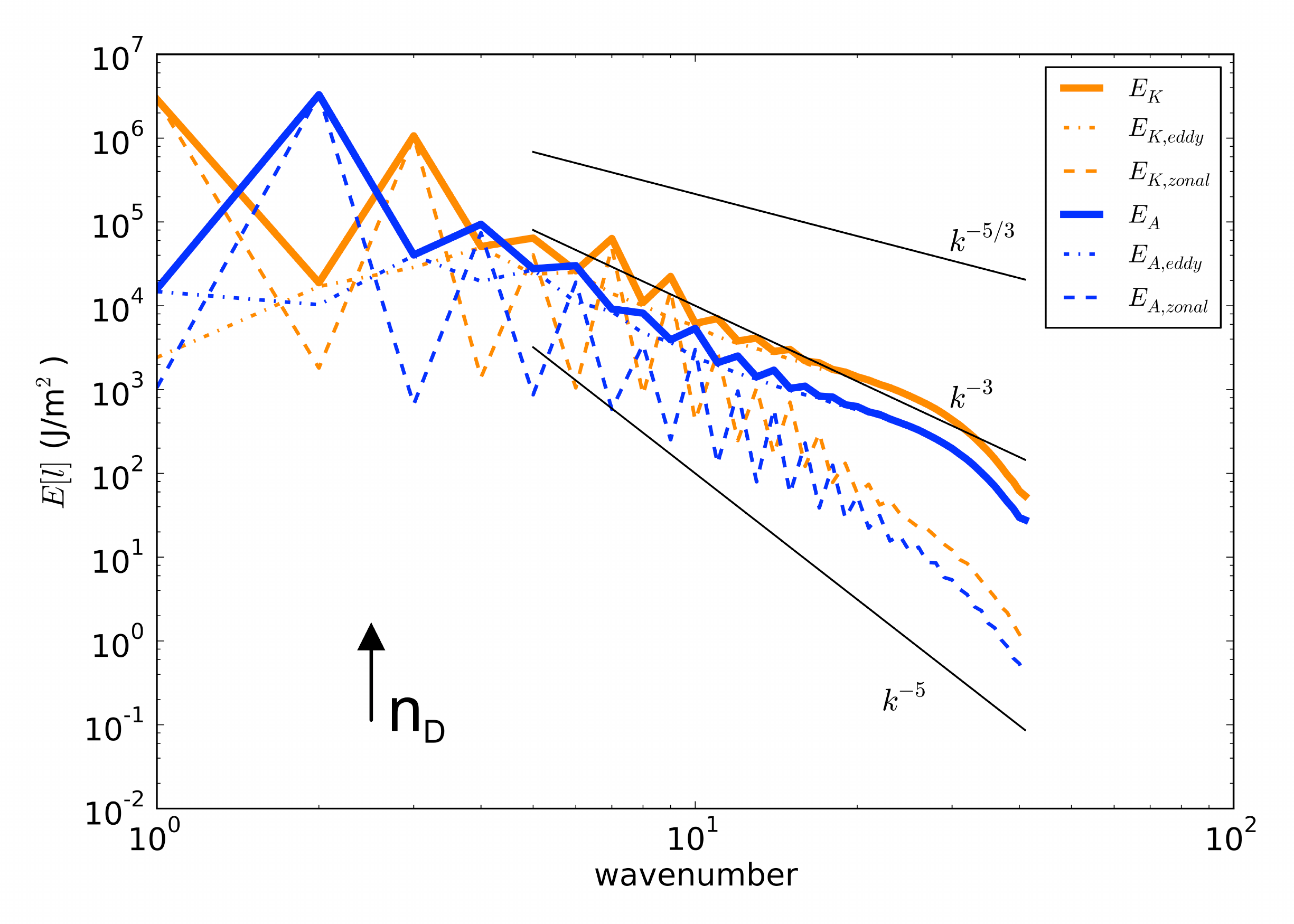}
\includegraphics[width=0.9\columnwidth]{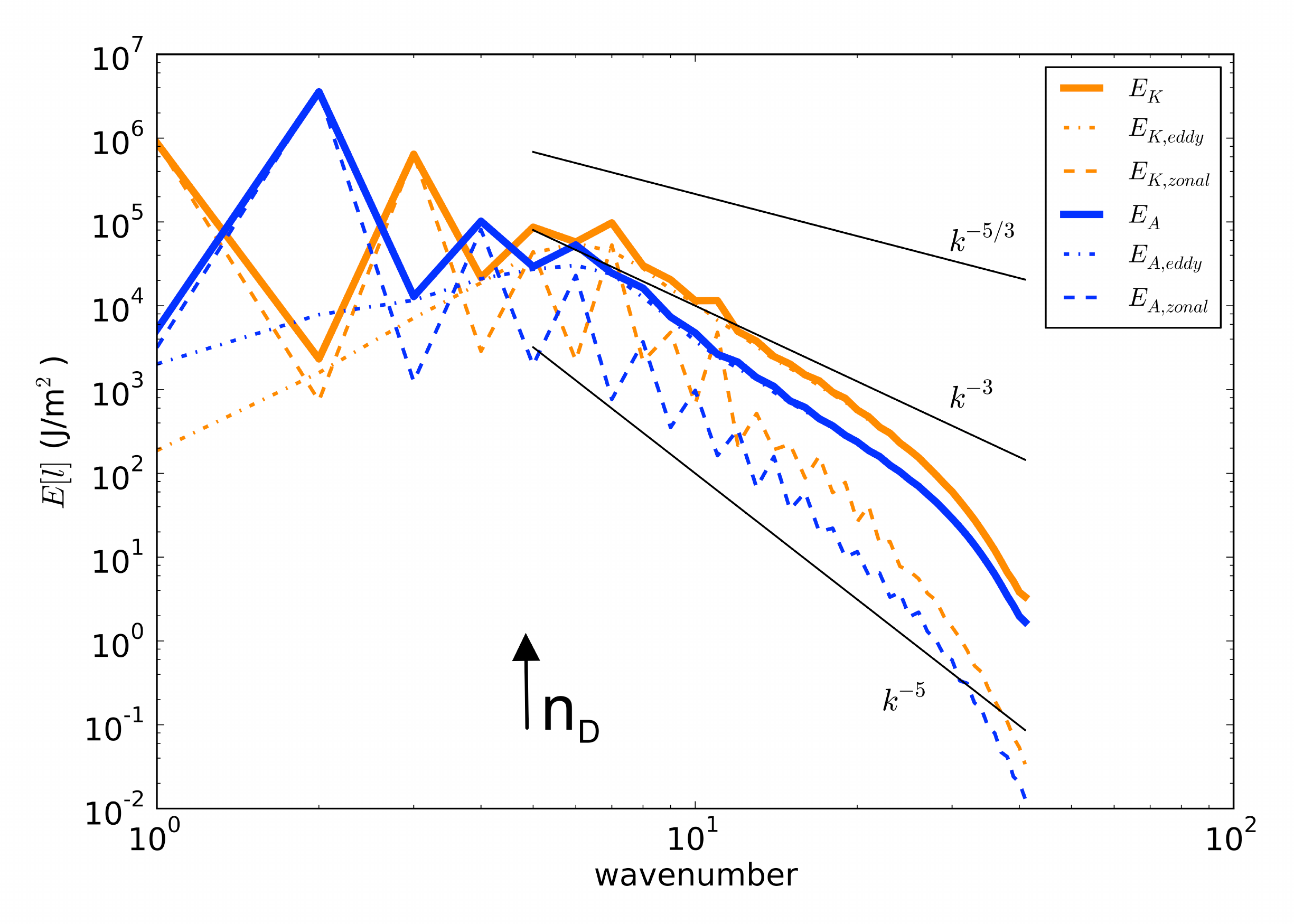}
}\\
\hspace{-4.5cm}\textbf{e)} $\Omega^*=1$, T127 Resolution \hspace{5cm} \textbf{f)} $\Omega^*=2$, T170 Resolution \\
\makebox[1.0\textwidth][c]{
\includegraphics[width=0.9\columnwidth]{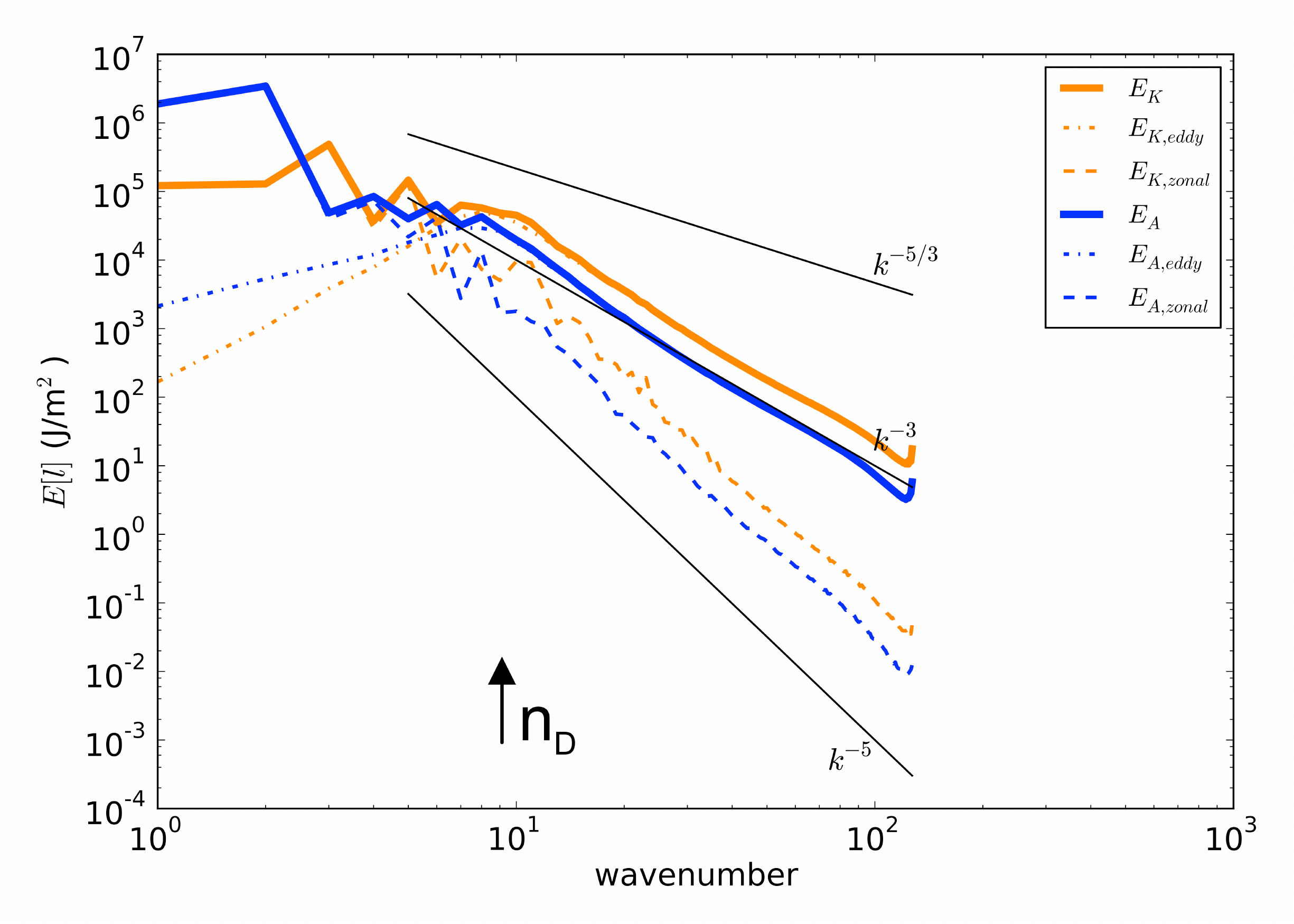}
\includegraphics[width=0.9\columnwidth]{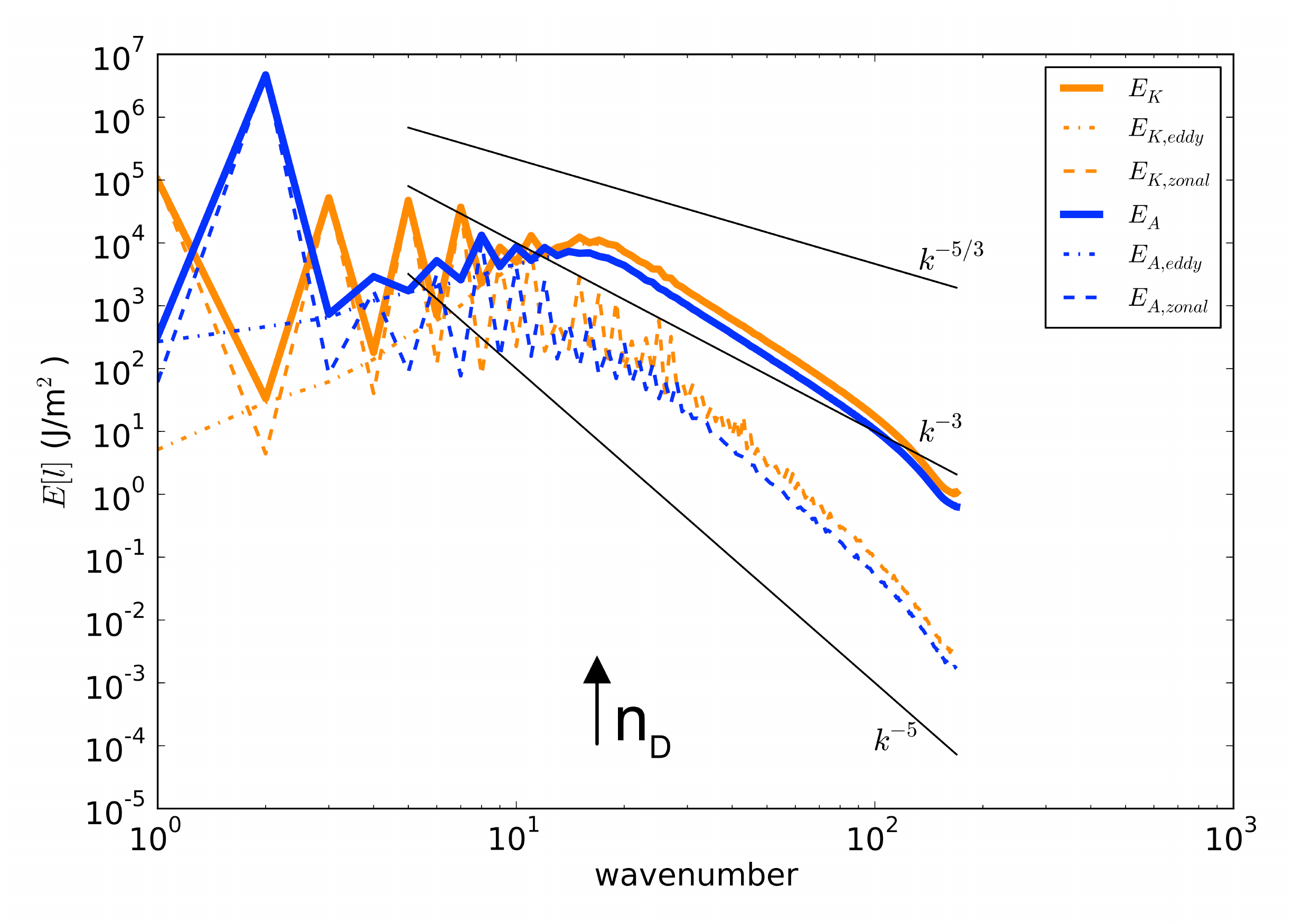}
}\\
\hspace{-4.5cm}\textbf{g)} $\Omega^*=4$, T170 Resolution \hspace{5cm} \textbf{h)} $\Omega^*=8$, T170 Resolution \\
\makebox[1.0\textwidth][c]{
\includegraphics[width=0.9\columnwidth]{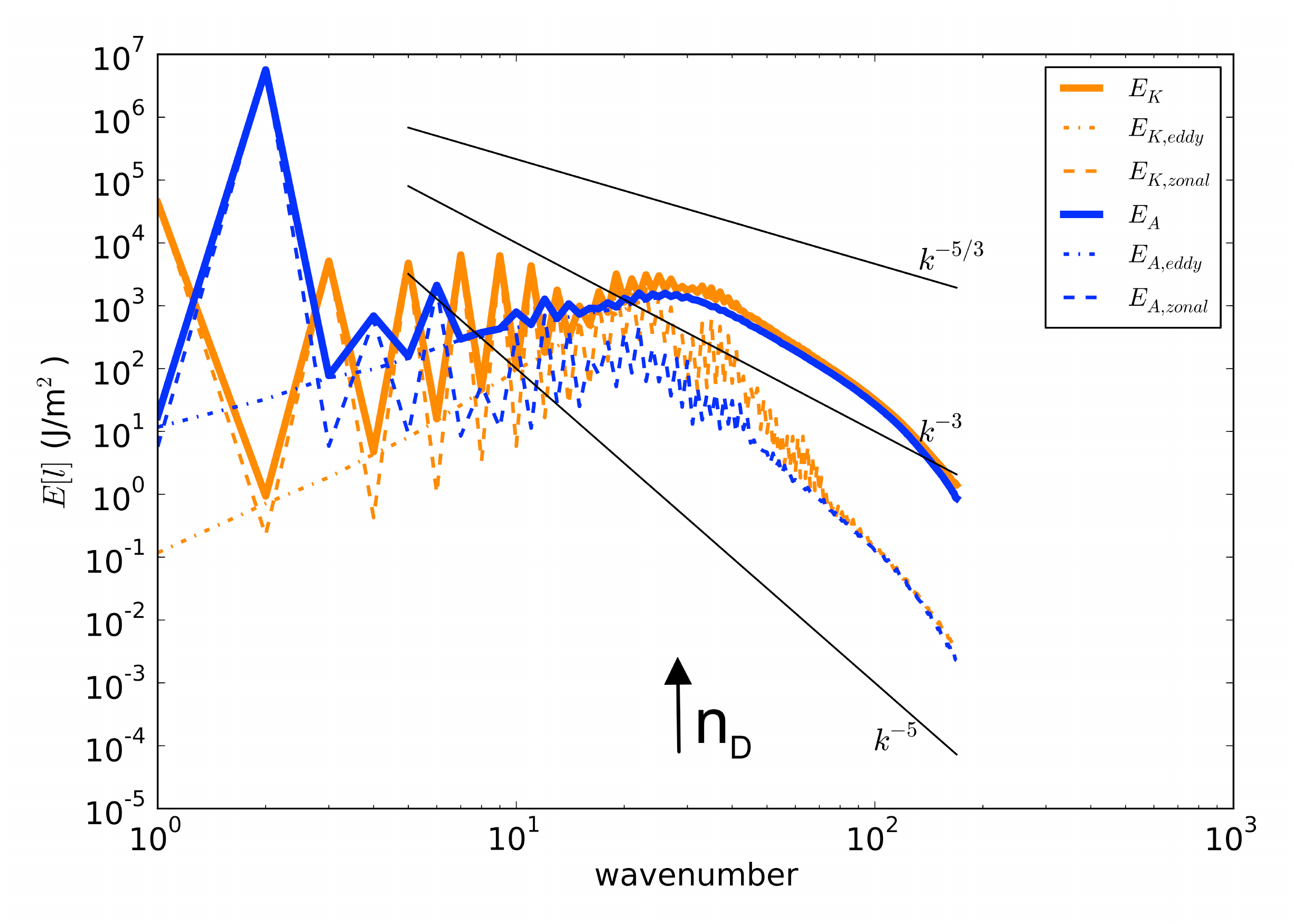}
\includegraphics[width=0.9\columnwidth]{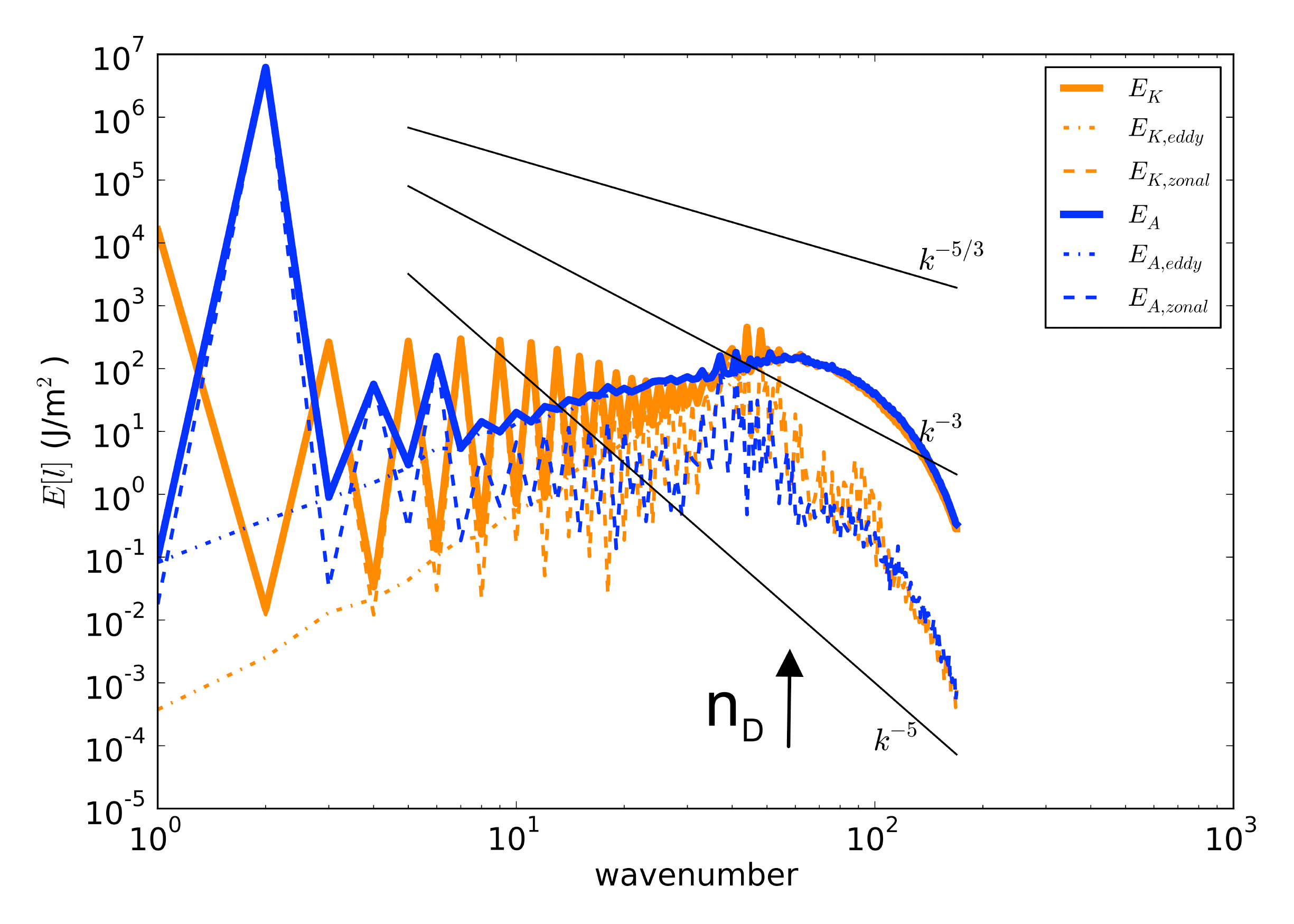}
}
\caption{Globally-averaged KE (orange) and APE (blue) spectra (each decomposed into zonal and eddy components) for PUMA-S runs  with $\Omega^\ast=1/16, 1/8, 1/4, 1/2$ (at horizontal resolution T42), $\Omega^\ast = 1$ (at resolution T127) and $\Omega^\ast=2, 4, 8$ (at resolution T170).}
\label{fig:enspec1}
\end{figure*}

\subsection{Slow rotation ($\Omega^\ast < 1$)}
The energy spectra for rotation rates $\Omega^\ast < 1$ are shown in Fig.~\ref{fig:enspec1}(a)-(d). These simulations were carried out at a horizontal resolution of T42, so it is likely that there are some artifacts in the spectra due to model diffusion at the highest wavenumbers. This is apparent as a steepening of the spectrum for wavenumbers $k \gtrsim 30$. But for the rest of the spectrum it appears to trend towards a self-similar form with a slope close to $n^{-5/3}$ from a $n^{-3}$ spectrum as $\Omega^\ast$ decreases. This trend towards a KBK-like slope would appear to suggest a trend towards a spectrum dominated by an energy-dominated cascade, though it is not immediately apparent whether this would entail upscale or downscale transfers of KE. This will be discussed further in the next section on spectral fluxes, {but assuming that kinetic energy is converted from potential energy by baroclinic instabilities at scales close to the Rossby deformation wavenumber, $n_D$, and given the trend in $n_D$ towards small $n$ as} $\Omega^\ast$ decreases (e.g. see \cite{wang2016,kaspi2015}), it would seem likely that the cascade would be predominantly downscale if injection of KE is taking place mainly at near-planetary scales. 

{Even numbered wavenumbers generally appear stronger in KE than odd numbered modes}, with the opposite trend in APE, which reflects the symmetries of the winds and temperature structure about the equator with annual mean forcing that is symmetric about the equator. The spectra are also dominated on average by kinetic energy for these rotation rates with a typical ratio {$E_K(n)/E_A(n) \sim 2-3$}. 

In contrast to these changes to the eddy part of the spectrum, the zonal components of both of the energies follows an $n^{-5}$ slope for the mid-range in wavenumber space. This seems to be a generic feature of the spectrum at almost all rotation rates and is discussed further below. {Most spectra (at all values of $\Omega^\ast$) show some evidence of a steep drop-off in energy at the highest wavenumbers, indicative of a region where the model dissipation is active, although this is more apparent in some runs than in others. This may indicate that some runs (notably the case for $\Omega^\ast = 1$) could be somewhat under-dissipated, which may have possible implications for the shape of the spectrum at other wavenumber ranges. Given limitations on computational resources available to us for this study, however, some compromises were necessary in our attempts to capture both realistic inertial ranges and an adequate range of dissipation (we are very far, of course, from the conditions appropriate for Direct Numerical Simulation). Such problems are not unique to this study \citep[see, e.g.][for examples at much higher resolution]{hamilton2008}, though the sensitivity of the KE and APE spectra on dissipation should ideally be investigated further in future work.}

%For lower rotation rates (see Fig.~\ref{fig:enspec2})
%£££

\subsection{Spectra at $\Omega^\ast \geq 1$}

%In this section we present KE and APE spectra of simulations with $\frac{1}{16}\Omega_E\le\Omega\le8\Omega_E$. 
Figure~\ref{fig:enspec1}(e)-(h) shows the KE and APE spectra of simulations with $1\Omega_E\le\Omega\le8\Omega_E$. These simulations were performed at T170 resolution (except for the $\Omega^*=1$ simulation at T127). The Earth-like run at $\Omega^\ast = 1$ {(Figure~\ref{fig:enspec1}e)} exhibits a $n^{-3}$ slope between wavenumbers 20 and 90 as well as a fairly consistent $n^{-5}$ slope in the zonal component. It is interesting to note that both KE and APE behave fairly similarly in this region. At lower wavenumbers ($k \lesssim 10$) the spectrum flattens with a hint of a segment tending towards the KBK $n^{-5/3}$ form between $2-3 \lesssim k \lesssim 10$, suggestive of an energy-dominated upscale cascade.
%Fig. \ref{fig:1omg-kesp-t127} shows the KE spectrum of the run with $\Omega^{\ast}=1$, the terrestrial scenario. 
%The total KE spectrum (the black curve) demonstrates the classic $-3$ slope between $n=10-30$, 

This is broadly consistent
with various previous studies based on observational/reanalysis datasets of Earth's atmosphere (e.g. see \cite{baer1974}, 
\cite{Boer1983}, \cite{Koshyk1999}), indicating the probable existence of a forward enstrophy cascade inertial 
range. The zonal spectrum in both APE and KE is characterised by a much steeper $-5$ slope, however, which is still not well-understood despite the prediction of a $-5$ slope in the early work e.g. of \cite{Rhines1975}. Rhines showed that, near the cross-over scale from Rossby waves to turbulence (i.e. near the Rhines wavenumber $k_R \simeq (\beta/U)^{1/2}$, where $k$ represents the total dimensional wavenumber $k = n/a$), the typical wind speed is 
$U\sim \beta/k^2$. Since $E=kE(k)\simeq1/2U^2$, the $-5$ power law can then be revealed as $E(k)\sim \beta^2k^{-5}$. 
But this does not explain the extended $-5$ slope solely in the zonal KE spectrum. A study by \cite{huang2001} {\citep[and further developed e.g. by][]{Sukoriansky2002,Galperin2004,Galperin2006,Galperin2010} identified the $-5$ slope associated with the zonal KE spectrum with a so-called zonostrophic regime, prevalent with strong planetary rotation and weak dissipation. They further suggested that the shape of the zonal spectrum} can be qualitatively explained by the stabilising 
effect of $\beta$ on the zonal jets,  According to \cite{huang2001}, {\citet{Sukoriansky2002}, \citet{Galperin2006} and others,} the $\beta$-effect modifies the necessary 
condition for barotropic instability from $\partial_{yy}[u]=0$ to $\beta-\partial_{yy}[u]=0$ somewhere within the flow domain. This 
means the stability criterion is eased by the $\beta$-effect from $\partial_{yy}[u]\neq0$ to $\beta-\partial_{yy}[u]\neq0$, which
allows the existence of velocity inflection points ($\partial_{yy}[u]$=0) and enables more energy to reside in the zonal 
modes without violating the stability criterion. But this %does not fully explain why a $-5$ slope is favoured over
%any other steep slope as observed here, which leads to the speculation of 
{argument is somewhat heuristic, leading some \citep[e.g.][]{Danilov2004} to question whether a well-defined power-law scaling
relationship even exists for the zonal KE spectrum.}%  (\cite{Danilov2004}). 
%Curiously, the $-5$ power-law actually corresponds to a characteristic wind velocity $U$ such that 
%$\beta\sim U_{yy}$, implying the atmosphere is possiblly in a state of marginal 
%barotropic instability. 
% further explanation and references.

Figs. \ref{fig:enspec1}f-h show the APE and KE spectra of the regime of multiple zonal jets. The strong zigzag feature of the zonal KE spectrum at small spherical wavenumbers is due to the hemispheric symmetry of the predominantly zonal structures across the globe. The classic $-5/3$ slope of an inverse energy cascade in KE cannot be found, indicating that {neither the `classical' picture of 2-D isotropic turbulence, nor the zonostrophic regime of \citet{Sukoriansky2002} and \citet{Galperin2010}, are} applicable to the multiple jet flows found in this regime (at least under the conditions explored here). As shown by \cite{wang2016}, the energy-containing wavenumber, Rossby deformation wavenumber, and the Rhines wavenumber are not widely separated in this regime, {although the KE spectrum is likely to be energised on scales close to $L_D$ through conversion of APE by baroclinic instabilities.} Such a lack of scale separation might suggest that the inverse energy cascade, {initiated around the deformation wavenumber, $n_D$ \citep{wang2016},} through eddy-eddy interactions, has been significantly suppressed because of a lack of room to develop an energy-conserving inertial range, although this will be further investigated below with respect to the computed spectral fluxes. {Such closeness of the Rhines and deformation scales also suggests that the simulated flows are far from the conditions necessary to observe fully developed zonostrophic dynamics \citep[e.g.][]{Galperin2006,Galperin2010}.}

With increasing rotation rate (see Figs.~\ref{fig:enspec1}f-h), the maximum of the zonal component moves to higher wavenumbers and the $n^{-3}$ slope that could be so well identified at $\Omega^*=1$ becomes inclined towards an even steeper slope at higher wavenumbers. This is likely due to the effect of the model-inherent hyperdiffusion,  %(see Sect.~\ref{sec:hyp}). Consequently, 
as a result of which the region over which a $n^{-3}$ slope can be discerned becomes smaller and smaller. %\cite{wang2014} calculated the spectral enstrophy flux \citep[see ][]{burgess2013} for the $\Omega_E$ simulations to show that the $k^{-3}$ slope is indeed consistent with a downscale eddy enstrophy flux and an inertial range. 
The same hyperdiffusion effect can be identified for the  {$n^{-5}$} slope of the  zonal component {(see also the discussion of this power law above).} %The $k^{-5}$ slope is discussed by \cite{wang2014}, who mentions that 
%\cite{Rhines1975} predicted a $n^{-5}$ energy slope near the Rhines scale, but in our cases here the $n^{-5}$ slope appears to extend down to the smallest modeled scales that are unaffected by model diffusion. As argued by \cite{huang2001}, when including the $\beta$-effect, the Rayleigh-Kuo criterion, one of the necessary criteria for barotropic instability, is expanded from $\partial_{yy}[u]=0$ to $\beta-\partial_{yy}[u]=0$. This enables the atmosphere to remain stable even for inflection points in the zonal wind velocity (i.e. $\partial_{yy}[u]=0$). This may allow more energy to be stored in the zonal energy component at smaller scales \citep{huang2001}. 
Overall, however, the slope of the zonal kinetic energy spectrum in these regimes is not well understood. Nevertheless, our work can report 
that the zonal component of the APE spectrum does have the same slope. 

The ratio of KE to APE also varies significantly with wavenumber in this regime, with APE dominating over KE at {$n = 2$} and with APE/KE ranging from $\sim 30$ at $\Omega^\ast = 1$ to more than 10$^8$ at $\Omega^\ast = 8$. At higher wavenumbers, however, within the {$n^{-3}$} region then KE is seen to dominate with a KE/APE ratio that ranges from around 3 at $\Omega^\ast = 1$ down to O(1) at $\Omega^\ast = 8$. %This reasoning, however, does not explain why the zonal energy spectrum has a $k^{-5}$ slope 

\section{Spectral transfer fluxes of energy and enstrophy for varying rotation rates}\label{sec:turb-fluxes}

In this section, we present enstrophy and energy spectral fluxes of the PUMA-S simulations discussed previously. This is intended to provide a more detailed view of the general spectral transfer pathways within our simulated atmospheric circulations across a range of parameter space, in particular using the spectral energy budget formulation of \citet{augier2013}. 
From such a spectral energy budget, we can answer the question of how the energy of macroturbulent fluid motion is transported between scales and converted between APE and KE. More specifically, we are interested in seeing at which scale kinetic energy is inserted into the system and where this energy ends up. We distinguish between two modes of transfer between scales, depending upon whether transfer is spectrally local, between nearby scales (representing a conventional energy \emph{cascade}), or non-local,  in which energy is directly transferred from one scale to another across large wavenumber intervals (akin to a ``waterfall'' - M. E. McIntyre, priv. comm.).The latter can occur, for instance, between disturbances of arbitrary wavenumber and the ($m=0$) zonal flow. %An example of this is zonal jet formation
%Atmospheric macroturbulence differentiates between 

To identify this interaction between eddies and the zonal flow (the eddy-mean flow interaction) %\comment{explain eddy, zonal decomposition. zonal means residual component i.e. eddy-mean and mean-mean interactions, zonal interaction term!}, 
we perform an additional decomposition into zonal and eddy components. 
This decomposition was achieved by taking the eddy component (via $X_{eddy}=X-[X]$ of each input variable (i.e. $u,v,\omega,\Phi,T$) and recalculate all the fluxes from this (for which the $\gamma$ value obtained from the initial flux calculation is used). The zonal component is then obtained as the residual. For spectral fluxes, the ``eddy'' component encompasses eddy-eddy interactions, while the ``zonal'' component consists of residual interactions with the zonal mean flow (i.e. combining eddy-zonal and zonal-zonal interactions). %From this point onward $k$ is used to represent the total wavenumber (instead of $l$).
%TODO ASK ROLAND WHY eddy is calculated
In the text below the terms wavenumber, total wavenumber and $k$ are used synonymously.
%\comment{wavenumber, total wavnumber and k are used synonomously}

\subsection{Spectral enstrophy fluxes}
\begin{figure*}[]
\centering
\hspace{-4.5cm}\textbf{a)} $\Omega^*=\frac{1}{16}$, T42 Resolution \hspace{5cm} \textbf{b)} $\Omega^*=\frac{1}{8}$, T42 Resolution \\
\makebox[1.0\textwidth][c]{
\includegraphics[width=\columnwidth,height=5cm]{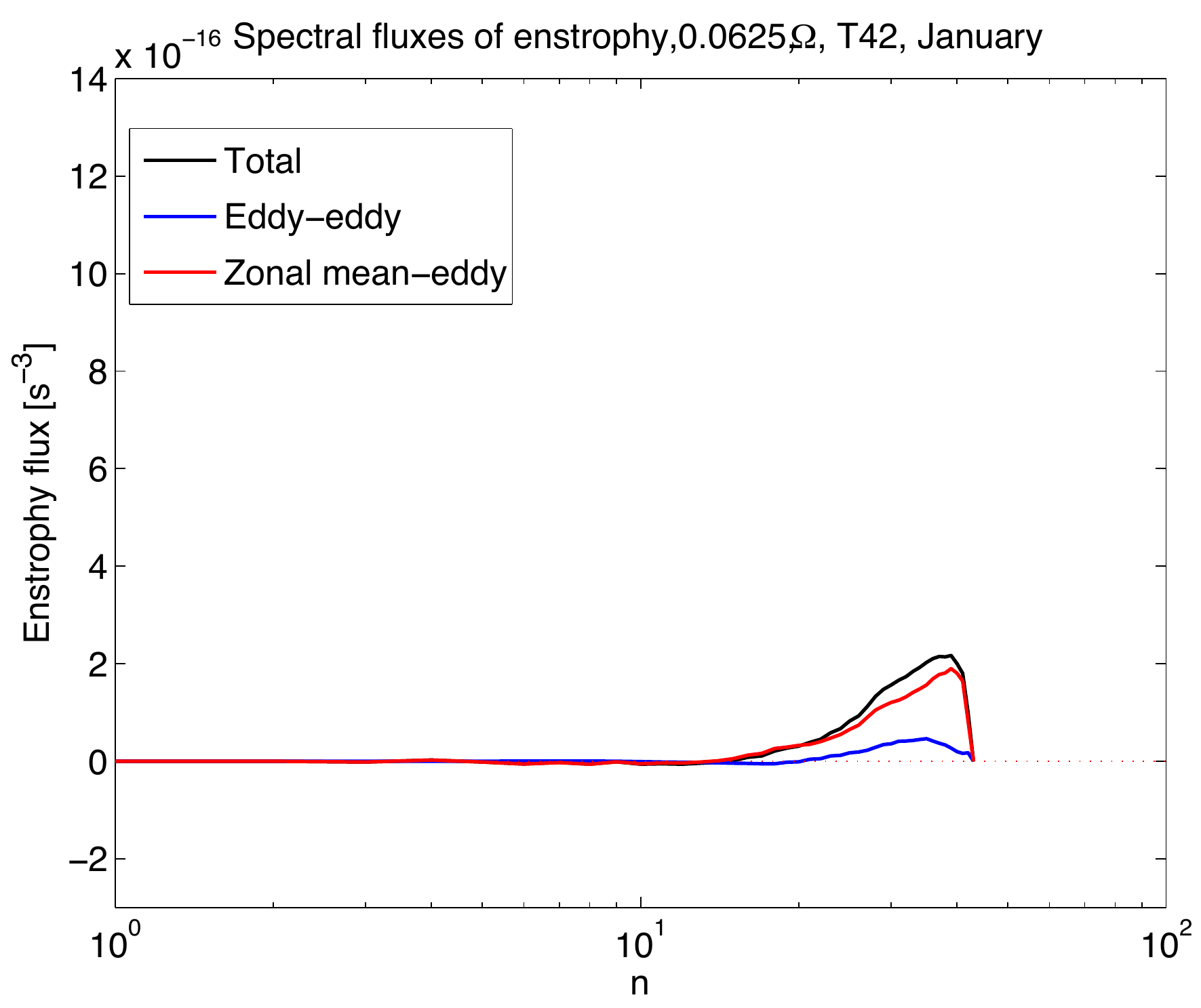}
\includegraphics[width=\columnwidth,height=5cm]{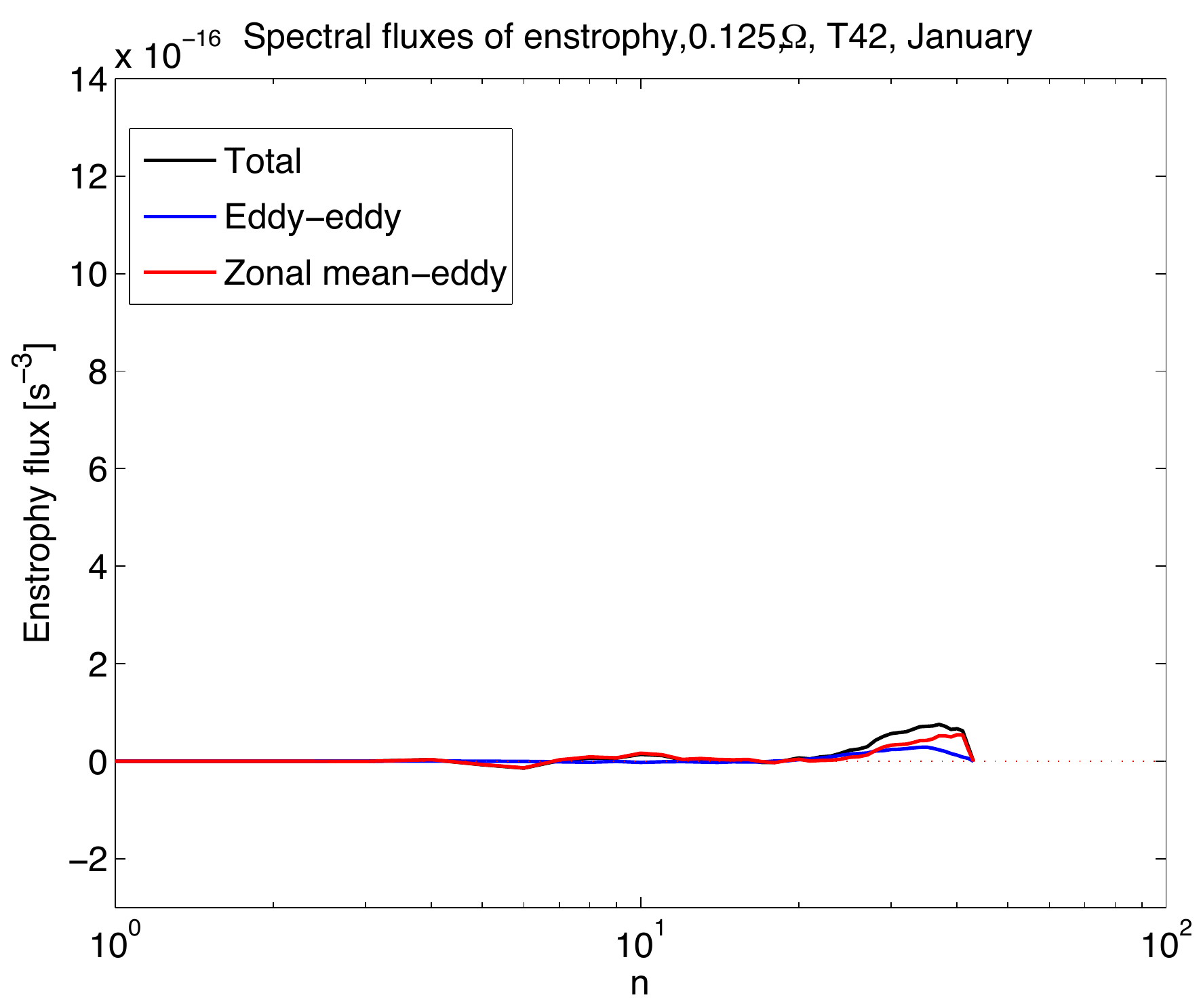}

}\\
\hspace{-4.5cm}\textbf{c)} $\Omega^*=\frac{1}{4}$, T42 Resolution \hspace{5cm} \textbf{d)} $\Omega^*=\frac{1}{2}$, T42 Resolution \\
\makebox[1.0\textwidth][c]{
\includegraphics[width=\columnwidth,height=5cm]{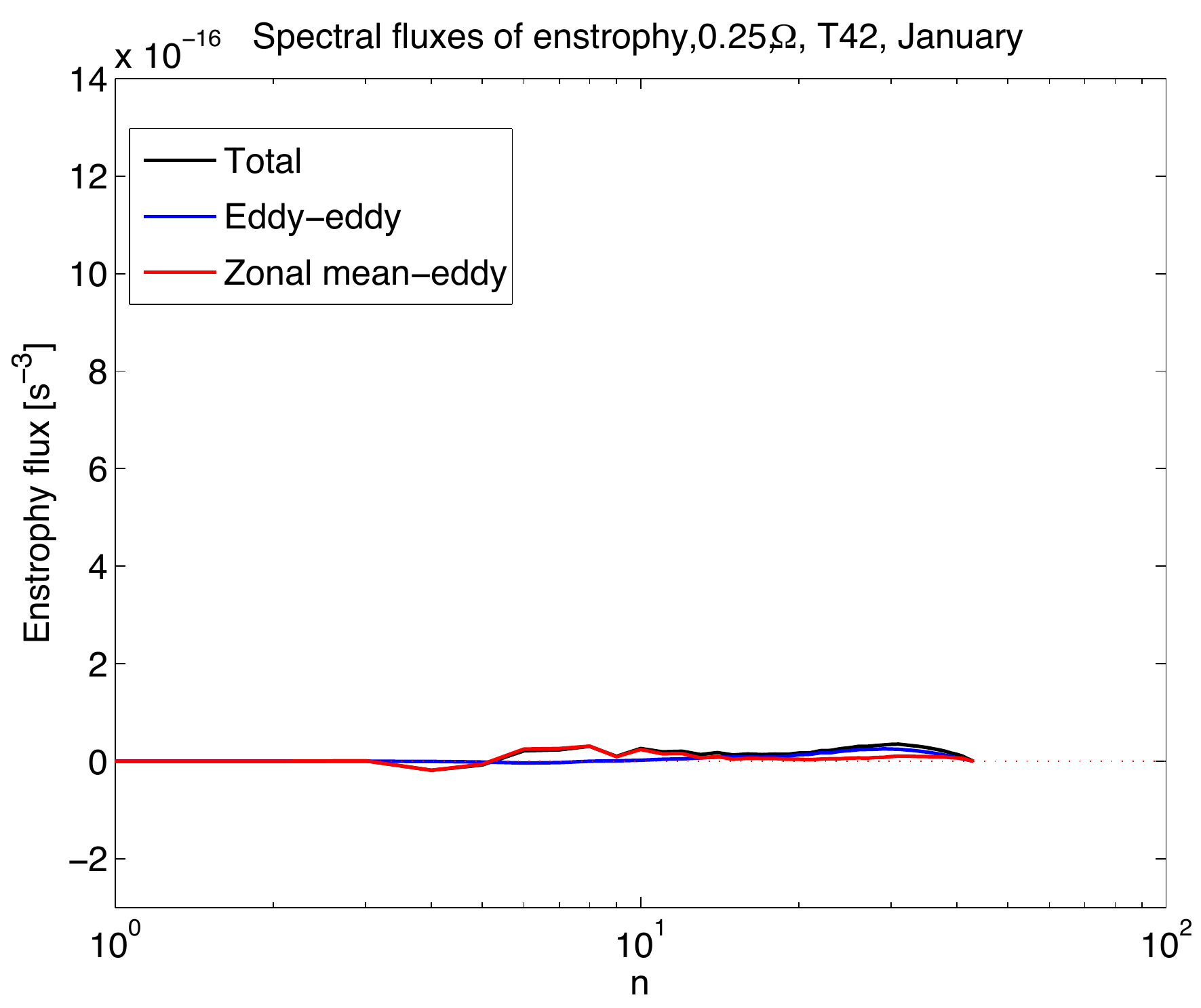}
\includegraphics[width=\columnwidth,height=5cm]{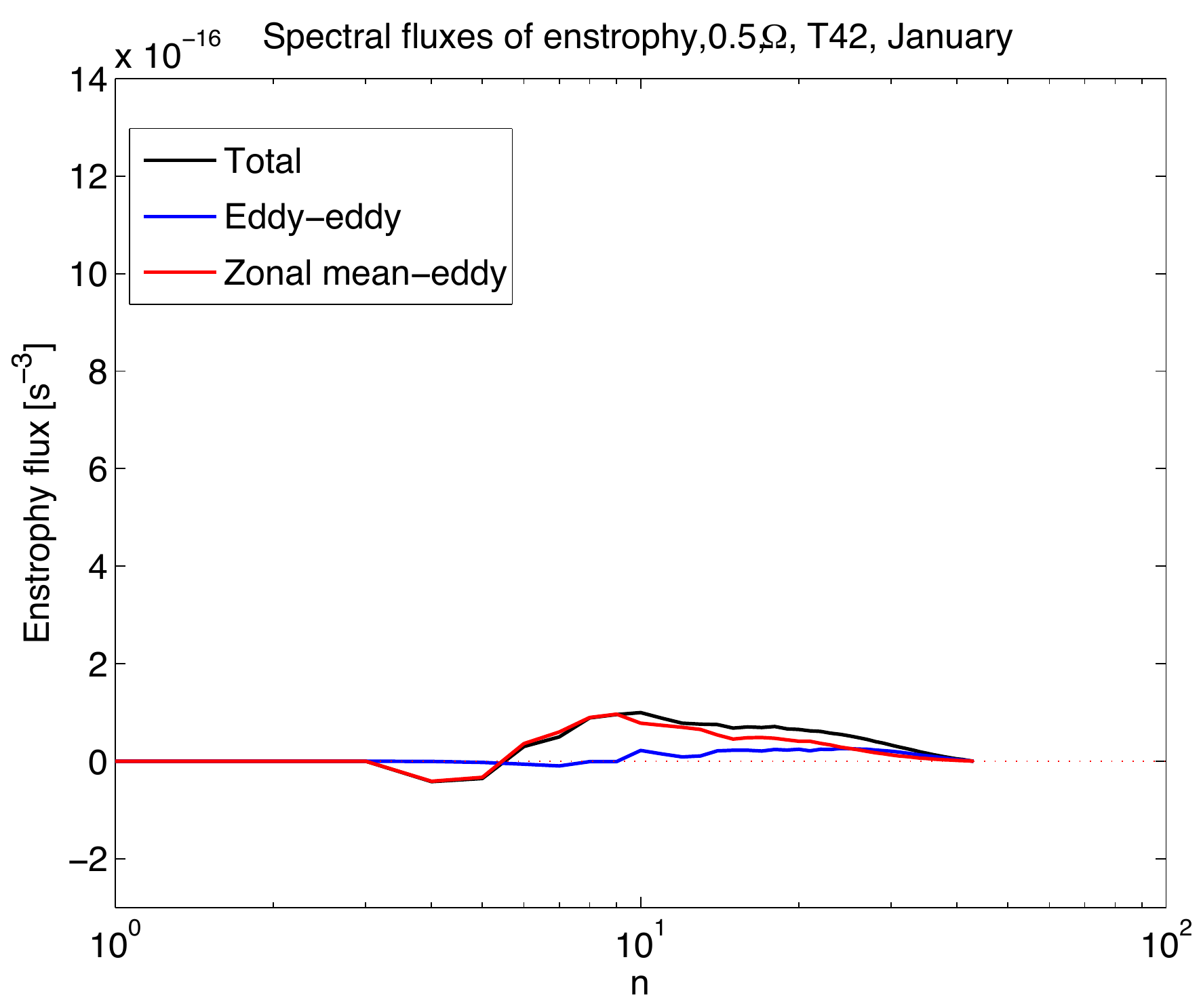}
}\\
\hspace{-4.5cm}\textbf{e)} $\Omega^*=1$, T42 Resolution \hspace{5cm} \textbf{f)} $\Omega^*=2$, T170 Resolution \\
\makebox[1.0\textwidth][c]{
\includegraphics[width=\columnwidth,height=5cm]{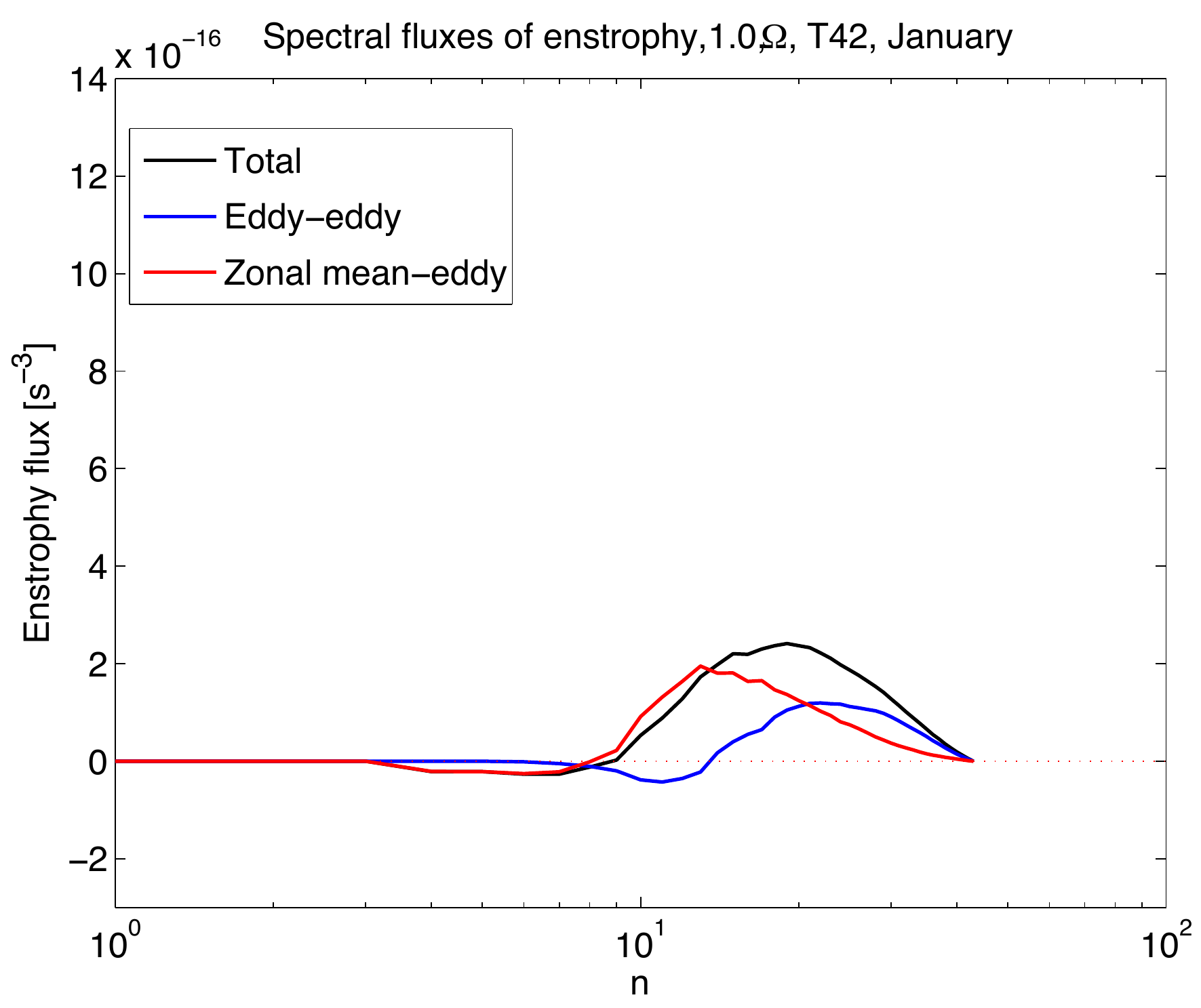}
\includegraphics[width=\columnwidth,height=5cm]{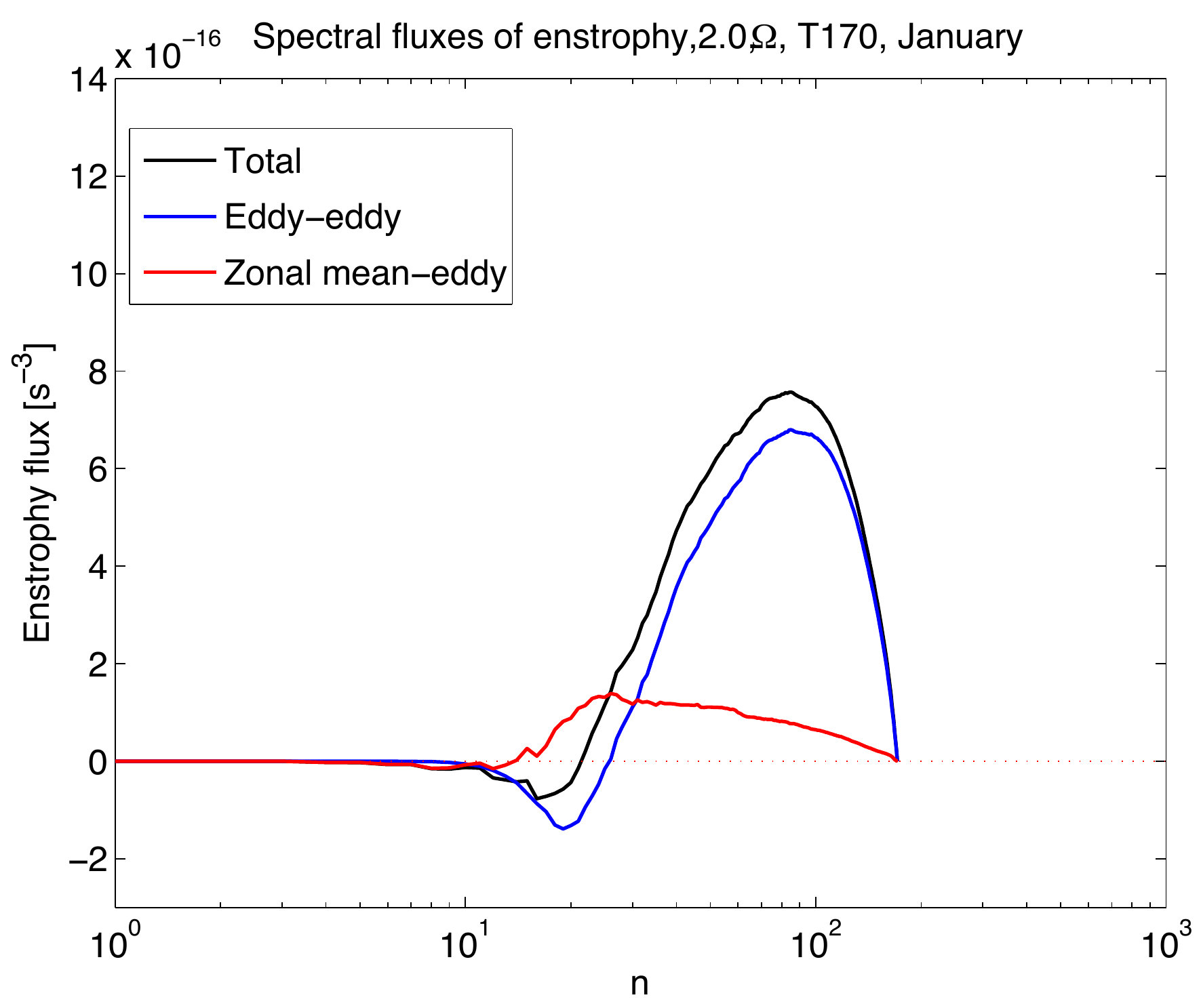}
}\\
\hspace{-4.5cm}\textbf{g)} $\Omega^*=4$, T170 Resolution \hspace{5cm} \textbf{h)} $\Omega^*=8$, T170 Resolution \\
\makebox[1.0\textwidth][c]{
\includegraphics[width=\columnwidth,height=5cm]{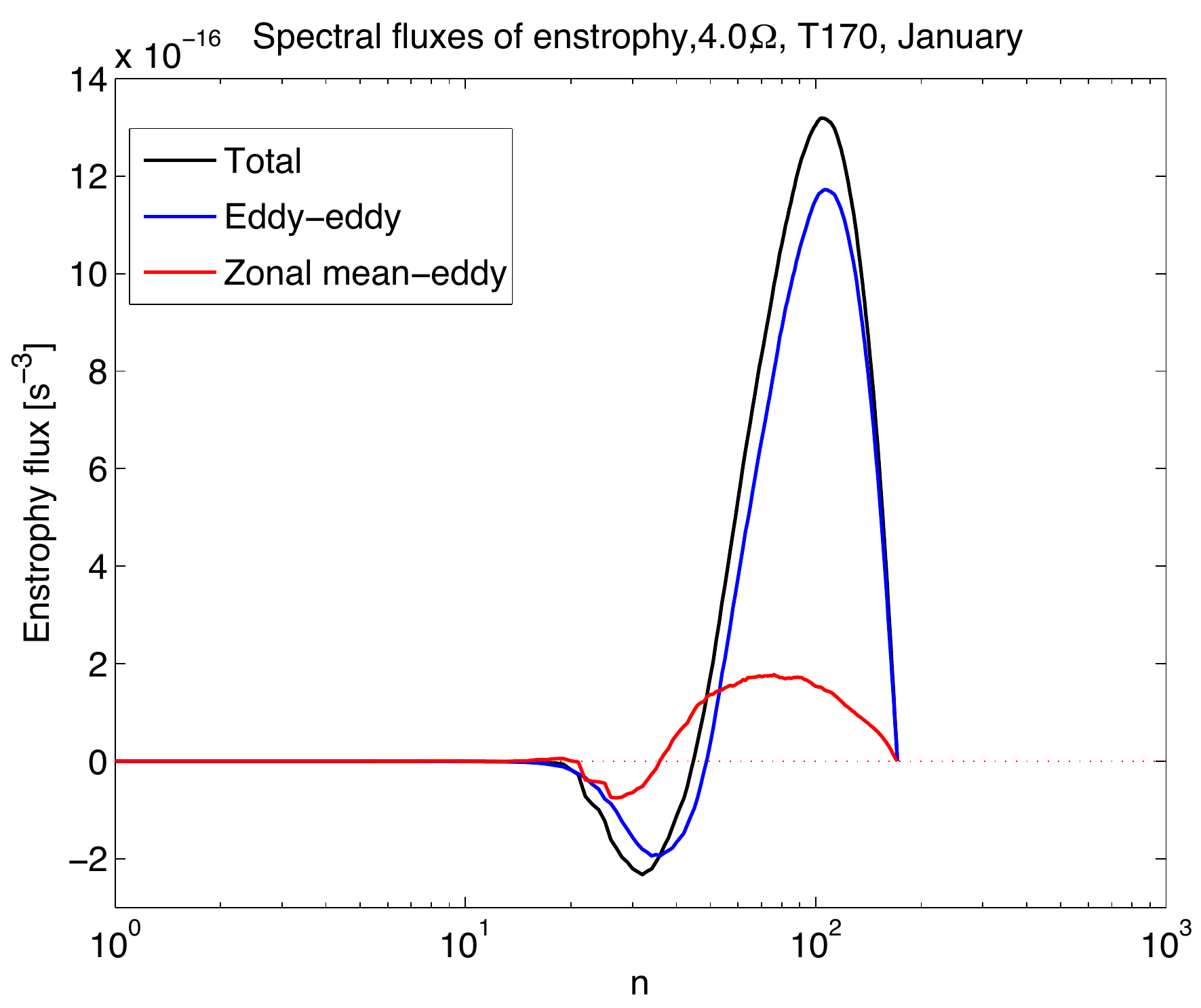}
\includegraphics[width=\columnwidth,height=5cm]{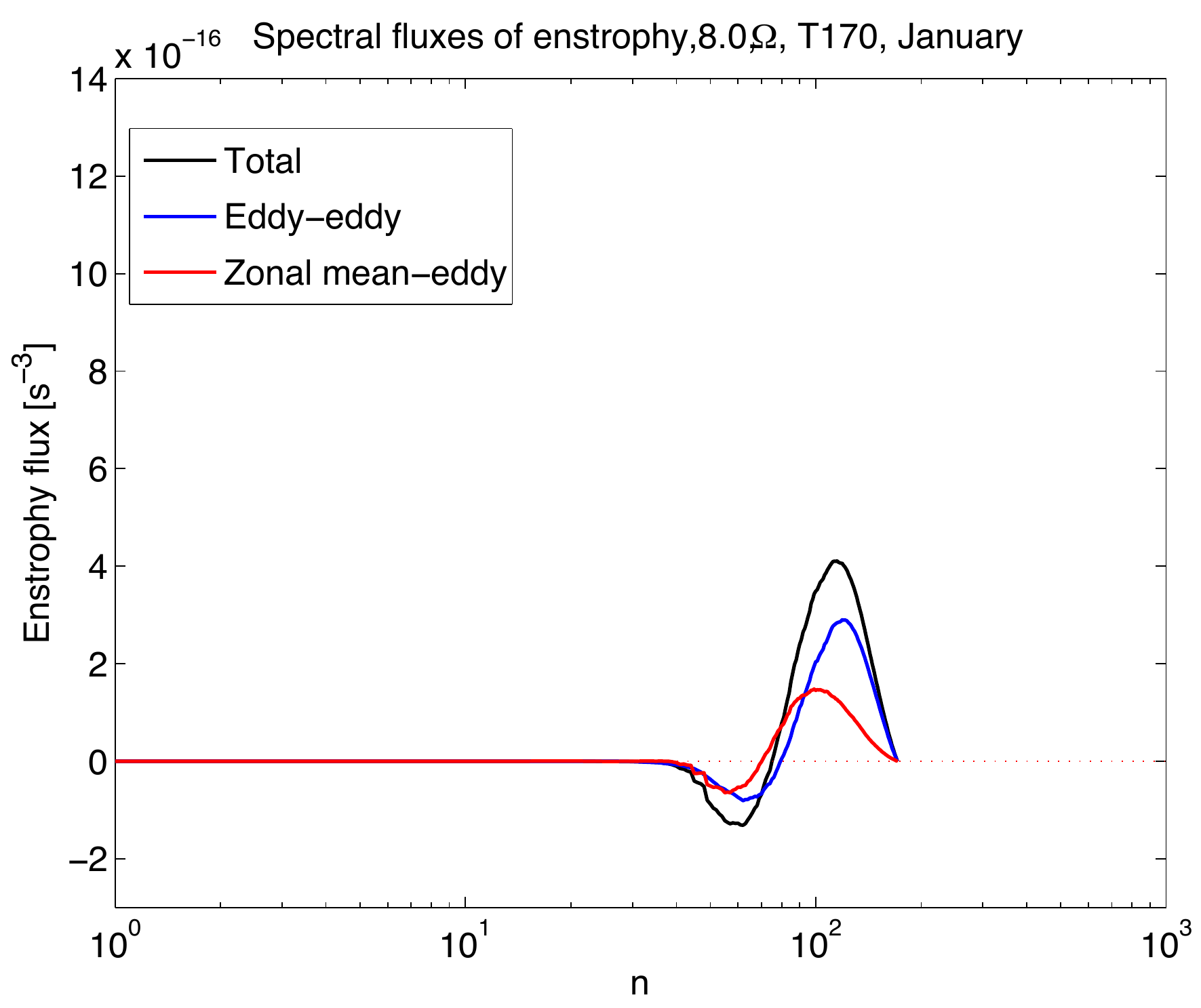}
}
\caption{Spectral fluxes of enstrophy $\mathcal{H}^k$, (also decomposed into eddy-eddy and residual zonal interaction components) for PUMA-S runs  with $\Omega^\ast=1/16, 1/8, 1/4, 1/2, 1$ (at horizontal resolution T42) and $\Omega^\ast=2, 4, 8$ (at resolution T170).}
\label{fig:enstflux1}
\end{figure*}

Figure \ref{fig:enstflux1} shows a sequence of profiles of spectral enstrophy fluxes, $\mathcal{H}_n$, covering the full range of $\Omega^*$. Theoretical discussions of quasi-geostrophic turbulence \citep[e.g.][]{Charney1971,salmon1978,salmon1980} suggest that the flux of enstrophy should be downscale throughout the range of scales, although this depends upon the flow satisfying conditions for quasi-geostrophy. In the cases shown in Fig. \ref{fig:enstflux1}, this trend is more or less consistent with this expectation, but with some exceptions. At low rotation rates ($\Omega^* \leq 1/4$ or $\mathcal{R}o_T \gtrsim 1$), where the quasi-geostrophic approximation is not likely to be valid, enstrophy fluxes are generally quite weak at all scales, though with a slight trend to increase towards the smallest resolved scales where dissipation becomes significant. At these small scales, the zonal-eddy interaction appears to dominate. 

For $\Omega^* \gtrsim 1/2$ (or $\mathcal{R}o_T \lesssim 1$), however, enstrophy fluxes become significantly larger and positive at moderate to small scales, as anticipated for quasi-geostrophic turbulence. At larger scales, however, there is a small tendency for $\mathcal{H}_n$ to become negative, indicating a weak upscale cascade range and implying a spectrally-local net source of enstrophy at the wavenumber $n_s$ at which $\mathcal{H}_n$ changes sign. This wavenumber gradually increases with $\Omega^*$ from around $n = 5-6$ at $\Omega^* = 1/2$ to $n \simeq 75$ at $\Omega^* = 8$. The magnitude and distribution of enstrophy flux between eddy-eddy and zonal-eddy terms also changes with $\Omega^*$. Enstrophy fluxes appear relatively weak for $\Omega^* \lesssim 1$ and dominated by zonal-eddy interactions. At higher rotation rates, the eddy-eddy interactions become more dominant, especially at higher wavenumbers, indicating a conventional, spectrally-local cascade of enstrophy, which becomes much stronger at $\Omega^* = 2$ and 4 and self-similar in shape as it moves towards higher wavenumbers as $\Omega^*$ increases. Fluxes become weaker at the highest $\Omega^*$ as $n_s$ approaches the resolution limit and dissipation presumably acts to damp the dynamics. This should be investigated further, though will require model resolutions that were beyond the scope of what was feasible in the present study. 

\subsection{Spectral energy fluxes}
\begin{figure*}[]
\centering
\hspace{-4.5cm}\textbf{a)} $\Omega^*=\frac{1}{16}$, T42 Resolution \hspace{5cm} \textbf{b)} $\Omega^*=\frac{1}{8}$, T42 Resolution \\
\makebox[1.0\textwidth][c]{
\includegraphics[width=\columnwidth,height=5cm]{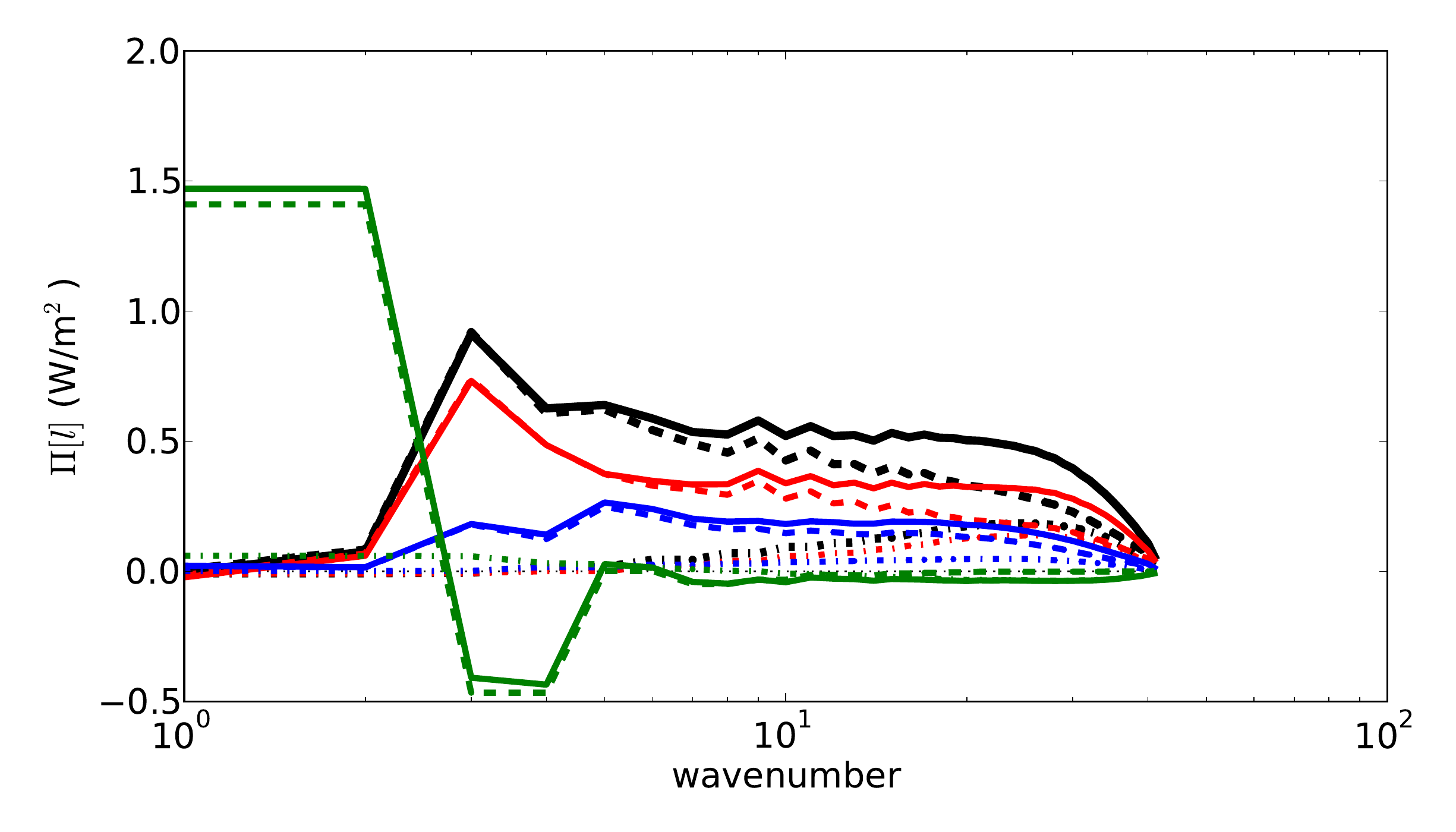}
\includegraphics[width=\columnwidth,height=5cm]{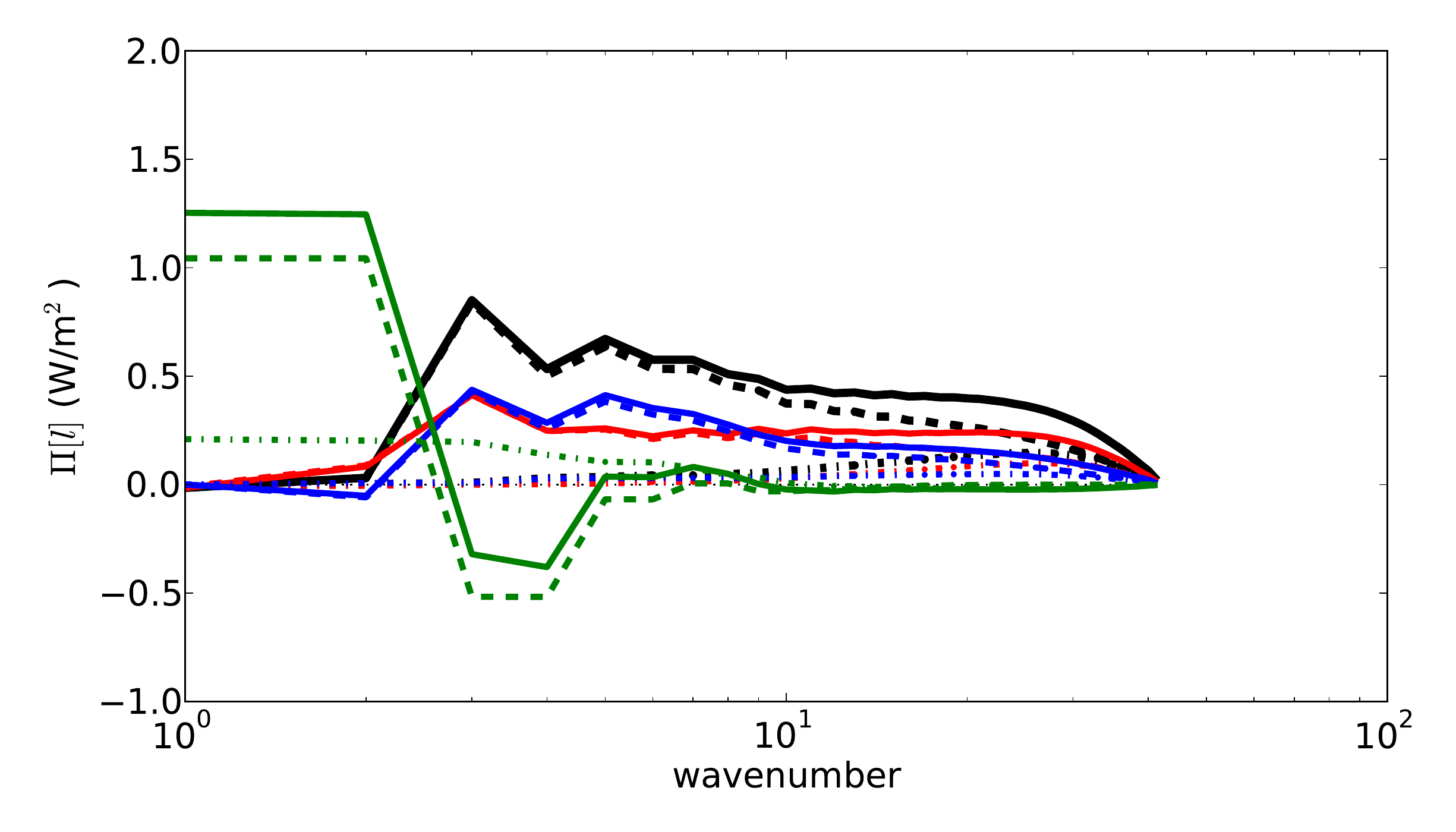}
}\\
\hspace{-4.5cm}\textbf{c)} $\Omega^*=\frac{1}{4}$, T42 Resolution \hspace{5cm} \textbf{d)} $\Omega^*=\frac{1}{2}$, T42 Resolution \\
\makebox[1.0\textwidth][c]{
\includegraphics[width=\columnwidth,height=5cm]{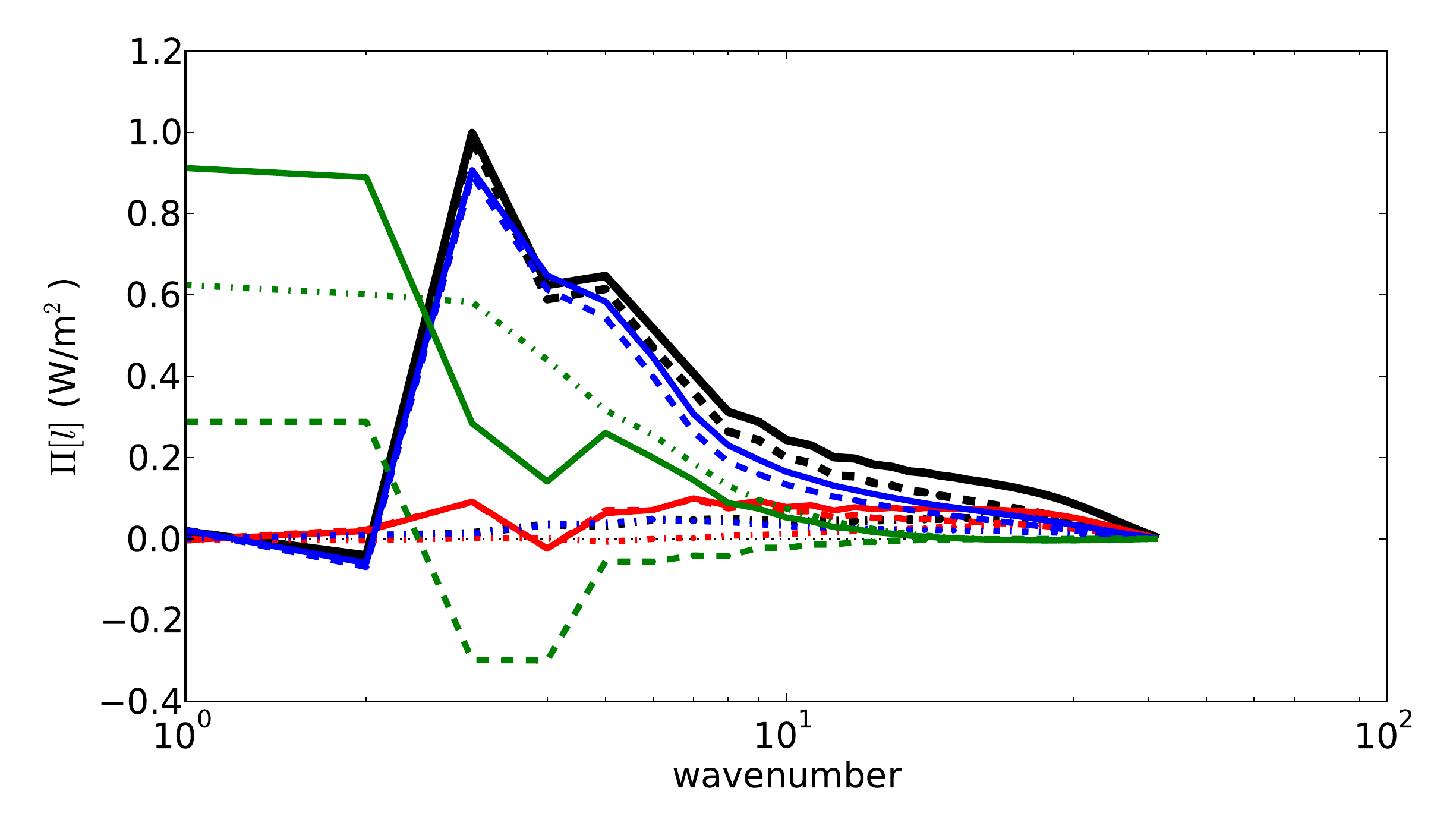}
\includegraphics[width=\columnwidth,height=5cm]{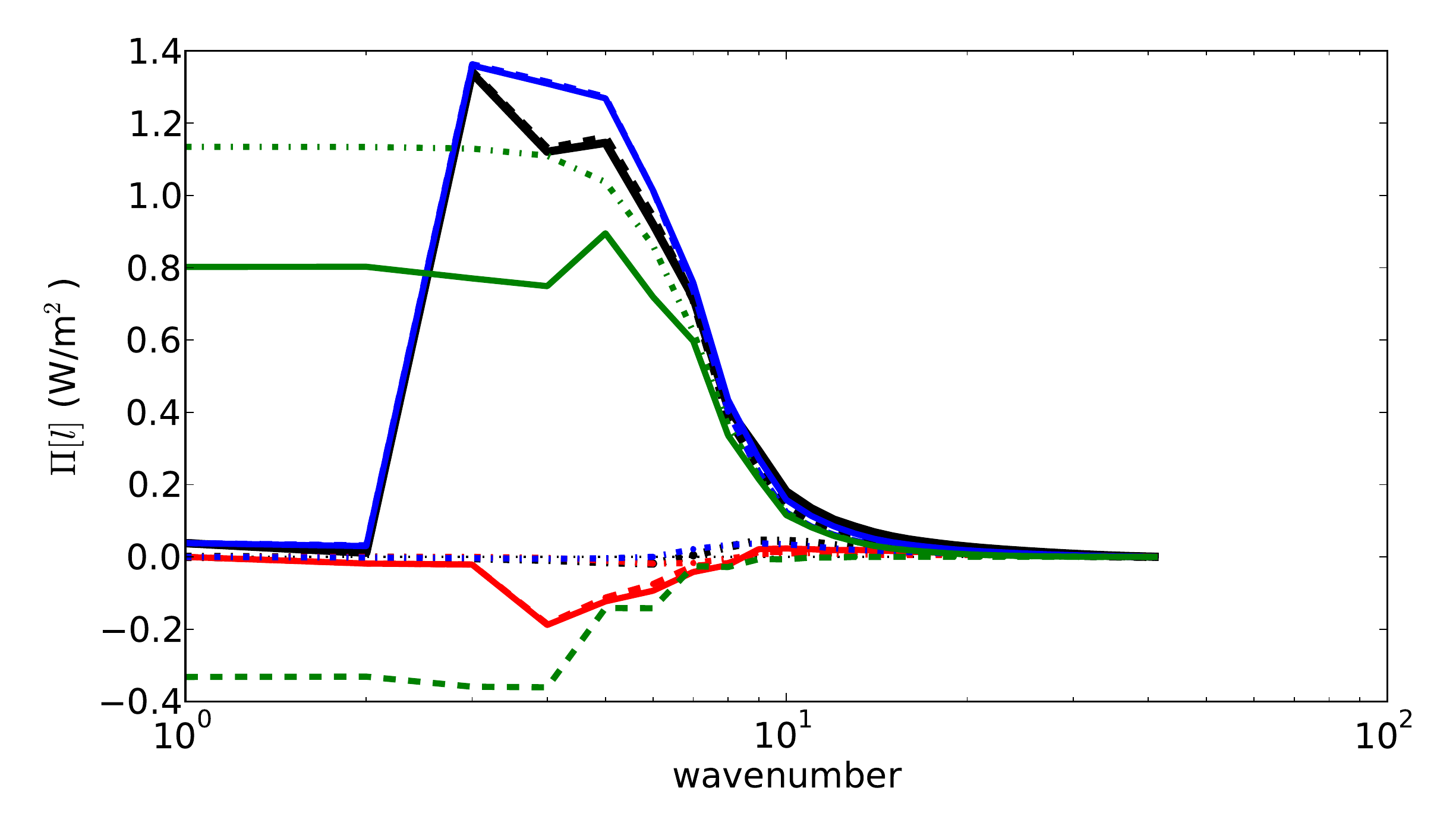}
}\\
\hspace{-4.5cm}\textbf{e)} $\Omega^*=1$, T127 Resolution \hspace{5cm} \textbf{f)} $\Omega^*=2$, T170 Resolution \\
\makebox[1.0\textwidth][c]{
\includegraphics[width=\columnwidth,height=5cm]{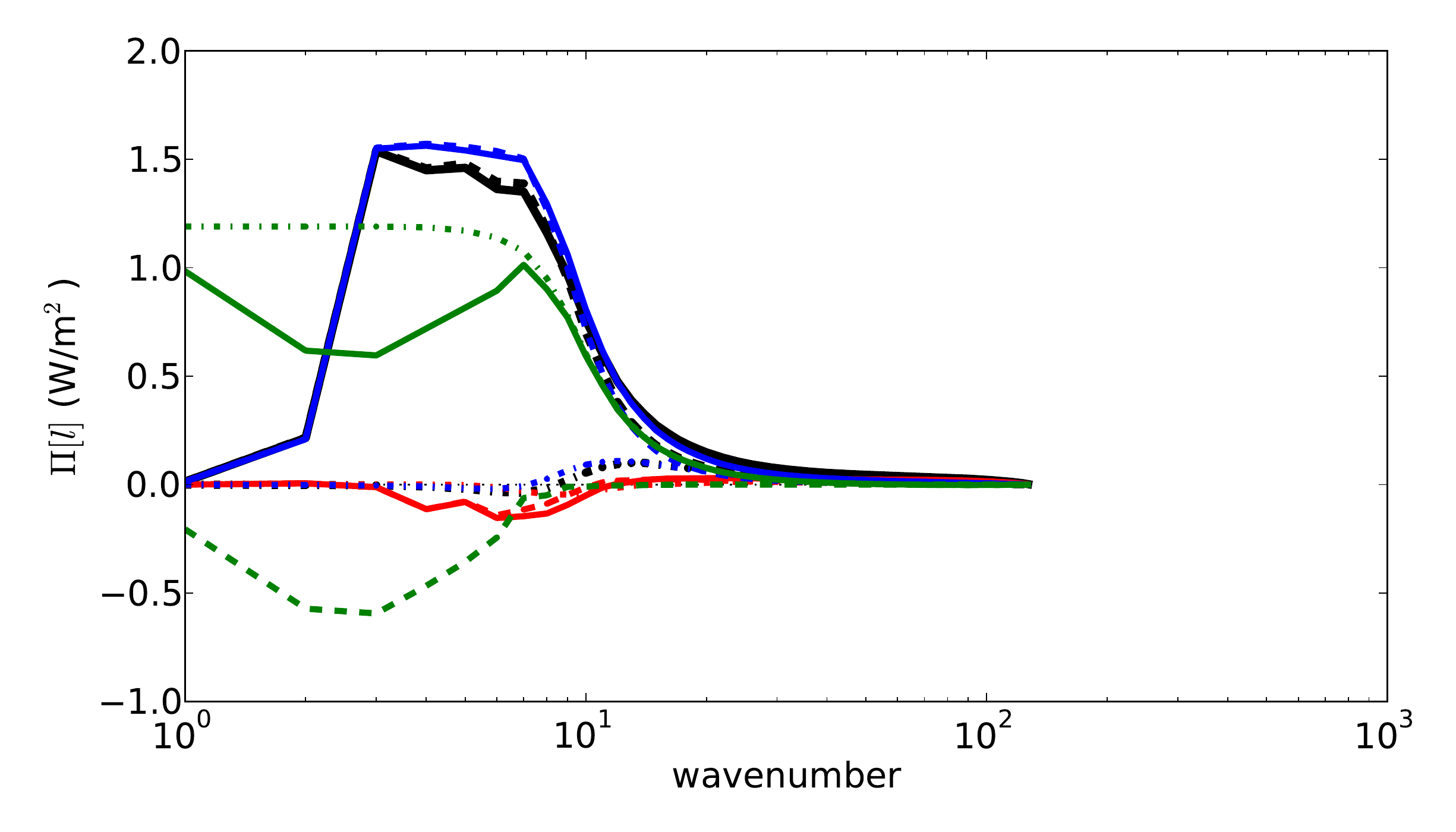}
\includegraphics[width=\columnwidth,height=5cm]{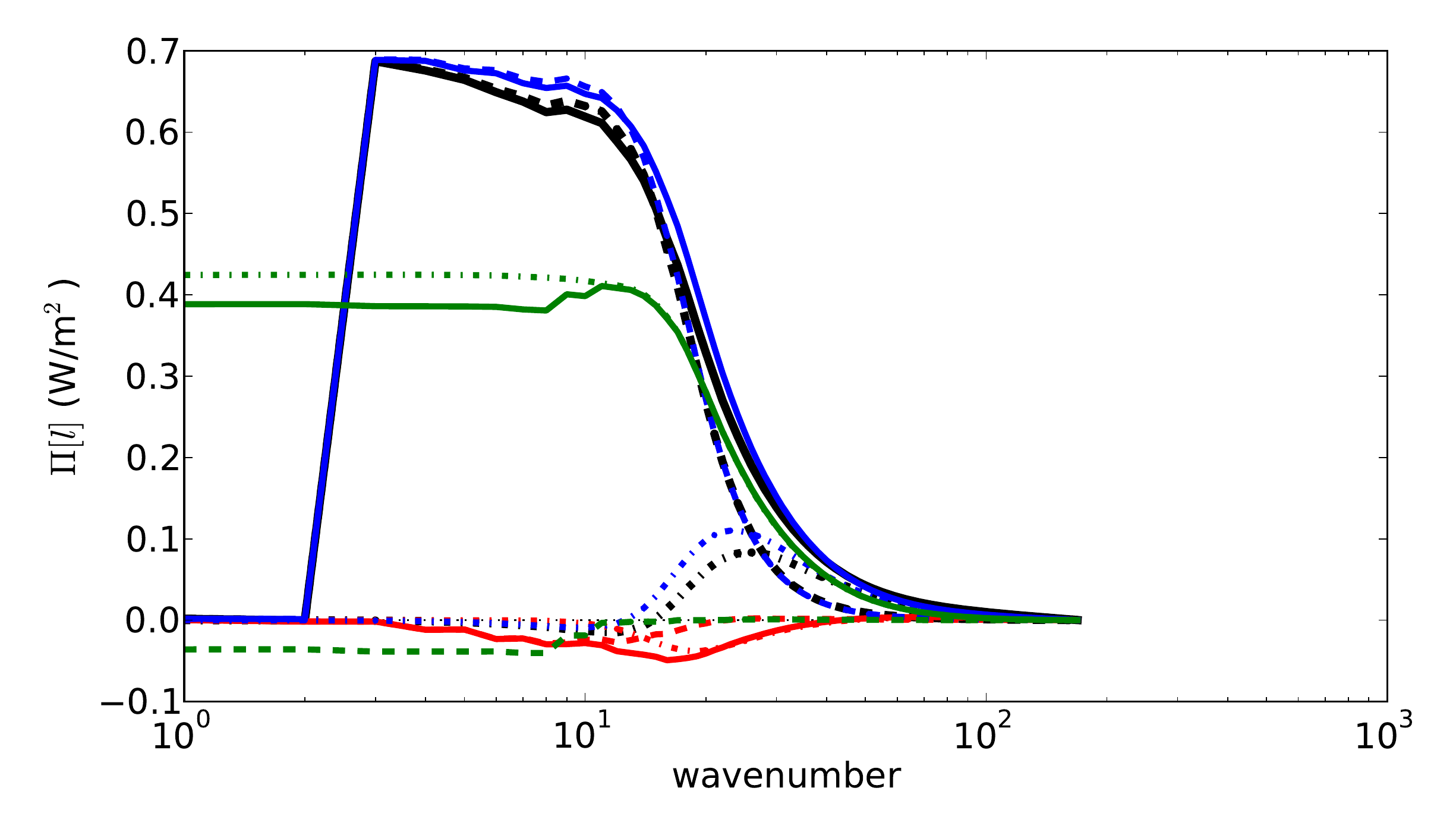}
}\\
\hspace{-4.5cm}\textbf{g)} $\Omega^*=4$, T170 Resolution \hspace{5cm} \textbf{h)} $\Omega^*=8$, T170 Resolution \\
\makebox[1.0\textwidth][c]{
\includegraphics[width=\columnwidth,height=5cm]{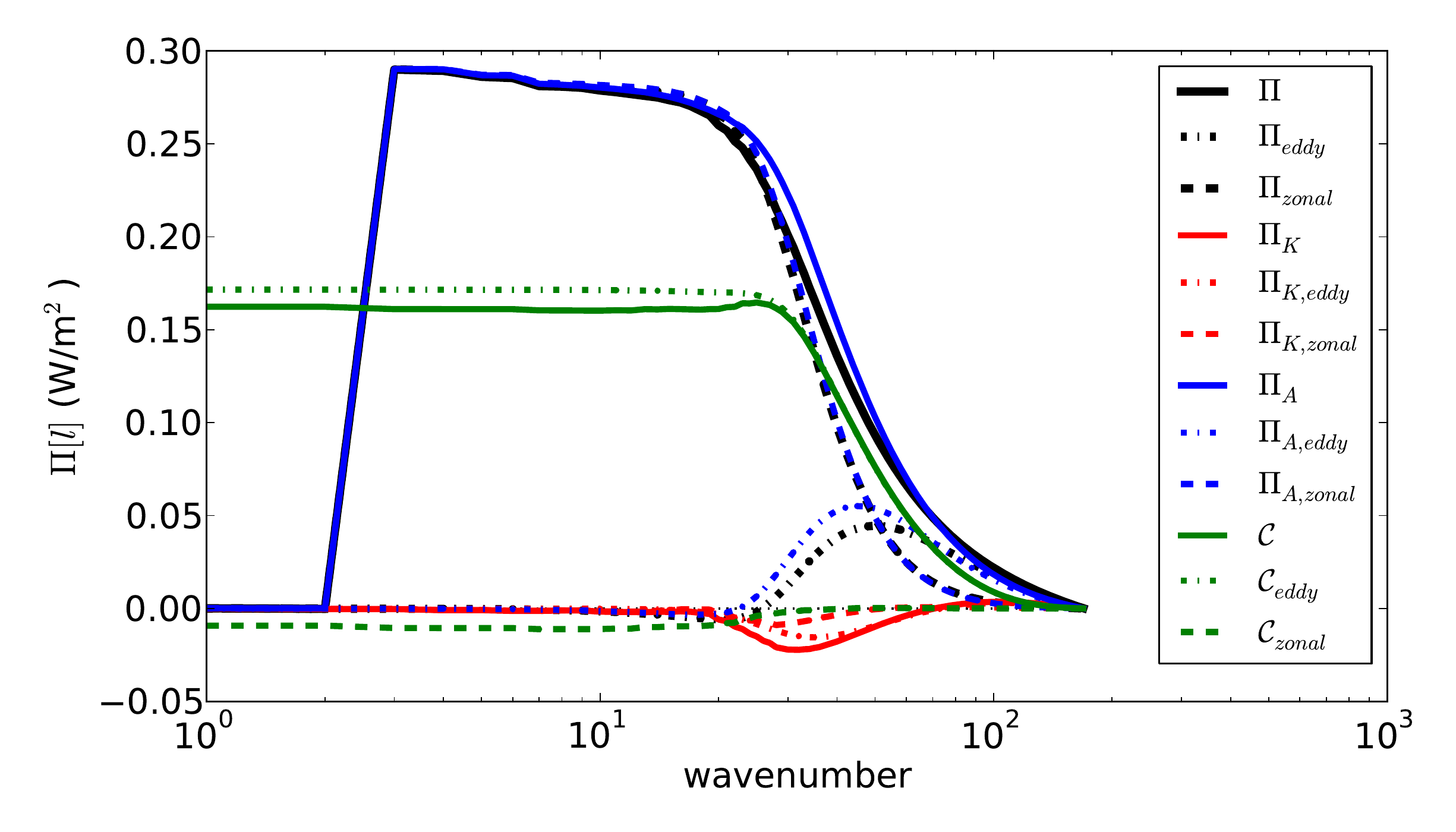}
\includegraphics[width=\columnwidth,height=5cm]{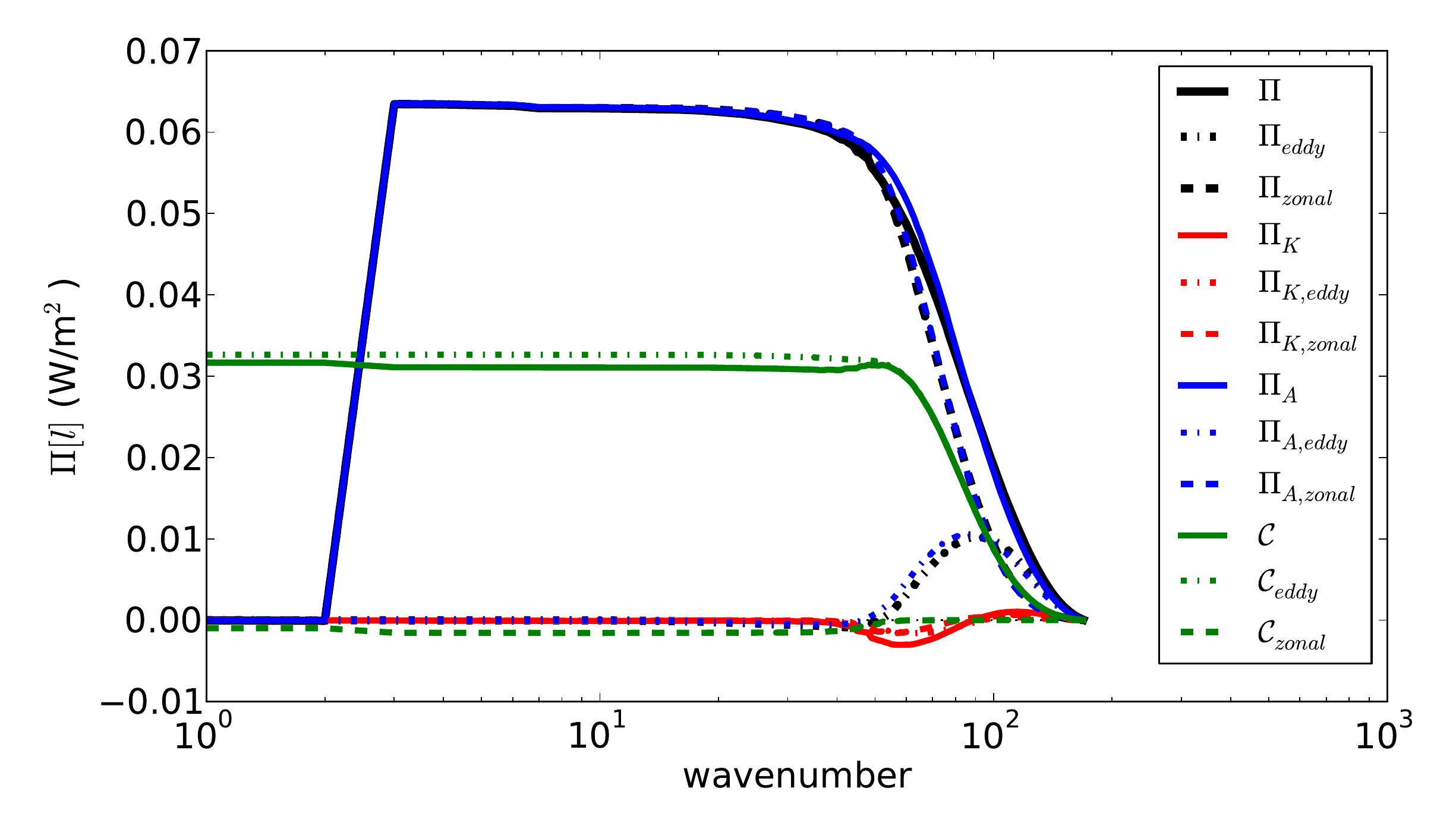}
}
\caption{Spectral fluxes of KE $\Pi_K$, APE $\Pi_A$ and total energy $\Pi=\Pi_A+\Pi_K$ as well as  conversion $\mathcal{C}$ (each decomposed into eddy-eddy and residual zonal interaction components) for PUMA-S runs  with $\Omega^\ast=1/16, 1/8, 1/4, 1/2$ (at horizontal resolution T42), $\Omega^\ast = 1$ (at resolution T127) and $\Omega^\ast=2, 4, 8$ (at resolution T170).}
\label{fig:flux1}
\end{figure*}

Figure~\ref{fig:flux1} shows spectral fluxes for KE ($\Pi_K$), APE ($\Pi_A$), and the total energy $\Pi=\Pi_K+\Pi_K$ as well as the cumulative conversion $\mathcal{C}$ from APE to KE for simulations across the full range of $\Omega^*$. The fluxes have been decomposed into eddy-eddy (``eddy'') and residual zonal (``zonal'')  interaction components {(as well as between rotational (non-divergent) and divergent components of the flow, shown in more detail for KE in Figure \ref{fig:flux1k} below)}. The terms presented here are integrated over the whole pressure range of the simulated atmospheres (see  Section~\ref{sec:sum}). The figure shows that in all cases the total energy flux $\Pi$ is always positive, signifying a downscale transfer (towards higher wavenumbers) of total energy. Potential energy fluxes $\Pi_A$ are also uniformly positive, indicative of downscale transfers, with some indications of inertial ranges (with fluxes independent of wavenumber) in some cases. 

\subsubsection{Spectral energy fluxes: $\Omega^*=1$}
For the Earth equivalent simulation at $\Omega^*=1$ with T127 resolution and normal friction (Fig.~\ref{fig:flux1}(e)), % and Figure \ref{fig:flux2}), 
the total energy flux $\Pi$ (black solid line) rises sharply at wavenumbers $n=2$ and 3 to a value of $\sim 1.6$ W m$^{-2}$, stays roughly constant until $n= 7$ before falling rapidly between wavenumbers 8 and 12 and then decreasing more slowly towards zero at the highest wavenumbers. $\Pi$ consists of two main components, of which the APE component, $\Pi_A$, dominates over the KE component, $\Pi_K$, up to a wavenumber of 50. Because of its larger magnitude, the trend of the APE flux $\Pi_A$ is similar to that of $\Pi$, except that its slope within the $\sim$constant region between $n = 3$ and 8 is less steep. This difference between $\Pi$ and $\Pi_A$ is the result of an upscale %(towards lower wavenumbers) 
energy transfer, $\Pi_K$, of KE  between $n = 3$ and $\sim$10-12. At around $n = 11$ there is an inflection point in the KE spectrum where $\Pi_K$ changes sign. This implies that kinetic energy is being transported towards smaller scales (i.e. larger wavenumbers) for  $n \gtrsim10-12$ and towards larger scales for $n \lesssim10$. In this region of the spatial spectrum, the baroclinic conversion, $\mathcal{C}$, has a steeply descending slope with wavenumber. This is a cumulative term (c.f. Eqn.~ \ref{eqn:cumu}) and so such a strong negative slope in $\mathcal{C}$ denotes a conversion of APE to KE in this wavenumber range of magnitude $ C_n=\mathcal{C}(n=15)-\mathcal{C}(n=7)\simeq0.9\:\text{W m$^{-2}$}$.

% schematic cascade plots
%\begin{figure*}
% \centering
%\includegraphics[width=\columnwidth]{new3_augier360_pumas/new3_augier360_withlabels/1omg_t127_zg_tg_ez_noKrotdiv.pdf}
%\includegraphics[width=\columnwidth]{new3_augier360_pumas/new3_augier360_withlabels/1omg_t127_zg_tg_ez_onlyK.pdf} \\
% (a) \hskip 8cm (b)
% \caption{Spectral energy fluxes for the experiment with $\Omega^{\ast}=1$ and normal friction at T127 resolution: (a) showing total energy fluxes and (b) showing just kinetic energy exchanges}
% \label{fig:flux2}
%\end{figure*}

%here halllo

%\comment{call it eddy-eddy and zonal (define at the beginning that zonal is residual and encompasses zonal-eddy and zonal-zonal things)   HOW CAN IT REFER TO ZONAL-ZONAL? should these not take place at wavenumber0?}
Regarding the partitioning between eddy-eddy and residual zonal components, the zonal components evidently dominate $\Pi_K$ and $\Pi_A$ at smaller wavenumbers ($1 \leq n \lesssim 20$) in Fig. \ref{fig:flux1}(e), while eddy-eddy components gain in
%for $\Pi_K$ and $\Pi_A$ zonal components dominate at smaller wavenumbers (1-20) and eddy components gain
 relative importance at higher wavenumbers ($n >20$), although the total fluxes there are relatively low. On the other hand, the main component of the conversion term occurs in the eddy-eddy component. Taking all of these points together the Earth-like case is evidently consistent with the defining behaviour of idealized baroclinic turbulence \citep[see e.g.][]{vallis2006}. At the injection wavenumber for KE (around the Rossby deformation radius wavenumber $n_D \sim 8-12$; see Fig. \ref{fig:flux1k}(e)), APE is converted into KE via the eddy-eddy component of $C_n$ (which is related to the baroclinic $C_E$ component of the Lorenz energy budget; see Fig.~\ref{lorenz_profiles}). The resulting KE is transported mostly upscale into the zonal component by an inverse barotropic conversion (cf $C_K$ in Lorenz budget), with a smaller amount of KE being transported downscale where the eddy-eddy interaction component dominates (c.f. Fig~\ref{fig:flux1k}(e)). %and Fig.\ref{fig:flux1k}(e)). %That the upscale KE flux resides in the eddy-mean component, shows that the energy transfer behaves like a waterfall  from  

At smaller wavenumbers, we see a zonal component also contributes to $\mathcal{C}$. The entire cumulative sum of $\mathcal{C}$ (i.e. the value depicted at wavenumber $n=1$) is comparable in sign and magnitude to $C_Z$ in the Lorenz energy budget (Fig.~\ref{lorenz_profiles}).%\footnote{Note that the data for the $\Omega^*=1$ plot in Fig.~\ref{lorenz_profiles} come from the run with T42 resolution.} 
This conversion shows that $C_Z$ is negative for wavenumbers $n=4-7$ and positive %$C_Z$ 
at the smallest wavenumbers. It is likely that the zonal-zonal components (associated with the Eulerian mean Hadley and Ferrell circulations) dominate at low wavenumbers, while zonal-eddy components are more significant at higher wavenumbers. The thermally-direct Hadley circulation, which dominates at low latitudes, leads to positive $C_Z$, which may account for the behaviour of $C_n$ for $n \leq 4$. But the Ferrell cell is generally thermally indirect, so would be expected to make a negative contribution to $C_Z$, as apparent for $n=4-7$. %\comment{Not sure what this means...  PLR: It is a bit unclear, I agree. But does your zonal component of C also include Zonal-zonal interactions? That will also directly include CZ terms from the Lorenz budget, but decomposes it spatially into spherical harmonic contributions. I suspect that the zonal-zonal components dominate at low wavenumbers and zonal components at higher wavenumbers?}

Segments of the spectrum where the spectral fluxes $\Pi_K$, $\Pi_A$ or $\Pi$ itself are approximately constant are identified as inertial ranges, and two such regions can be discerned in this case. Firstly, the region between wavenumbers $n=3$ and 8, where $\Pi_A$ is constant (and $\Pi$ and $\Pi_K$ are almost constant) may describe an inertial range characterised by forward baroclinic APE transfers and an inverse (rotational) barotropic KE flux. The second region lies at $n\simeq30-80$ (see Fig~\ref{fig:flux1k}e for a close-up of $\Pi_K$) with both forward APE and KE cascades. This second wavenumber region can be identified in the energy spectrum with an $n^{-3}$ slope in Fig.~\ref{fig:enspec1}e. The first region, however, is not so easily identifiable with features in the spectrum. %It seems that the energy spectra in Fig.~\ref{fig:enspec1}a have a slope steeper than $k^{-3}$ in this region \comment{hm?}.
%the first, however, can hardly be discerned and has an uncharacteristic slope of $k^{<-1}$. 
Further comparison with Fig.~\ref{fig:enspec1}e, however, confirms an association of the {$n^{-5}$} slope in zonal energies with a downscale flux of both KE and APE. {The narrow inertial range around $3 \leq n \leq 8$ in KE occurs mainly in the zonal-eddy component, which means that most of the energy jumps directly between the zonal mean wind and a range of energy-significant non-axisymmetric wavenumbers (more like a ``waterfall'' than a ``cascade''?). For the other inertial range for $30 \leq n \leq 80$, however, the transfers are more dominated by eddy-eddy interactions.  This means that the latter inertial range involves not only scales at which no dissipation occurs while cascading, but also scales where only weak interactions between the zonally symmetric flow and the respective eddy-scales occur.} %Thus, energy transport

\subsubsection{Spectral energy fluxes: rapidly rotating cases ($\Omega^* > 1$)}
\begin{figure*}[]
\centering
\hspace{-4.5cm}\textbf{a)} $\Omega^*=\frac{1}{16}$, T42 Resolution \hspace{5cm} \textbf{b)} $\Omega^*=\frac{1}{8}$, T42 Resolution \\
\makebox[1.0\textwidth][c]{
\includegraphics[width=\columnwidth,height=5cm]{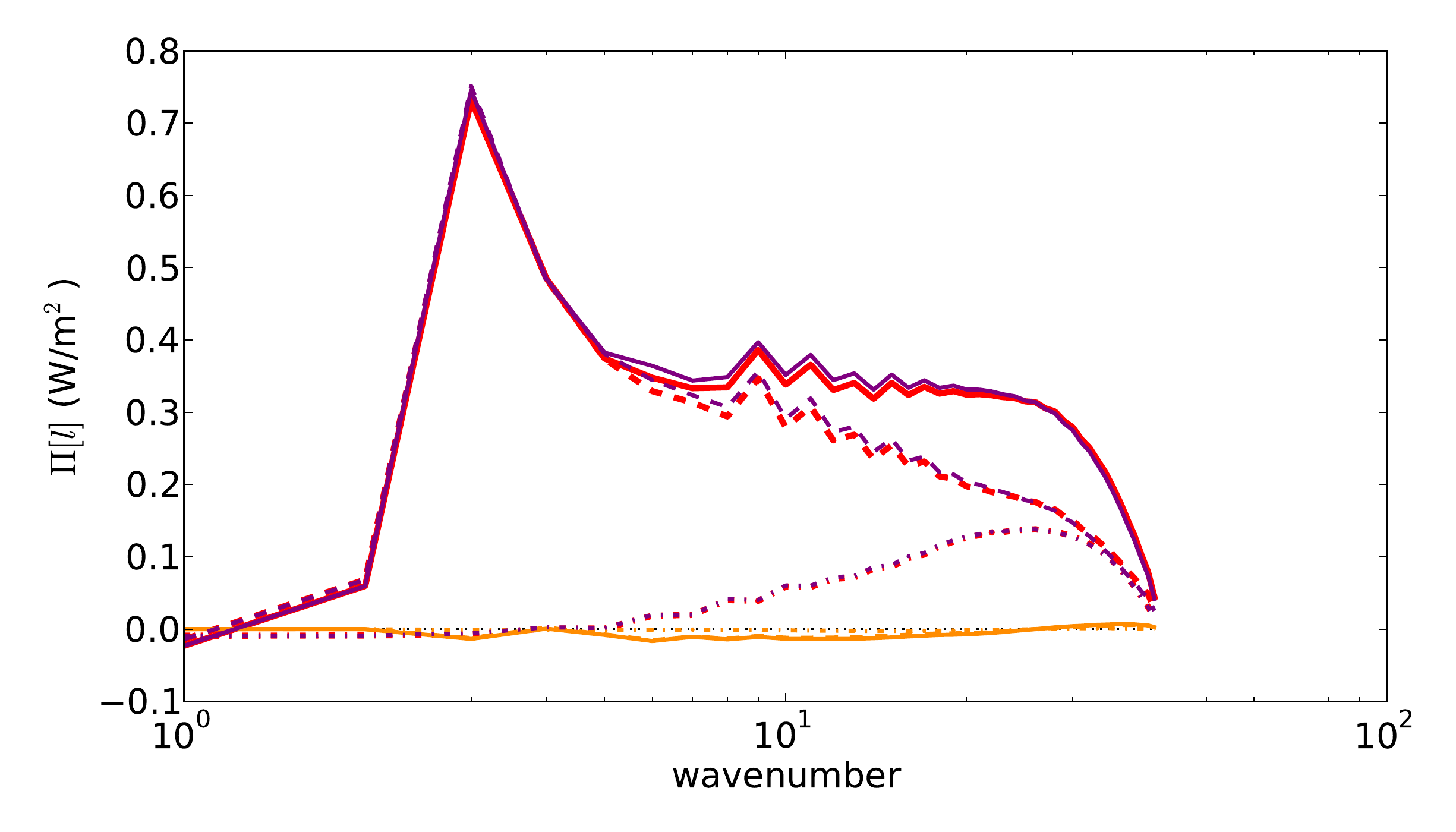}
\includegraphics[width=\columnwidth,height=5cm]{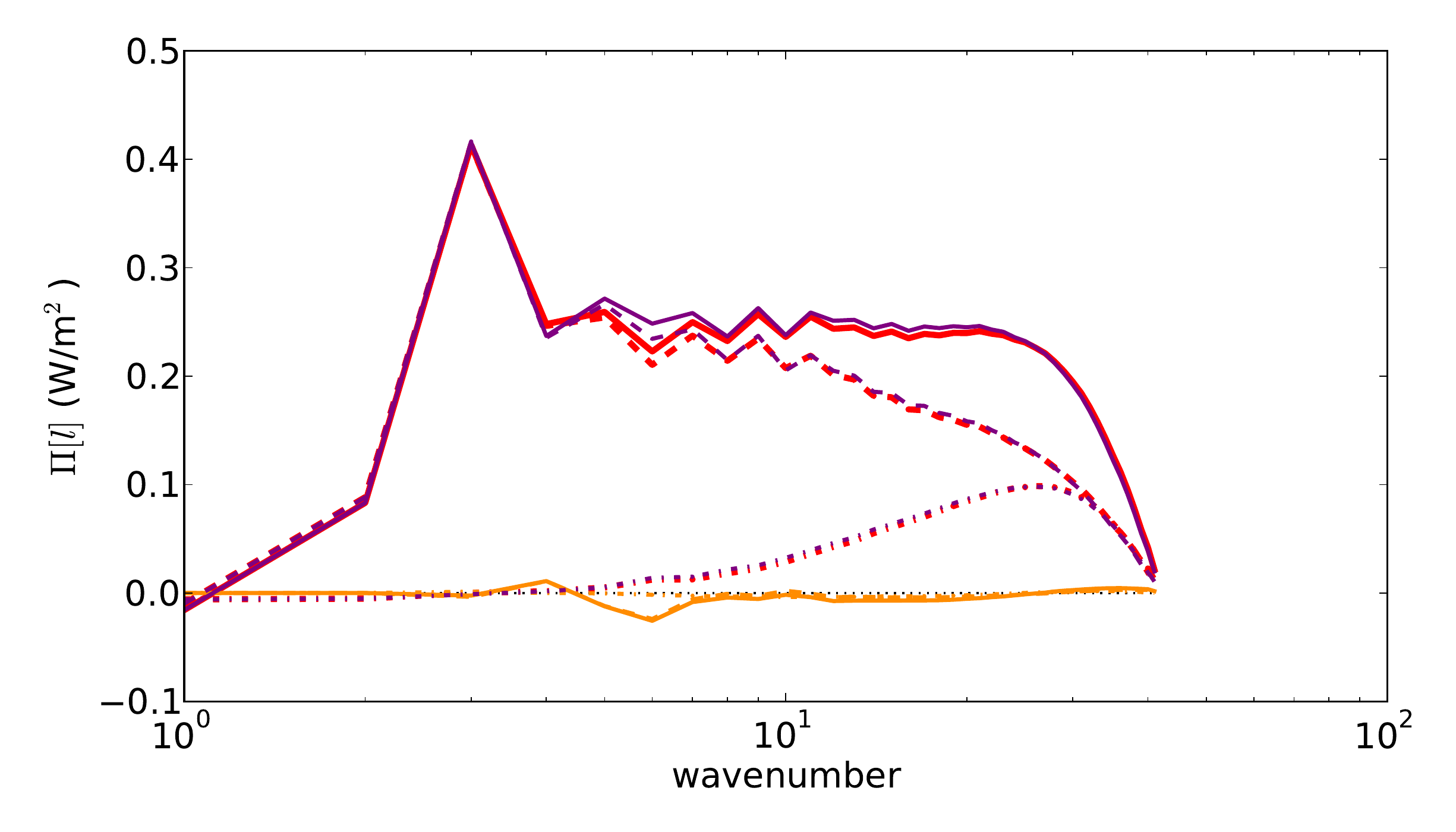}
}\\
\hspace{-4.5cm}\textbf{c)} $\Omega^*=\frac{1}{4}$, T42 Resolution \hspace{5cm} \textbf{d)} $\Omega^*=\frac{1}{2}$, T42 Resolution \\
\makebox[1.0\textwidth][c]{
\includegraphics[width=\columnwidth,height=5cm]{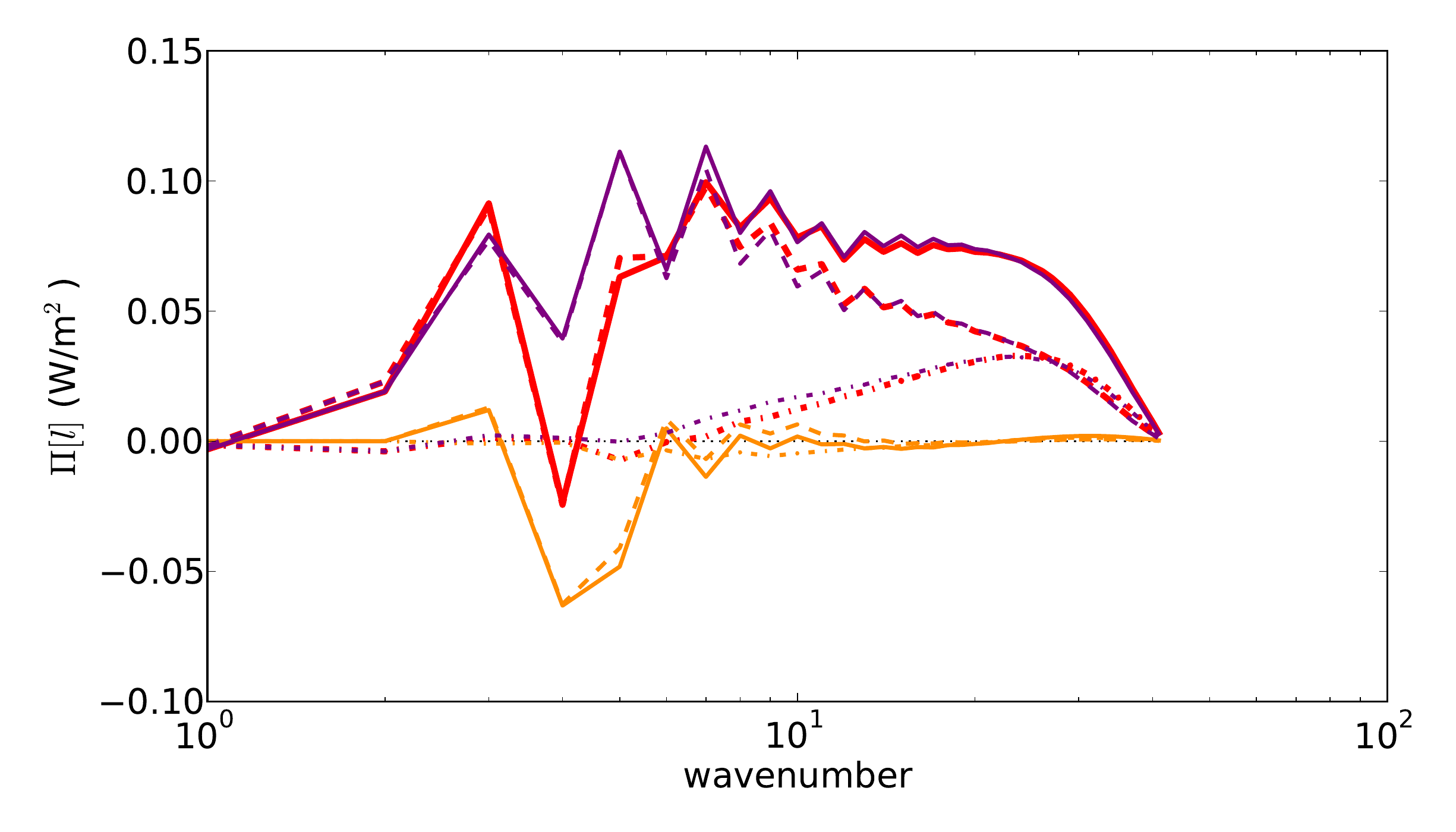}
\includegraphics[width=\columnwidth,height=5cm]{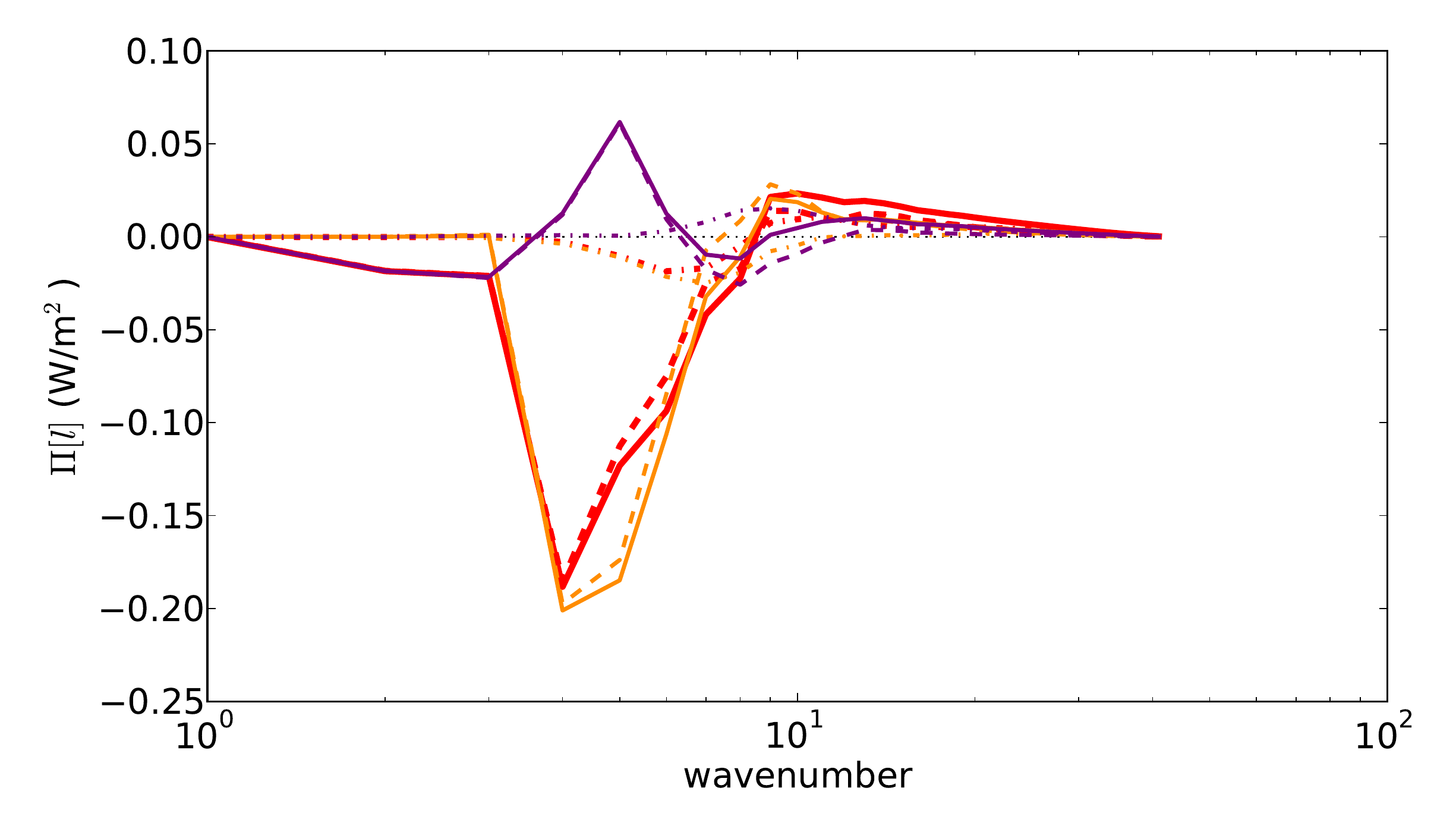}

}\\
\hspace{-4.5cm}\textbf{e)} $\Omega^*=1$, T127 Resolution \hspace{5cm} \textbf{f)} $\Omega^*=2$, T170 Resolution \\
\makebox[1.0\textwidth][c]{
\includegraphics[width=\columnwidth,height=5cm]{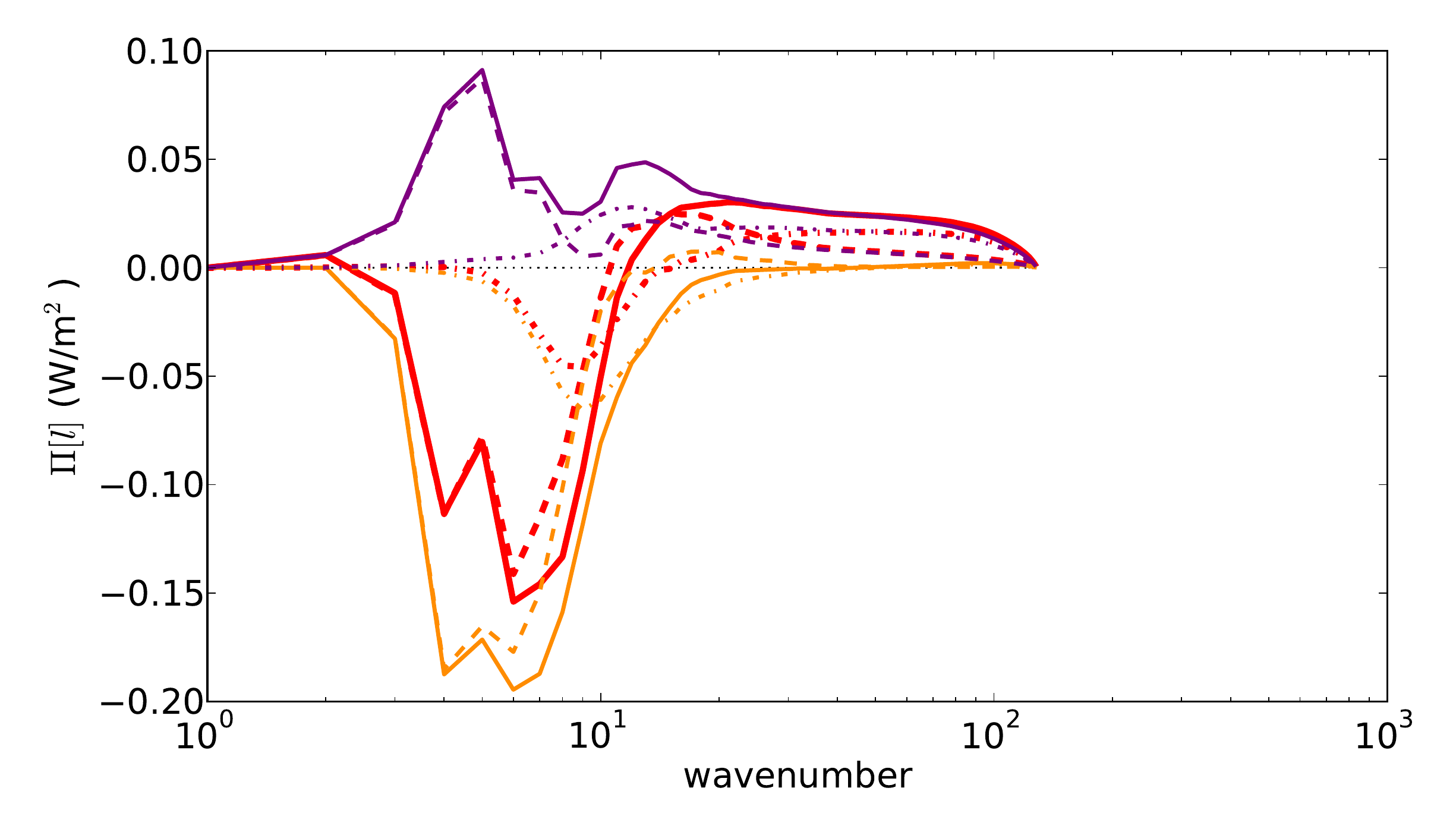}
\includegraphics[width=\columnwidth,height=5cm]{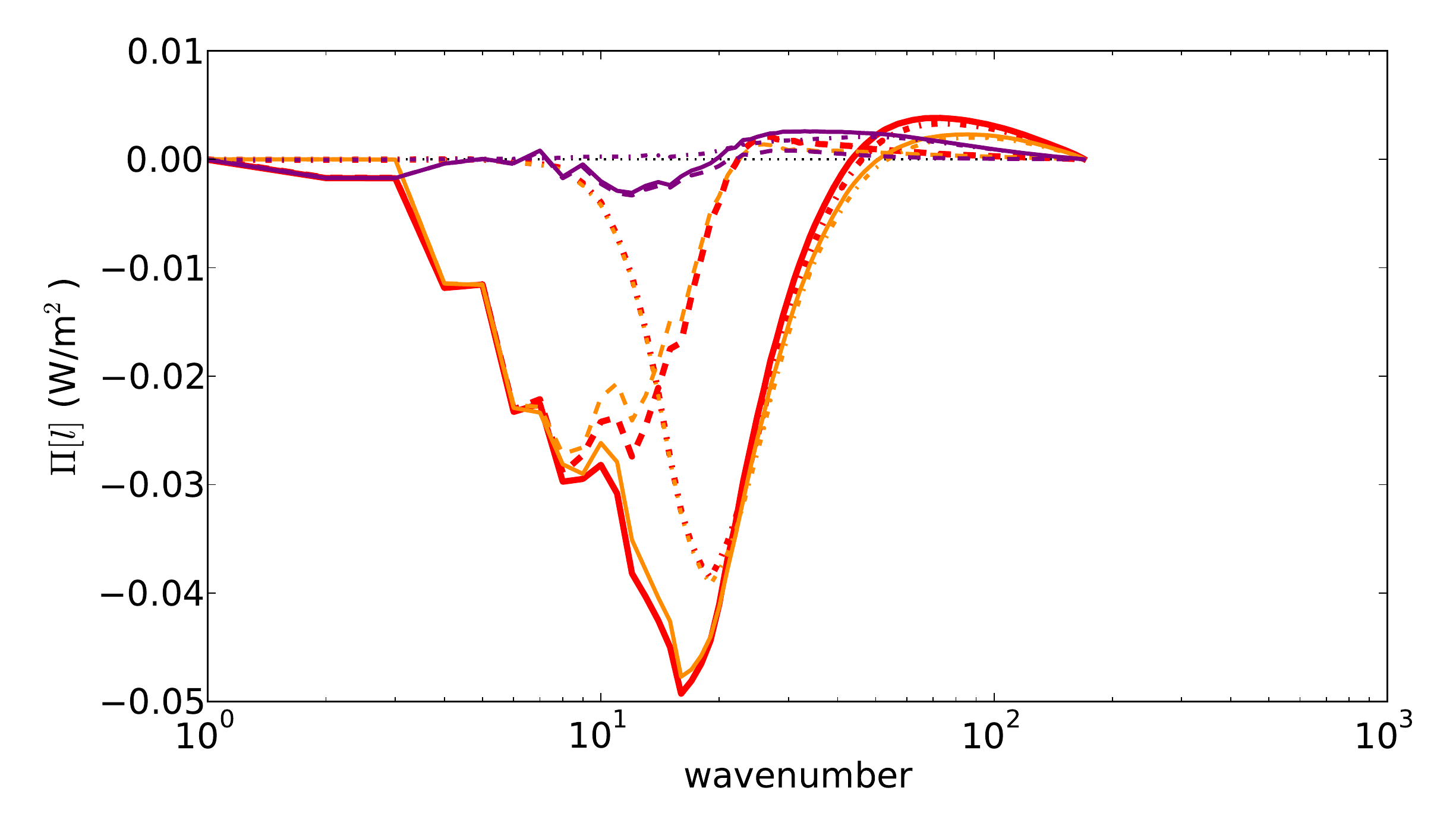}
}\\
\hspace{-4.5cm}\textbf{g)} $\Omega^*=4$, T170 Resolution \hspace{5cm} \textbf{h)} $\Omega^*=8$, T170 Resolution \\
\makebox[1.0\textwidth][c]{
\includegraphics[width=\columnwidth,height=5cm]{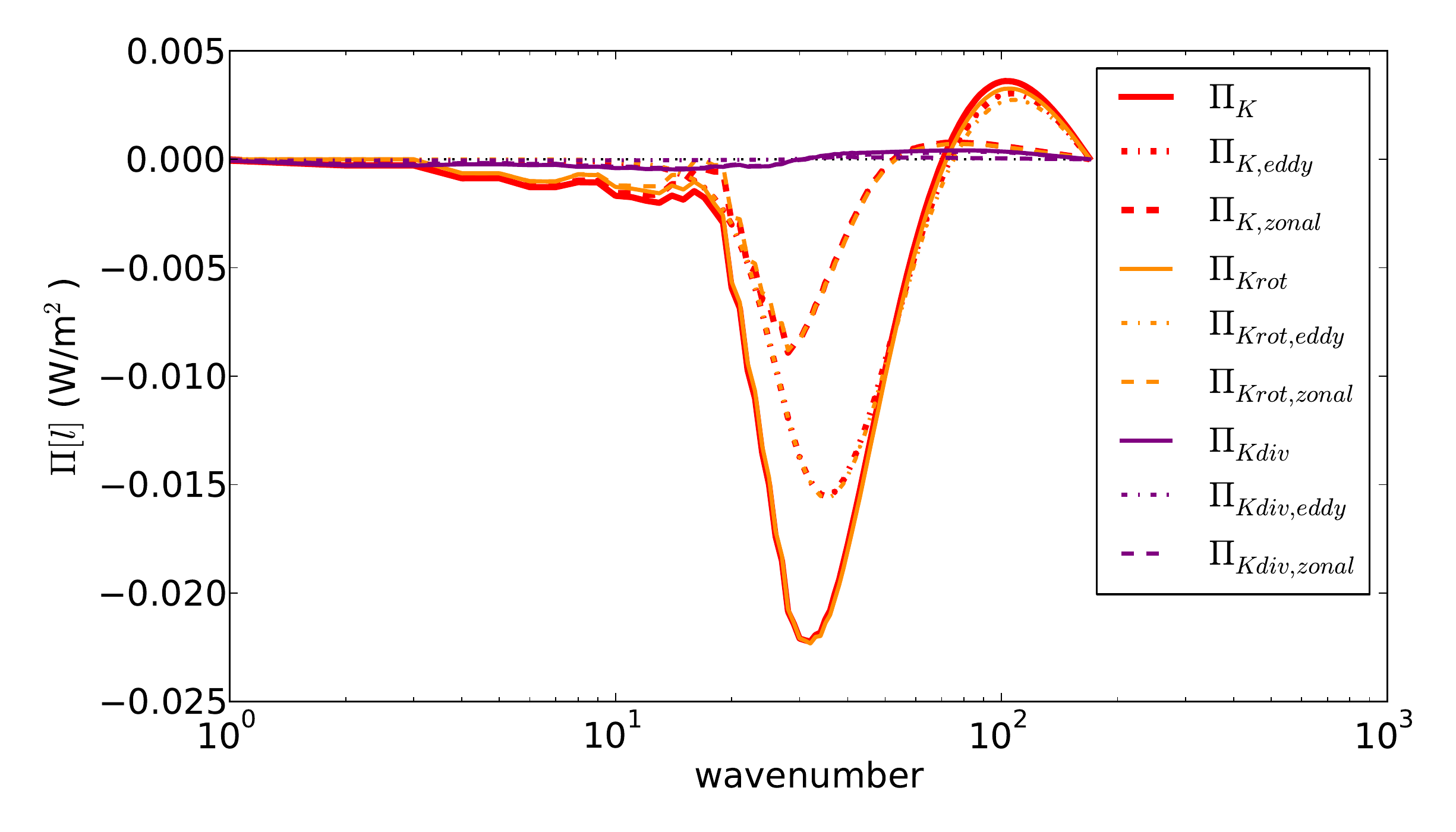}
\includegraphics[width=\columnwidth,height=5cm]{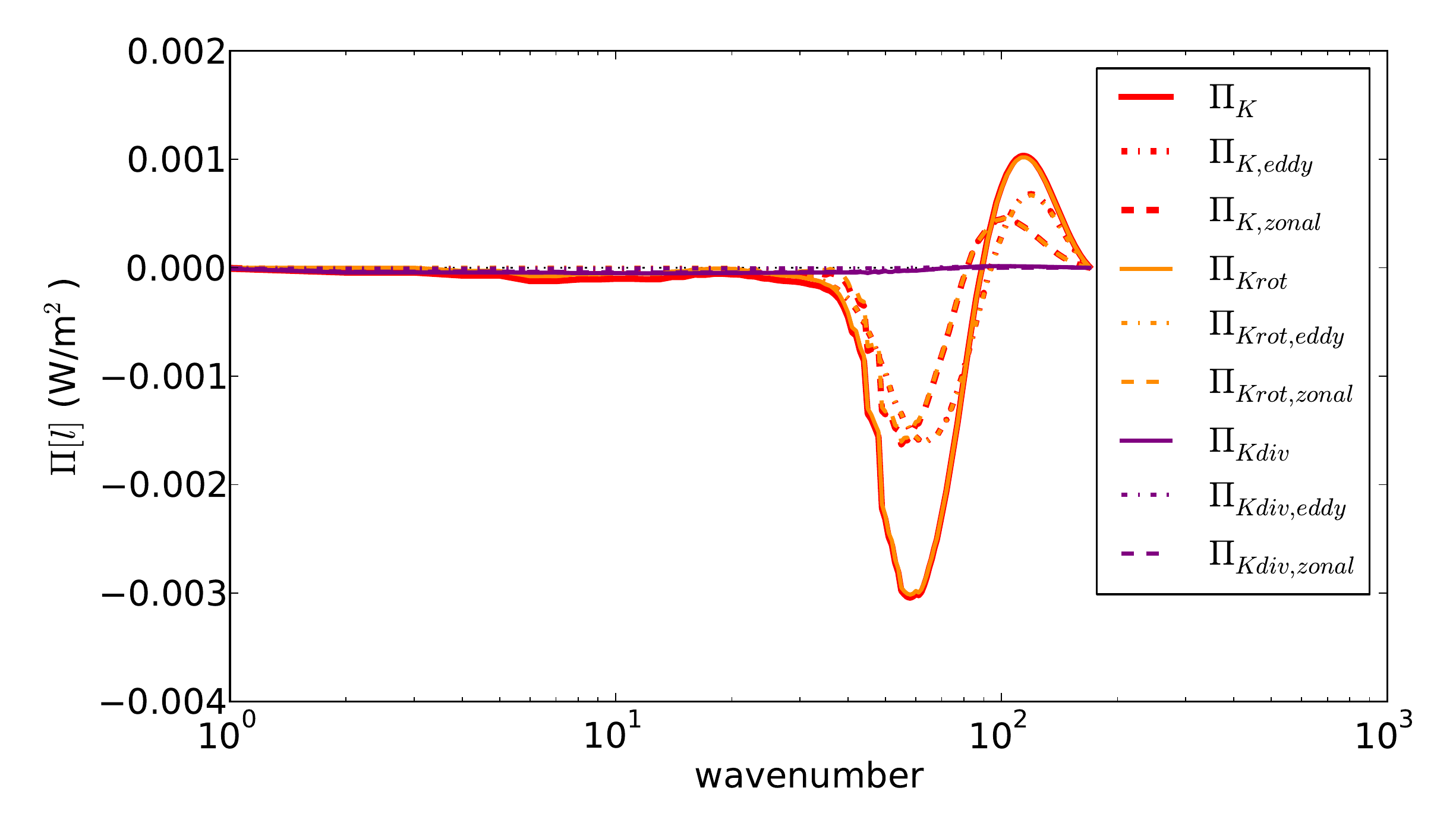}
}
\caption{Spectral fluxes of KE $\Pi_K$, decomposed into divergent and rotational component (each decomposed into eddy-eddy and residual zonal interaction components) for PUMA-S runs  with $\Omega^\ast=1/16, 1/8, 1/4, 1/2$ (at horizontal resolution T42), $\Omega^\ast = 1$ (at resolution T127) and $\Omega^\ast=2, 4, 8$ (at resolution T170).}
\label{fig:flux1k}
\end{figure*}

With increasing rotation rate (Figs.~\ref{fig:flux1}f-h) the Rossby deformation radius decreases (and $n_D$ increases) so that the wavenumber at which most baroclinic conversion occurs increases commensurately. At $\Omega^*\ge2$, the KE inertial range at large wavenumbers, seen for $\Omega^* = 1$ (see Fig. \ref{fig:flux1k}(e)) can no longer be discerned because it starts to close in on the resolution cut-off in these simulations at wavenumber $n=170$ where it becomes affected by the hyperdiffusion. However, the inertial range in APE flux widens and flattens with increasing rotation rate in a region that corresponds to a positive slope in the energy spectra (cf Figs.~\ref{fig:enspec1}(f)-(h)).

% (meaning KE is transported from larger wavenumbers to wavenumber 3-5).

%as their zonal  and

Figure~\ref{fig:flux1k}(e)-(h) shows the spectral kinetic energy flux for simulations with $\Omega^*=1,2,4,8$ in detail. As mentioned above, for $\Omega^*=1$ (Fig.~\ref{fig:flux1k}(e)) %and \ref{fig:flux2}(b)) 
we can see that the energy injected at the Rossby deformation lengthscale {through conversion from APE by baroclinic instability} is transferred both upscale and downscale (indicated by the red solid line). The upscale component can be identified with an upscale barotropic transfer of KE by the rotational part of the flow, which is dominated by the zonal interaction components. The downscale component at higher wavenumbers, however, is dominated by the divergent eddy-eddy interactions. 
%In this case a further decomposition into rotational and divergent components of the KE flux has been performed \citep[c.f.][their Fig.~1]{augier2013}.  %TODO ASK ROLAND WHATS UP WITH THAT?
%This separation can give information between the type of wave present in the atmosphere. For example, a pure Rossby wave only has a rotational component and a pure gravity wave would only have a divergent component. 
%We find that the purely rotational component  of the flux provides an approximately flat slope in the KE flux at the range of upscale transfer (wavenumbers 3-8), which makes this component an inertial range. However, the total KE flux at this range is offset by a downscale divergent component. In addition, this region is dominated by zonal interactions. 
%However, the spectral flux is not very flat at this range (wavenumbers 3-8), so calling this an inertial range may be farfetched. 
%For the downscale flux at larger wavenumbers (10-100) the divergent part of the flux becomes negligible  and only the rotational part contributes towards the total KE flux. In this range the flux is influenced by both the zonal and eddy-eddy interaction terms. This means that our simulation at $\Omega^*=1$ has both rotational and divergent waves.
With increasing rotation rate (Figs.~\ref{fig:flux1k}(f)-(h)), however, in contrast to the $\Omega^* = 1$ case, the divergent mode decreases sharply in magnitude so that, at the highest rotation rates, only the rotational part of the flux transfers energy in either direction. In addition, the contribution of the eddy-eddy interaction terms at larger wavenumbers becomes stronger. At the highest values of $\Omega^*$, therefore, the macroturbulent interactions are almost entirely dominated by the rotational flow, with the divergent eddies playing little role.

%conservative, so adds up to zero

%maximum value

%dominated by APE
\subsubsection{Spectral energy fluxes: slowly rotating cases ($\Omega^* < 1$)}

%\begin{figure*}[]
%\centering
%\hspace{-4.5cm}\textbf{a)} $\Omega^*=\frac{1}{2}$, T42 Resolution \hspace{5cm} \textbf{b)} $\Omega^*=\frac{1}{4}$, T42 Resolution \\
%\makebox[1.0\textwidth][c]{
%\includegraphics[width=\columnwidth]{for_peter/pumas_1omg_zg_tg_ez_noKrotdiv}
%\includegraphics[width=\columnwidth]{for_peter/pumas_05omg_zg_tg_ez_noKrotdiv}
%\includegraphics[width=\columnwidth]{for_peter/pumas_025omg_zg_tg_ez_noKrotdiv}
%}\\
%\hspace{-4.5cm}\textbf{c)} $\Omega^*=\frac{1}{8}$, T42 Resolution \hspace{5cm} \textbf{d)} $\Omega^*=\frac{1}{16}$, T42 Resolution \\
%\makebox[1.0\textwidth][c]{
%\includegraphics[width=\columnwidth]{for_peter/pumas_0125omg_zg_tg_ez_noKrotdiv}
%\includegraphics[width=\columnwidth]{for_peter/pumas_00625omg_zg_tg_ez_noKrotdiv}
%}\\
%\caption{Spectral fluxes of KE $\Pi_K$, APE $\Pi_A$ and total energy $\Pi=\Pi_A+\Pi_K$ as well as cumulative conversion $\mathcal{C}$ (each decomposed into  eddy-eddy and residual zonal interaction components) for PUMA-S runs with $\Omega^*=1,\frac{1}{2},\frac{1}{4},\frac{1}{8},\frac{1}{16}$ at T42 resolution.}
%\label{fig:flux2}
%\end{figure*}

Figure~\ref{fig:flux1}(a)-(d) shows the spectral energy fluxes for slowly-rotating simulations ($\Omega^*=\frac{1}{16},\frac{1}{8},\frac{1}{4},\frac{1}{2}$). %Figure~\ref{fig:flux2}a shows the Earth-like case again, but at the lower model resolution of T42. Although the magnitudes are slightly diminished compared to the T127 simulation (Fig.~\ref{fig:flux1}a), it is remarkable how well the qualitative behaviour of the fluxes match. This qualitative similarity shows that, even though the T42 simulation does not feature a properly identifiable slope in the energy spectra (Fig.~\ref{fig:enspec1}a) due to model inherent diffusion, the same spectral fluxes occur nonetheless.
With decreasing rotation rate, %(Figs.~~\ref{fig:flux1}b-e),  
the baroclinically active region (i.e. with downscale $\Pi_A$, and negative slope in $\mathcal{C}$), identified in the previous section, moves towards smaller wavenumbers. Between $\Omega^\ast = frac{1}{8}$ and $\frac{1}{4}$, however, this baroclinically dominated behaviour is suppressed, giving way to a quite different pattern of fluxes at the lowest values of $\Omega^*$. This trend is consistent with that found by \cite{mitchell2010}, who observed that their super-rotating simulated circulations, unlike Earth-like cases, were not dominated by baroclinic zonal-eddy interactions, as indicated by a lack of divergence of the vertical component of the EP-fluxes \citep[c.f. ][their Fig.~7]{mitchell2010}.  In addition, the zonal components of $\mathcal{C}$, which were comparatively small at higher values of $\Omega^*$, now begin to dominate at all lengthscales. This occurs because, at smaller rotation rates (larger values of $\mathcal{R}o_T \gtrsim 10$), the Rossby deformation lengthscale exceeds the planetary radius and APE is then injected directly into the KE reservoir at very low wavenumbers directly via interactions with the zonal mean flow. $\mathcal{C}_{zonal}$ at $n = 1$ is again very similar to $C_Z$ in the corresponding Lorenz budget (cf Fig. \ref{lorenz_profiles}), which points towards a strong influence of zonal-zonal interactions in this conversion term.  %\comment{PLR: Could this again be due to zonal-zonal conversion (i.e. the CZ term in Lorenz)?}. 

At $\Omega^\ast =\frac{1}{16}$ and $\frac{1}{8}$, the qualitative structure of the fluxes is therefore entirely different to the more quasi-geostrophic cases at higher $\Omega^\ast$. Conversion from APE to KE now occurs at the smallest wavenumbers, principally via zonal interactions. In addition both $\Pi_K$ and $\Pi_A$ now feature a well developed inertial range in the form of a forward transfer with an approximately constant spectral flux between wavenumbers $n=6$ and 30. This is indicative of a forward barotropic ``waterfall''. In both cases the zonal-eddy interactions dominate. However, the influence of eddy-eddy interactions is still evident and still increases in magnitude at larger wavenumbers.

%\begin{figure*}[]
%\centering
%\hspace{-4.5cm}\textbf{a)} $\Omega^*=\frac{1}{2}$, T42 Resolution \hspace{5cm} \textbf{b)} $\Omega^*=\frac{1}{4}$, T42 Resolution \\
%\makebox[1.0\textwidth][c]{
%\includegraphics[width=\columnwidth]{for_peter/pumas_1omg_zg_tg_ez_onlyK}
%\includegraphics[width=\columnwidth]{for_peter/pumas_05omg_zg_tg_ez_onlyK}
%\includegraphics[width=\columnwidth]{for_peter/pumas_025omg_zg_tg_ez_onlyK}
%}\\
%\hspace{-4.5cm}\textbf{c)} $\Omega^*=\frac{1}{8}$,  T42 Resolution \hspace{5cm} \textbf{d)} $\Omega^*=\frac{1}{16}$, T42 Resolution \\
%\makebox[1.0\textwidth][c]{
%\includegraphics[width=\columnwidth]{for_peter/pumas_0125omg_zg_tg_ez_onlyK}
%\includegraphics[width=\columnwidth]{for_peter/pumas_00625omg_zg_tg_ez_onlyK}
%}\\
%\includegraphics[width=0.7\columnwidth]{for_peter/empty}
%}
%\caption{Spectral fluxes of KE $\Pi_K$, decomposed into divergent and rotational component (each decomposed into eddy-eddy and residual zonal interaction components) for PUMA-S runs with $\Omega=1,\frac{1}{2},\frac{1}{4},\frac{1}{8},\frac{1}{16}\Omega_E$ at T42 resolution.}
%\label{fig:flux2k}
%\end{figure*}

Figure~\ref{fig:flux1k}(a)-(d) again features the kinetic energy flux in detail. In the case of decreasing $\Omega^\ast$, it is the rotational component that diminishes and the divergent component of the flux that controls the forward energy cascade. This suggests a much greater role for gravity and equatorial inertia-gravity planetary waves as these do not possess a rotational component. This would not be unduly surprising given that the equatorial waveguide grows in width at low rotation rates to span much of the planet.  %\comment{PLR: What does this mean? Greater role for gravity and equatorial inertia-gravity planetary waves?}

The behaviour identified in this section fits well with other results obtained for the large thermal Rossby number regime ($\mathcal{R}o_T >>1$). For $\Omega^\ast = \frac{1}{8}$ and $\frac{1}{16}$, the baroclinic conversion becomes weak and barotropic effects become stronger (also apparent in the corresponding Lorenz energy budgets; Fig.~\ref{lorenz_profiles}), such that $\Pi_K > \Pi_A$ in this regime. The flow becomes largely zonal and super-rotating flow emerges. Unfortunately, this analysis does not help directly in identifying the mechanism of formation and maintenance of the equatorial superrotation as this occurs mostly in the zonal component in a specific region of the globe, whereas this analysis computes over a global mean and focusses on the non-zonal spherical wavenumber spectrum. What we can learn, however, is that the kinetic energy in the zonal mode of super-rotating cases dissipates via a downscale cascade that involves both zonal-eddy and eddy-eddy interactions, with the latter dominating at high wavenumbers. 

%\subsection{Conclusion}

%We have shown that our simulations behave largely like the geostrophic turbulence theory in the fast-rotating regime ($\Omega^*\gtrsim\frac{1}{2}$). However, instead of incrementally cascading flow, we mostly find direct ``waterfalls'' between the zonal flow and different wavenumbers. 

%At fast rotation rates the rotational component of the KE spectral flux  dominates, while at slow rotation rates the divergent component dominates. The turnover point between these two components occurs between $\Omega^*=\frac{1}{2}$ and $\Omega^*=\frac{1}{4}$. Hence one could argue that a fundamental change in wave activity occurs here, which is likely as this is also roughly the cross-over point between geostrophic and cyclostrophic balance.  In addition, Mars lies near this point in parameter space, which would be interesting to explore in future work.

%This region in is parameter space is of great interest because this is roughly where Mars lies.  

%The fluxes presented in this section will act as a point of reference for the analysis of seasonal and diurnal effects in the following chapters.

\section{Discussion}\label{sec:conclusions}

This study has explored how the dynamical transfers of energy and vorticity between different horizontal scales depend upon the planetary rotation rate, at least as represented in a highly simplified, but nevertheless fully nonlinear and generic, numerical circulation model of a prototypical terrestrial planetary atmosphere. Such explorations are important sources of insight into the factors that determine the form, structure and intensity of atmospheric circulations under various conditions, thereby helping us to understand and quantify the similarities and differences between different planets of our own Solar System (and beyond), as well as indicating how aspects of any atmospheric circulation will scale with key planetary parameters. 

\subsection{Lorenz energy budgets}
Heat transport by both eddies and zonally-symmetric meridional overturning form important contributions to the overall energy budget of an atmosphere. This is commonly analysed using the framework originally developed by \citet{lorenz1967} and is still used as a source of insight for understanding the atmospheres of Earth and other planets \citep[e.g.][]{Peixoto1974,James1995,schubert2014,tabataba2015}. 

In the present work, we have computed how the various terms in the Lorenz energy budget for a simple, dry, Earth-like atmosphere vary with $\Omega^\ast$. Although the magnitudes of the zonal mean energy reservoirs vary monotonically with $\Omega^\ast$, with increasing dominance of APE over KE as $\Omega^\ast$ increases, the eddy energies rise to a maximum around the value of $\Omega^\ast$ where $\mathcal{R}o_T\sim 1$. This is also reflected in most of the conversion rates, which also peak in magnitude around a value of $\mathcal{R}o_T$ between 1 and 0.1. The trends in energy conversion rate also demonstrate the change in character of the dominant eddy generation processes from mainly barotropic processes at low rotation rates towards predominantly baroclinic processes at more rapid rotation rates, consistent with the onset of strong and deep baroclinic instabilities when $\mathcal{R}o_T \lesssim 1$. At much higher rotation rates, however, even baroclinic instability %is inhibited by rotation 
{becomes less effective at energy conversion} as the Rossby deformation radius becomes much smaller than the planetary radius, leading to a decrease in the intensity of the whole Lorenz energy cycle as $\Omega^\ast \rightarrow \infty$. 

This tendency of the Lorenz energy cycle to peak in intensity around conditions not too far from the Earth has been noted before, e.g. by \citet{pascale2013}. In their study, this was associated with a maximum in entropy production rates, although the precise conditions were found to depend not just on rotation rate but also on the strength of dissipation in the system. We have not sought to explore this in detail in the present study, but this would be of interest to investigate further in future work.

\subsection{Spectral energy budgets}
The results shown in Section \ref{sec:turb-fluxes} present for the first time a reasonably comprehensive overview of how the pattern of spectral fluxes of enstrophy and various forms of energy changes between different planetary circulation regimes. The simulation span a broad range of parameter space, extending from an extreme quasi-geostrophic limit through to a highly ageostrophic, super-rotating regime at very low rotation rates, over which the pattern of enstrophy and energy cascades changes significantly. 

Despite the use of a highly simplified global circulation model, the results for Earth-like conditions capture a circulation regime with a pattern of enstrophy and energy cascades that compares reasonably well with results from much more realistic models \citep[e.g.][]{burgess2013,augier2013,malardel2016}, at least qualitatively. The enstrophy fluxes at $\Omega^\ast = 1$ indicate a predominantly forward cascade over most wavenumbers with a flux that increases towards high wavenumbers in the baroclinically active troposphere. The magnitude of the enstrophy flux in the PUMA simulations is generally smaller than found e.g. by \citet{burgess2013} in their reanalysis data by a factor of $\sim 5$, but this likely reflects differences in the way the models are energised as well as effects of finite spatial resolution. %The fluxes of both energy and enstrophy in the PUMA simulations were also found to be sensitive to the assumed bottom friction parameterization, with weaker friction favouring the concentration of fluxes towards the smallest resolved scales.  
Energy fluxes at $\Omega^\ast = 1$ are broadly comparable to those found by \citet{augier2013} in their analyses of AFES and ECMWF simulations, though again somewhat smaller in magnitude. Vertically integrated, upscale rotational KE fluxes are around half the magnitude of those in both AFES and ECMWF simulations, while the forward cascade for $n > 20$, which is dominated in all simulations by divergent components, is weaker in the PUMA simulations by a factor $\sim 5$.  Available potential and total energy spectral fluxes, however, were quite comparable in magnitude in the PUMA Earth-like simulation to the NWP models at low-moderate wavenumber, though follows the AFES model more closely at high wavenumbers with positive (downscale) fluxes down to the resolution limit. This is consistent with the results of \citet{malardel2016}, who also found spectral fluxes to be significantly weaker in their cases with Held-Suarez (linear relaxation) forcing, suggesting that this approach underestimates the realistic energetic forcing of the simulated circulation. 

Resolution is likely to be a limiting factor for various features in the circulation. The KE and APE spectra for $\Omega^\ast = 1$ exhibit some features in common with the Earth's spectra \citep[e.g.][]{nastrom1985,burgess2013} in following an $n^{-3}$ trend over most of the spectrum for $n \lesssim 100$ in both KE and APE. With normal levels of surface drag, however, there is little evidence in the vertically-averaged spectrum from the PUMA simulations for the mesoscale break in the KE spectrum towards $n^{-5/3}$ around $n \gtrsim 20$. %With weaker surface friction, however, the results in Fig. \ref{fig:1omg-200tauf-jan-kesp} do indicate a flattening in the vertically integrated spectrum around $n = 20$ in association with enhanced amplitudes of divergent components of KE, much as inferred for the Earth. 
Dissipation may also play an important role in these simulations in removing energy at scales similar to where baroclinic energy conversion is taking place. These and similar effects from sub-gridscale parameterizations were noted in the study by \citet{malardel2016} and it would be of significant interest to explore this further under conditions that are significantly different from Earth.

As $\Omega^\ast$ is increased, the results show a gradual transition from the Earth-like pattern of spectral fluxes, in which both rotation and divergent KE components contribute to the cascades, towards a more rotationally dominated KE cascade at all scales. The magnitude of such fluxes quickly become quite weak as $\Omega^\ast$ is increased, probably due to relatively strong bottom friction. Nevertheless, the upscale segment at relatively low wavenumbers is dominated by the rotational flow, but at the highest rotation rates (and smaller values of $Ro_T$) the forward KE cascade also becomes dominated by rotational components. Such a pattern resembles more closely the distribution of spectral fluxes found in both the Earth's oceans \citep{scott2005,scott2007} and in Jupiter's atmosphere \citep{young2017}. Limited spatial resolution probably restricts the ability of the simulated flows to develop fully inertial ranges, so the KE and APE spectra are only marginally consistent with expectations of observing clear enstrophy and KE-dominated cascades. But the results are broadly consistent in this regime with some of the predictions of classical geostrophic turbulence theory, as summarised schematically in Figure \ref{fig:cascades1}(a) (though with some modifications discussed further below). These results include uniformly downscale transfers of APE and total energy, excitation of the KE spectrum around the deformation scale $n_D$ in association with baroclinic instabilities and barotropization, and near-equipartition between the APE and KE spectra at the highest values of $\Omega^\ast$. Under Earth-like conditions, however, some of these classical predictions are not borne out, in particular because of the significant role of divergent components of KE and because the deformation scale is not sufficiently well separated from the planetary scale. 

\subsection{Zonal jet formation}
\begin{figure}
 \centering
 \includegraphics[width=\columnwidth]{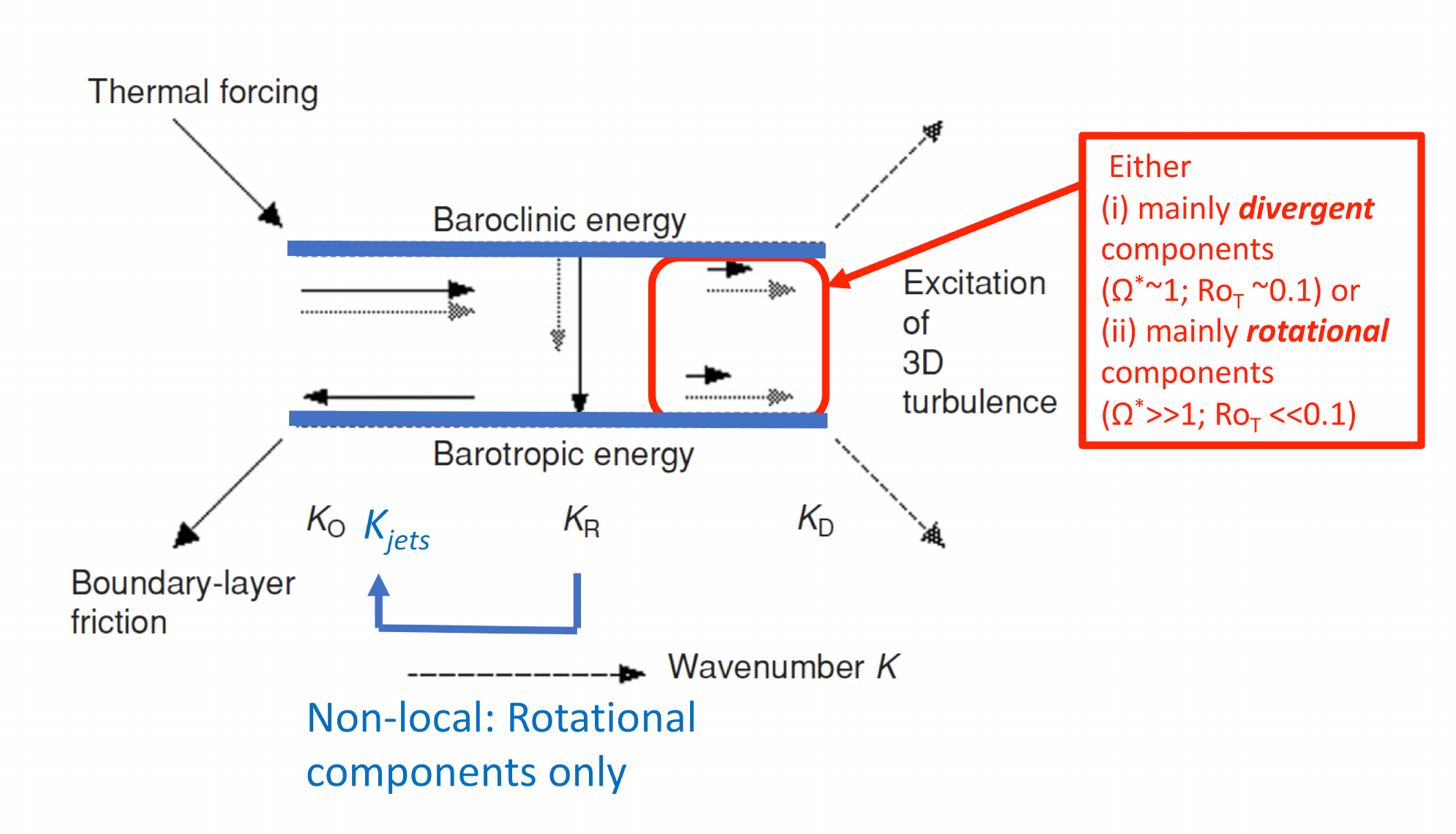}\\
 (a) \\
 \includegraphics[width=\columnwidth]{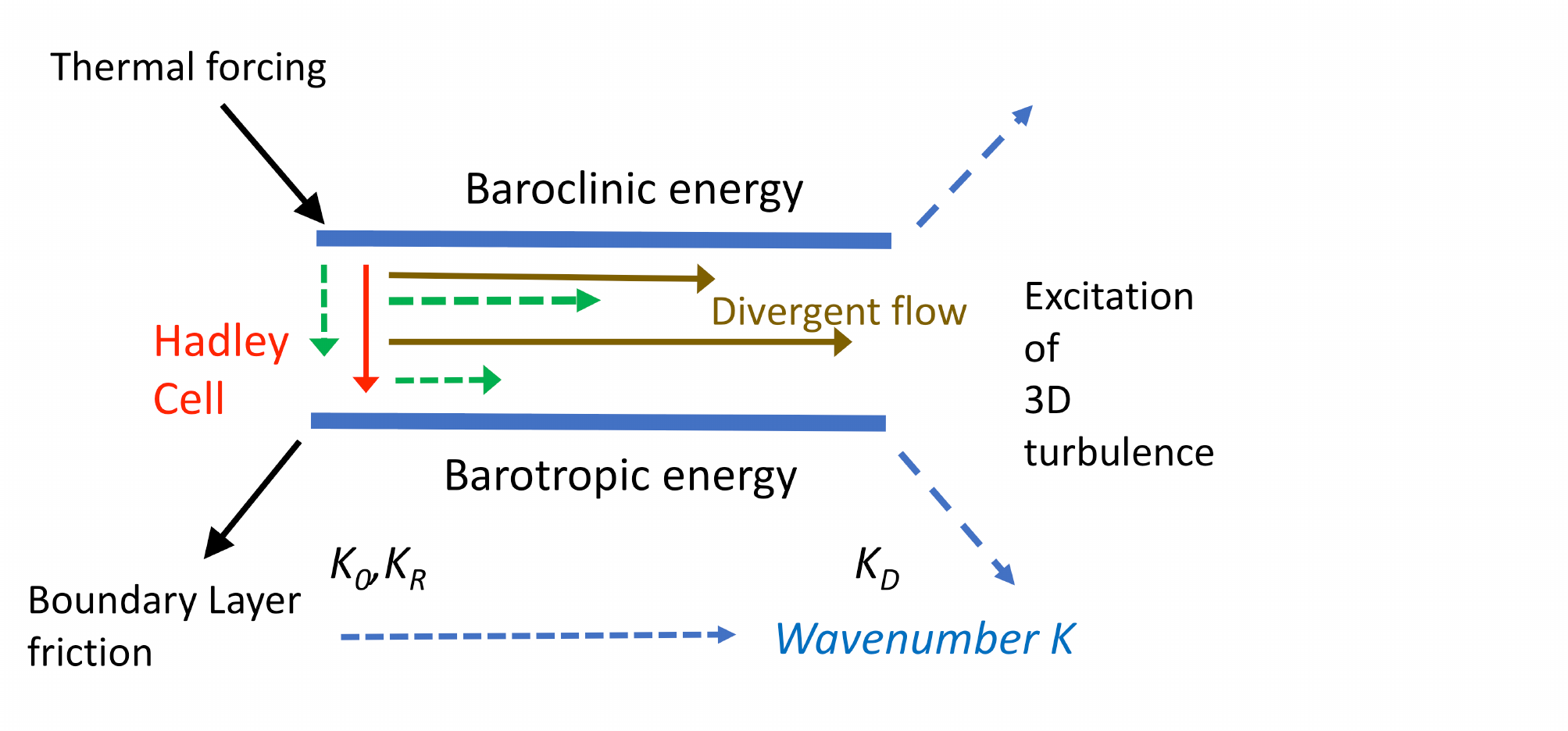} \\
 (b) \\%\hskip 8cm (b)
 \caption{Schematic representations of the cascades of APE (baroclinic energy) and KE (barotropic energy), following \citet{salmon1978} and \citet{salmon1980}, for (a) rapidly rotating, quasi-geostrophic atmospheres and (b) slowly rotating, stratified atmospheres. }
 \label{fig:cascades1}
\end{figure}

The results shown in Sections \ref{sec:turb-ke-spectra} and \ref{sec:turb-fluxes} also examine the jet formation mechanism in terms of KE spectra, with particular attention to the paradigm of \lq\lq zonostrophic turbulence\rq\rq\ recently proposed by 
Galperin and coworkers as a potential candidate for a universal regime for jet formation in various geophysical
fluids, including planetary atmospheres \citep[e.g.][]{Sukoriansky2002,Galperin2006,Galperin2010}. The experiments presented here demonstrate that, provided surface friction is not too strong, the atmosphere does develop strongly coherent zonal jets with a highly anisotropic KE spectrum that shares at least some features in common with the idealised zonostrophic turbulence regime (e.g. with $-5$ and $-5/3$ slopes respectively in the zonal and eddy KE spectra; \citet{Sukoriansky2002,Galperin2006,Galperin2010}). However, with relatively stronger surface friction, zonal jets appear in a weaker and more meandering/wavy form. This is consistent with the \lq\lq barotropic governor\rq\rq\ mechanism \citep[e.g.][]{James1995} which indicates that weak frictional damping leads to stronger barotropic shear 
in the atmosphere, thereby suppressing the growth of baroclinic instability, and making the circulation equilibrate into a more zonally symmetric state. This also evidently results in the accumulation of KE in predominantly barotropic zonal flows, mainly through non-local upscale transfers of KE directly from eddies into zonally symmetric flow components. This is an aspect of the transfer of energy between scales that was not considered in the early work of \citet{Charney1971} or \citet{salmon1980}, but is indicated in Fig. \ref{fig:cascades1}(a) to make the point that the upscale cascade is more complicated and anisotropic than initially considered.

% schematic cascade plots

\subsection{An Inertio-stratified regime}
At values of $\Omega^\ast < 1$, both the spectra and pattern of energy transfers undergo a significant change of regime, taking place around $Ro_T \simeq 1$. The resulting regime at the highest values of $Ro_T$, illustrated schematically in Fig. \ref{fig:cascades1}(b), has an entirely different character to any of the quasi-geostrophic sub-regimes. In this regime (which might be termed \emph{inertio-stratified turbulence}), the system is energized by differential heating at the planetary scale in zonally symmetric modes, which immediately begin cascading energy and enstrophy uniformly towards smaller scales following a direct conversion of zonally symmetric APE into KE. For $n \gtrsim 4$, the overall flow develops a clear inertial range that even our low spatial resolution simulations are able to represent quite well, whereby APE and KE cascade uniformly towards small scales with a near-constant spectral flux. In this regime, divergent KE components dominate the KE spectral flux, which in turn dominates over the APE flux. Both the KE and APE spectra also adopt a ``classical'' KBK form with a clear $n^{-5/3}$ slope until the resolution limit is approached. KE dominates the total energy spectrum, but the ratio of KE to APE appears to tend towards a fixed value $\simeq 2-3$. This is clearly distinct from any of the quasi-geostrophic regimes, but the precise details of which wave modes govern the properties of the cascade in both the horizontal and vertical directions remain to be explored. {This regime exhibits a number of similarities to the mesoscale and sub-mesoscale regimes in the Earth's atmosphere and oceans, recently identified by \citet{callies2014} and \citet{callies2013} respectively, in which relatively fast inertia-gravity waves take over the role of Rossby waves in classical QG turbulence. The KE and APE spectra, however, do not conform very closely to what we find in our model spectra.} As discussed also by \citet{wang2016}, these very slowly rotating circulations are dominated by strongly super-rotating zonal flows. Like the rapidly rotating, quasi-geostrophic regimes, therefore, they are likely to be characterized by highly anisotropic spectra and energy transfers which should be explored in more detail in future work.

Finally, together with the results presented on analyses of NWP models by \citet{augier2013}, our results demonstrate that spectral fluxes of energy and enstrophy provide a very clear and insightful approach to diagnosing the performance of numerical models. Even the limited results shown so far indicate some significant differences between different model formulations, which might be expected to lead to some helpful advances in model design, based on sound physical principles.

\ack PLR, FT-V, AV and RMBY acknowledge support  from the UK Science and Technology Facilities Council during the course of this research under grants ST/K502236/1, ST/I001948/1 and ST/K00106X/1. The  collaboration with Drs Pierre Augier and Erik Lindborg was made possible through the International Network on Waves and Turbulence, funded by the Leverhulme Trust. We are also grateful to Dhruv Balwada and an anonymous referee for their constructive comments on an earlier version of this paper.%[Others?]

%\href{http://www.sunrise-setting.co.uk}{\texttt{www.sunrise-setting.co.uk}}
\balance
%\nolinenumbers
\bibliographystyle{wileyqj}
\bibliography{atmosphere_EB,mars,refs}\label{refs}
\end{document}